\documentclass[longauth]{aa}  

\usepackage{comment}
\usepackage{graphicx}
\usepackage{multirow}
\usepackage{booktabs}
\usepackage{textgreek}
\usepackage{xargs}
\usepackage[dvipsnames]{xcolor}
\usepackage{xspace}
\usepackage{txfonts}
\usepackage{hyperref}
\hypersetup{
    colorlinks=true,
    linkcolor=blue,
    citecolor=blue,
    filecolor=magenta,      
    urlcolor=cyan,
    }
\usepackage{subcaption}
\usepackage{adjustbox}

\newcommand{\Te}{\ensuremath{T_\text{e}}\xspace}
\newcommand{\Tiii}{\ensuremath{t_3}\xspace}
\newcommand{\Tii}{\ensuremath{t_2}\xspace}

\newcommand{\target}{JADES-GS-z9-0\xspace}
\newcommand{\targetshort}{GS-z9-0\xspace}

\graphicspath{{./}{figs/}}

\usepackage{etoolbox}
\makeatletter
\newcommand\sendemail[3]{%                %\newcommand\tpj@compose@mailto[3]{%
\edef\@tempa{mailto:#1?subject=#2 }%
\edef\@tempb{\expandafter\html@spaces\@tempa\@empty}%
\href{\@tempb}{#3}}

\catcode\%=11
\def\html@spaces#1 #2{#1%20\ifx#2\@empty\else\expandafter\html@spaces\fi#2}
\catcode\%=14
\makeatother

\newcommand{\citationneeded}{\textcolor{ForestGreen}{$^{\rm citation\;needed}$}}
\let\oldtextsigma\textsigma
\renewcommand{\textsigma}{\oldtextsigma\xspace}
\let\oldtextalpha\textalpha
\renewcommand{\textalpha}{\oldtextalpha\xspace}
\let\oldAA\AA
\renewcommand{\AA}{\text{\oldAA}\xspace}
\let\oldtextdegree\textdegree
\renewcommand{\textdegree}{\oldtextdegree\xspace}

\newcommand{\orcid}[2]{\href{http://orcid.org/#2}{#1{\includegraphics[height=10pt]{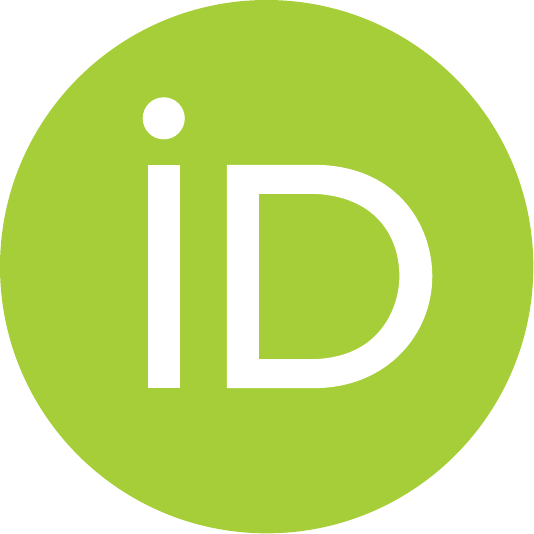}}}}

\newcommand{\kms}{\ensuremath{\mathrm{km\,s^{-1}}}\xspace}
\newcommand{\MSun}{\ensuremath{{\rm M}_\odot}\xspace}
\newcommand{\yr}{\ensuremath{{\rm yr}}\xspace}
\newcommand{\Myr}{\ensuremath{{\rm Myr}}\xspace}
\newcommand{\Gyr}{\ensuremath{{\rm Gyr}}\xspace}
\newcommand{\peryr}{\ensuremath{{\rm yr^{-1}}}\xspace}
\newcommand{\Lsun}{\hbox{\,${\rm L}_\odot$}}
\newcommand{\mum}{\text{\textmu m}\xspace}
\newcommand{\dex}{\text{dex}\xspace}
\newcommand{\kpc}{\text{kpc}\xspace}
\newcommand{\ZH}{\text{[Z/H]}\xspace}
\newcommand{\CO}{\text{[C/O]}\xspace}
\newcommand{\FeH}{\text{[Fe/H]}\xspace}
\newcommand{\percm}[1]{\ensuremath{\rm cm^{#1}}\xspace}

\newcommand{\eps}{\ensuremath{\epsilon}\xspace}
\newcommand{\mstar}{\ensuremath{M_\star}\xspace}
\newcommand{\mgas}{\ensuremath{M_\mathrm{gas}}\xspace}
\newcommand{\re}{\ensuremath{R_\mathrm{e}}\xspace}
\newcommand{\NH}{\ensuremath{N_\mathrm{H}}\xspace}
\newcommand{\tauv}{\ensuremath{\tau_\mathrm{V}}\xspace}
\newcommand{\AV}{\ensuremath{A_\mathrm{V}}\xspace}
\newcommand{\xid}{\ensuremath{\xi_\mathrm{d}}\xspace}
\newcommand{\logoh}{\ensuremath{12 + \log\,(\mathrm{O/H})}\xspace}

\newcommand{\nelec}{\ensuremath{n_\mathrm{e}}\xspace}
\newcommandx{\Mout}[2][1=,2=]{\ensuremath{M_{\mathrm{out}{#2}}^{#1}}\xspace}
\newcommandx{\Mdotout}[2][1=,2=]{\ensuremath{\dot{M}_{\mathrm{out}{#2}}^{#1}}\xspace}

\newcommandx{\fluxdcgs}[1][1=-20]{\ensuremath{\mathrm{10^{#1}~erg~s^{-1}~cm^{-2}~\AA^{-1}}}\xspace}
\newcommandx{\fluxcgs}[1][1=-20]{\ensuremath{\mathrm{10^{#1}~erg~s^{-1}~cm^{-2}}}\xspace}
\newcommandx{\powercgs}[1][1=44]{$\times 10^{#1}$~erg~s$^{-1}$\xspace}
\newcommand{\Av}{\ensuremath{A_V}\xspace}

\newcommand{\jwst}{{JWST}\xspace}
\newcommand{\hst}{{HST}\xspace}
\newcommand{\ppxf}{{\sc ppxf}\xspace}
\newcommand{\beagle}{{\sc beagle}\xspace}
\newcommand{\forcepho}{{\sc forcepho}\xspace}
\newcommand{\prospector}{{\sc prospector}\xspace}

\newcommand{\Lyalpha}{\text{Ly\textalpha}\xspace}
\newcommand{\Halpha}{\text{H\textalpha}\xspace}
\newcommand{\Hbeta}{\text{H\textbeta}\xspace}
\newcommand{\Hgamma}{\text{H\textgamma}\xspace}
\newcommand{\Hdelta}{\text{H\textdelta}\xspace}
\newcommand{\Pabeta}{\text{Pa\textbeta}\xspace}
\newcommand{\Hepsilon}{\text{H\textepsilon}\xspace}

\newcommandx{\permittedEL}[6][1=O,2=III,3=,4=,5=,6=]{\text{{#1}\,{\sc {#2}}{#3}{#4}{#5}{#6}}\xspace}
\newcommandx{\semiforbiddenEL}[6][1=O,2=III,3=,4=,5=,6=]{\text{{#1}\,{\sc{#2}}]{#3}{#4}{#5}{#6}}\xspace}
\newcommandx{\forbiddenEL}[6][1=O,2=III,3=,4=,5=,6=]{\text{[{#1}\,{\sc{#2}}]{#3}{#4}{#5}{#6}}\xspace}

\newcommand{\EW}[1]{\text{EW(#1)}\xspace}

\newcommand{\NV}{\permittedEL[N][v]}
\newcommandx{\NVL}[1][1=1243]{\permittedEL[N][v][\textlambda][#1]}
\newcommandx{\NVall}{\permittedEL[N][v][\textlambda][\textlambda][1239,][1243]}

\newcommand{\NIV}{\semiforbiddenEL[N][iv]}
\newcommandx{\NIVL}[1][1=1486]{\semiforbiddenEL[N][iv][\textlambda][#1]}

\newcommand{\CIV}{\permittedEL[C][iv]}
\newcommandx{\CIVL}[1][1=1550]{\permittedEL[C][iv][\textlambda][#1]}
\newcommand{\CIVall}{\permittedEL[C][iv][\textlambda][\textlambda][1549,][1551]}

\newcommand{\HeII}{\permittedEL[He][ii]}
\newcommandx{\HeIIL}[1][1=1640]{\permittedEL[He][ii][\textlambda][#1]}

\newcommand{\OIII}{\semiforbiddenEL[O][iii]}
\newcommandx{\OIIIL}[1][1=1666]{\semiforbiddenEL[O][iii][\textlambda][#1]}
\newcommand{\OIIIall}{\semiforbiddenEL[O][iii][\textlambda][\textlambda][1661,][1666]}

\newcommand{\OIIIopt}{\forbiddenEL[O][iii]}
\newcommandx{\OIIIoptL}[1][1=5007]{\forbiddenEL[O][iii][\textlambda][#1]}
\newcommand{\OIIIoptall}{\forbiddenEL[O][iii][\textlambda][\textlambda][4959,][5007]}

\newcommand{\NIII}{\semiforbiddenEL[N][iii]}
\newcommandx{\NIIIL}[1][1=1750]{\semiforbiddenEL[N][iii][\textlambda][#1]}
\newcommand{\NIIIall}{\semiforbiddenEL[N][iii][\textlambda][\textlambda][1747--][1754]}

\newcommandx{\CIII}{\semiforbiddenEL[C][iii]}
\newcommandx{\CIIIL}[1][1=1909]{\semiforbiddenEL[C][iii][\textlambda][#1]}
\newcommand{\CIIIall}{\semiforbiddenEL[C][iii][\textlambda][\textlambda][1907,][1909]}

\newcommandx{\SiIII}{\semiforbiddenEL[Si][iii]}
\newcommandx{\SiIIIL}[1][1=1883]{\semiforbiddenEL[Si][iii][\textlambda][#1]}
\newcommand{\SiIIIall}{\semiforbiddenEL[Si][iii][\textlambda][\textlambda]
[1883,][1892]}

\newcommand{\NeIV}{\forbiddenEL[Ne][iv]}
\newcommandx{\NeIVL}[1][1=2424]{\forbiddenEL[Ne][iv][\textlambda][#1]}
\newcommand{\NeIVall}{\forbiddenEL[Ne][iv][\textlambda][\textlambda][2422,][2424]}

\newcommand{\MgII}{\permittedEL[Mg][ii]}
\newcommandx{\MgIIL}[1][1=2803]{\permittedEL[Mg][ii][\textlambda][#1]}
\newcommand{\MgIIall}{\permittedEL[Mg][ii][\textlambda][\textlambda][2796,][2803]}

\newcommand{\NeV}{\forbiddenEL[Ne][v]}
\newcommandx{\NeVL}[1][1=3426]{\forbiddenEL[Ne][v][\textlambda][#1]}
\newcommand{\NeVall}{\forbiddenEL[Ne][v][\textlambda][\textlambda][3346,][3426]}

\newcommand{\OII}{\forbiddenEL[O][ii]}
\newcommandx{\OIIL}[1][1=3727]{\forbiddenEL[O][ii][\textlambda][#1]}
\newcommand{\OIIall}{\forbiddenEL[O][ii][\textlambda][\textlambda][3726,][3729]}

\newcommand{\NeIII}{\forbiddenEL[Ne][iii]}
\newcommandx{\NeIIIL}[1][1=3869]{\forbiddenEL[Ne][iii][\textlambda][#1]}
\newcommand{\NeIIIall}{\forbiddenEL[Ne][iii][\textlambda][\textlambda][3869,][39xx]}

\newcommand{\OIV}{\permittedEL[O][iv]}
\newcommandx{\OIVL}{\permittedEL[O][iv][\textlambda][\textlambda][1402,][1404]}

\newcommand{\hda}{\ensuremath{\mathrm{H\text{\textdelta}_A}}\xspace}
\newcommand{\hga}{\ensuremath{\mathrm{H\text{\textgamma}_A}}\xspace}

\begin{document} 

   \title{JADES: The star-formation and chemical enrichment history of a luminous galaxy at $z\sim9.43$ probed by ultra-deep \jwst/NIRSpec spectroscopy}
  % \title{Unveiling the nature of GS-z9, a luminous galaxy at $z\sim9.43$, with ultra-deep \jwst/NIRSpec spectroscopy: a young system with top-heavy IMF embedded in a fully neutral IGM}

   \titlerunning{\jwst ultra-deep spectroscopy of \target}

\author{Mirko Curti \inst{1}\fnmsep\thanks{E-mail: mirko.curti@eso.org}   
\and
Joris Witstok \inst{2,3}
\and
Peter Jakobsen \inst{4,5}
\and
Chiaki Kobayashi \inst{6}
\and
Emma Curtis-Lake \inst{6}
\and
Kevin Hainline \inst{7}
\and
Xihan Ji \inst{2,3}
\and
Francesco D'Eugenio \inst{2,3,8}
\and
Jacopo Chevallard \inst{9}
\and
Roberto Maiolino \inst{2,3,10}
\and
Jan Scholtz \inst{11,12}
\and
Stefano Carniani \inst{13}
\and
Santiago Arribas \inst{14}
\and
William M. Baker \inst{2,3}
\and
Rachana Bhatawdekar \inst{15}
\and
Kristan Boyett \inst{16,17}
\and
Andrew J.\ Bunker \inst{9}
\and
Alex Cameron \inst{9}
\and
Phillip A. Cargile \inst{18}
\and
Stéphane Charlot \inst{19}
\and
Daniel J.\ Eisenstein \inst{18}
\and
Zhiyuan Ji \inst{7}
\and
Benjamin D.\ Johnson \inst{18}
\and
Nimisha Kumari \inst{20}
\and
Michael V. Maseda \inst{21}
\and
Brant Robertson \inst{22}
\and
Maddie S. Silcock \inst{6}
\and
Sandro Tacchella \inst{2,3}
\and
Hannah \"Ubler \inst{2,3}
\and
Giacomo Venturi \inst{13}
\and
Christina C. Williams \inst{23}
\and
Christopher N. A. Willmer \inst{7}
\and
Chris Willott \inst{24}
}

\authorrunning{Curti et al.}
   \date{}

\institute{
European Southern Observatory, Karl-Schwarzschild-Strasse 2, 85748 Garching, Germany 
 \and 
Kavli Institute for Cosmology, University of Cambridge, Madingley Road, Cambridge, CB3 0HA, UK
 \and 
Cavendish Laboratory, University of Cambridge, 19 JJ Thomson Avenue, Cambridge, CB3 0HE, UK
 \and 
Cosmic Dawn Center (DAWN), Copenhagen, Denmark
 \and 
Niels Bohr Institute, University of Copenhagen, Jagtvej 128, DK-2200, Copenhagen, Denmark
 \and 
Centre for Astrophysics Research, Department of Physics, Astronomy and Mathematics, University of Hertfordshire, Hatfield AL10 9AB, UK
 \and 
Steward Observatory, University of Arizona, 933 N. Cherry Avenue, Tucson, AZ 85721, USA
 \and 
INAF -- Osservatorio Astronomico di Brera, via Brera 28, I-20121 Milano, Italy
 \and 
Department of Physics, University of Oxford, Denys Wilkinson Building, Keble Road, Oxford OX1 3RH, UK
 \and 
Department of Physics and Astronomy, University College London, Gower Street, London WC1E 6BT, UK
 \and 
Kavli Institute for Cosmology, University of Cambridge, Madingley Road, Cambridge, CB3 OHA, UK.
 \and 
Cavendish Laboratory - Astrophysics Group, University of Cambridge, 19 JJ Thomson Avenue, Cambridge, CB3 OHE, UK.
 \and 
Scuola Normale Superiore, Piazza dei Cavalieri 7, I-56126 Pisa, Italy
 \and 
Centro de Astrobiolog\'ia (CAB), CSIC–INTA, Cra. de Ajalvir Km.~4, 28850- Torrej\'on de Ardoz, Madrid, Spain
 \and 
European Space Agency (ESA), European Space Astronomy Centre (ESAC), Camino Bajo del Castillo s/n, 28692 Villanueva de la Cañada, Madrid, Spain
 \and 
School of Physics, University of Melbourne, Parkville 3010, VIC, Australia
 \and 
ARC Centre of Excellence for All Sky Astrophysics in 3 Dimensions (ASTRO 3D), Australia
 \and 
Center for Astrophysics $|$ Harvard \& Smithsonian, 60 Garden St., Cambridge MA 02138 USA
 \and 
Sorbonne Universit\'e, CNRS, UMR 7095, Institut d'Astrophysique de Paris, 98 bis bd Arago, 75014 Paris, France
 \and 
AURA for European Space Agency, Space Telescope Science Institute, 3700 San Martin Drive. Baltimore, MD, 21210
 \and 
Department of Astronomy, University of Wisconsin-Madison, 475 N. Charter St., Madison, WI 53706 USA
 \and 
Department of Astronomy and Astrophysics University of California, Santa Cruz, 1156 High Street, Santa Cruz CA 96054, USA 
 \and 
NSF’s National Optical-Infrared Astronomy Research Laboratory, 950 North Cherry Avenue, Tucson, AZ 85719, USA
 \and 
NRC Herzberg, 5071 West Saanich Rd, Victoria, BC V9E 2E7, Canada}

   \authorrunning{M. Curti et al.}
   \date{}

  \abstract

   \date{}

% \abstract{}{}{}{}{} 
% 5 {} token are mandatory
 
  \abstract
   {We analyse ultra-deep \jwst observations of the galaxy \target at $z=9.4327$, and derive detailed stellar and interstellar medium (ISM) properties of this luminous (M$_{\text{UV}}$=--20.43) high-redshift system.
   Complementary information from NIRCam imaging and NIRSpec (both low- and medium-resolution) spectroscopy reveal a compact system (R$_{e}\sim110$~pc) characterised by a steeply rising star formation history, which is reflected in the inferred young stellar age (t~$\sim3$~Myr, light-weighted), high star-formation rate surface density ($\Sigma_{\text{SFR}}\sim72$~\MSun yr$^{-1}$ kpc$^{-2}$), high ionisation parameter (log(U)~$\sim-1.5$), low metallicity (12+log(O/H)~$\sim7.5$), and low carbon-over-oxygen abundance ([C/O]~$=-0.64$). 
   Leveraging the detection of \NIIIL we derive nitrogen-over-oxygen abundance ([N/O]~$\sim0$) higher than the plateau followed by low-redshift galaxies of similar metallicity, possibly revealing the imprint from (very) massive stars on the ISM enrichment and favouring a top-heavy Initial Mass Function (IMF) scenario. Massive stars powering a hard radiation field are also required to explain the rest-frame UV line ratios, though the presence of the high-excitation  \NeVL emission line possibly hints at additional ionization from an AGN. 
   We also report the tentative detection of \Lyalpha emission in the G140M spectrum, shifted by $\sim450$~km/s redward of the systemic redshift.
   Combined with a modelling of the \Lyalpha spectral break, we rule out the presence of very-high column densities of neutral gas pertaining to local absorbers, as well as any extended surrounding ionised bubble, suggesting that \target has not yet significantly contributed to cosmic Reionization.}


   \maketitle
%
%________________________________________________________________

\section{Introduction}
%(Castellano et al. 2022a, 2023a; Finkelstein et al. 2022, 2023a,b; Harikane et al. 2023; Bouwens et al. 2023; Pe ́rez-Gonza ́lez et al. 2023; Chemerynska et al. 2023; McLeod et al. 2024)
The identification and characterisation of the earliest galaxies ever formed in the history of the Universe is one of the topics at the forefront of the current astrophysical research, and one of the main motivations behind the concept and development of the James Webb Space Telescope (\jwst).
Within already the first two cycles of operations, early results from extensive imaging and spectroscopic campaigns have not only pushed further the limits of the known redshift frontier \citep[e.g.][]{robertson_jades_2022, curtis-lake_2023, Arrabal_Haro_nature_2023, hainline_zgtr8_2024, castellano_ghz12_2024, carniani_z14_2024}, but marked unprecedented progress in the study of the physical properties of the early galaxy population, in terms of their number density \citep{harikane_uv_LF_2024,Chemerynska_LF_2023,robertson_JOF_LF_2023,mcleod_LF_2024}, star-formation histories (SFH) \citep{dressler_sfh_2023,endsley_sfh_2023, looser_SFH_2023, tacchella_eros_2022, clarke_SFMS_2024}, interstellar medium (ISM) conditions \citep{sanders_ceers_2023, cameron_jades_bpt_2023, reddy_ism_2023, calabro_ism_ghz2_2024}, incidence, growth, and impact of supermassive black holes \citep{harikane_agn_2023, greene_agn_uncover_2024, kokorev_agn_census_2024, maiolino_jades_agn_2023, scholtz_jades_agn_2023, Ubler_AGN_z5_2023, matthee_LRD_2024}
% (Greene et al. 2023; Kokorev et al. 2023; Larson et al. 2023; Maiolino et al. 2023c,a; Ubler et al. 2023; Scholtz et al. 2023).

One of the key advances provided by \jwst/NIRSpec resides in the possibility to simultaneously cover rest-frame UV and rest-optical spectra of galaxies at $z>6$.
Even prior to the advent of the \jwst, observations of rest-frame UV spectra in $z>6$ galaxies had exhibited large equivalent widths (EW), high-ionization emission lines, as seen in none but the most extreme galaxies in the local Universe \citep{berg_he2_civ_2019, izotov_2024}, suggesting that extreme radiation fields characterise galaxies in the Epoch of Reionization (EoR) \citep{stark_ciii_2015, stark_civ_2015,mainali_2017, senchyna_2017, hutchinson_2019}.
Leveraging the wide spectral coverage of \jwst/NIRSpec, it finally became feasible to combine diagnostic features pertaining to both spectral regions, to enable a more in-depth characterisation of the underlying ionising spectrum (aiding in deciphering whether this originates from metal-poor stellar populations or requires instead the hardness typical of active galaxies), as well as of the conditions of the ionised gas which produces the bright emission lines as seen in high-z galaxy spectra, in terms of its density, ionization structure, and chemical enrichment.
Despite the intrinsic weakness of rest-UV features still hampering the analysis of large galaxy samples, observations of some of the brightest sources have already revealed peculiar (and sometimes unexpected) ionization and chemical enrichment patterns \citep[e.g.][]{bunker_gnz11_2023, maiolino_gnz11_2023, cameron_gnz11_2023, isobe_CNO_2023, topping_z6_lens_2024, d_eugenio_gsz12_2023, Schaerer_nitrogen_z94_2024}. 
Indeed, chemical abundances provide some of the most relevant observational constraints for galaxy formation and evolution models \citep{maiolino_re_2019}.
% These heavy elements, formed through stellar evolution's nucleosynthesis, offer insights into the growth and history of galaxies (Matteucci, 2012; Maiolino & Mannucci, 2019). 
The different pathways in which heavy elements are produced by stars of different masses in fact are reflected in the differential timescales regulating the enrichment of the interstellar medium (ISM). 
Therefore, relative abundance ratios among different chemical species are powerful probes of the past history of mass assembly and star formation in galaxies.
% As elements with different nucleosynthetic origins are released into the interstellar medium (ISM) on different timescales, their relative abundance ratios are powerful probes of the past history of mass assembly and star formation in galaxies.

In a simple framework of galactic chemical evolution, $\alpha$-elements like oxygen or neon are primarily produced by massive stars (\mstar>~8~\MSun) and returned to the ISM relatively quickly through core-collapse supernovae (SNe), on timescales of approximately $10$ Myr. 
Although carbon is also generated in massive stars, the main production channel is associated with intermediate-mass asymptotic giant branch (AGB) stars (M $\approx$ 1–4 M$_{\odot}$) with lifetimes spanning from about 100 Myr to 10 Gyr \citep{kobayashi_isotopes_MW_2011, kobayashi_origin_2020}. %(Kobayashi et al., 2011, 2020).
Consequently, young, metal-poor galaxies with formation timescales of less than $\sim100$ Myr are expected to showcase a C/O abundance consistent with the predicted yields of core-collapse SNe, whereas C/O levels are expected to rise as galaxies evolve and become more metal-rich. 
The C/O abundance ratio is, therefore, a very valuable tracer of early galaxy formation due to its variations within relatively short timescales, and it is generally inferred from the ratios of rest-frame UV emission lines of carbon (\CIIIall, \CIVL) and oxygen (\OIIIL). 
Complementary information is provided by the ratio between nitrogen and oxygen abundance (N/O) which, in the average population of local galaxies, is observed to follow a plateau at low N/O and low O/H representative of `primary' nitrogen production mechanisms (i.e. with a yield independent of metallicity, e.g. \citealt{matteucci_considerations_1986, chiappini_stellar_rotation_2006}), while increasing at higher metallicity due to the onset of the CNO cycle in low- and intermediate-mass stars and also the `secondary' nitrogen production (i.e. where the nitrogen yield depends on the amount of carbon and oxygen already present within the star, e.g. \citealt{vincenzo_NO_simulations_2018}). 

While limited for decades almost exclusively up to intermediate redshifts ($z\lesssim3$), detailed studies of chemical abundances in early galaxies have seen an unprecedented development following the advent of the \emph{\jwst}. 
%which is delivering high-quality spectra covering the rest-frame UV and optical regimes in galaxies up to z$\sim10$.
This has enabled not only the characterisation of the metallicity scaling relations for the high-redshift galaxy population \citep[e.g.][]{nakajima_mzr_ceers_2023, curti_jades_mzr_2024, langeroodi_fmr_compactness_2023}, but has also allowed us to investigate the history of chemical enrichment in some of the earliest systems ever observed. 
% revealing a variety of chemical abundance patterns.
Although some of the analysed galaxies appear in agreement with the expected behaviour as predicted by standard galactic chemical evolution models for young galaxies \citep[e.g.][]{jones_CO_z6_2023, arellano-corodva_2023}, others have shown peculiar patterns in their C/O and N/O abundances, possibly revealing the signatures of enrichment processes occurring on short timescales in the earliest phases of galaxy formation, and which are not commonly observed in the typical galaxy population at lower redshift.
This includes evidence for super-solar nitrogen enrichment in $z>5$ galaxies \citep[e.g.][]{isobe_CNO_2023,ji_nitrogen_AGN_z5_2024}, and has been observed not only in the extremely luminous GN-z11 at $z=10.6$ \citep{bunker_gnz11_2023, cameron_gnz11_2023}, but even at higher redshift in the galaxy GHz2 \cite{Zavala_GHZ2_2024, castellano_ghz12_2024}. 
Such observations have been interpreted as the possible seeds of forming globular clusters \citep{Senchyna_gnz11_2023, marques-chaves_Nitrogen_2024, watanabe_2024}, or as the effect of enrichment confined within the small volume of the Broad Line Region (BLR) of AGN \citep{maiolino_gnz11_2023}. 
Furthermore, possible evidence for super-solar C/O has been instead reported in a galaxy at $z\sim12.5$ \citep[GS-z12]{d_eugenio_gsz12_2023}, and interpreted as the footprint of chemical enrichment from SNe explosions of the first populations of overly massive, extremely metal-poor (or even metal-free Population III) stars.
% The resonant scattering of Ly$\alpha$ photons by the neutral intergalactic medium (IGM) suppresses the observed fluxes at wavelengths shorter than 1215$\AA$ rest-frame. Furthermore, the Lorentzian wing cross-section attenuates the continuum redward of Ly$\alpha$, softening the shape of the observed break \citep[Gunn-Peterson damping wing attenuation,][]{miralda_escude_GP_1998}.
% Such effect has been directly observed already in \jwst/NIRSpec prism spectra of high-redshift galaxies \citep[e.g.,][]{bunker_gnz11_2023, Arrabal_Haro_nature_2023, heintz_DLA_2023, hsiao_macs0647_2023, d_eugenio_gsz12_2023, Heintz_primal_2024}, and it is known to significantly impact the predicted photometric redshifts \citep{hainline_zgtr10_jades_2024}
%(Bunker et al. 2023;  Arrabal Haro et al. 2023a,b; Heintz+23, D'Eugenio+23, Hainline+24).

One remarkable example of high-redshift system whose rich emission line spectrum has been unveiled by \jwst/NIRSpec is \target (hereafter \targetshort), a luminous galaxy spectroscopically confirmed at $z\sim9.43$ in the HUDF \citep{bunker_hst_deep_DR_2023}, and one of the most distant objects for which it is possible to simultaneously probe rest-frame UV and optical spectra from \Lyalpha to \OIIIoptL, as the latter leaves the NIRSpec coverage at $z\gtrsim9.55$.
This source was originally identified as a robust high-redshift galaxy candidate within the CANDELS GOODS-S field on the basis on its red J$_{125}$ - H$_{160}$ color by \cite{oesch_LF_highz_2014} (the source was known as GS-z10-1 in that work), who reported a photometric redshift of $z_{\text{phot}}=9.9\pm0.5$. Initial modelling of SED and morphology delivered log(\mstar/\MSun)$\approx$9 and a size of r$_{e}\approx0.5$~kpc, with an inferred star-formation rate (SFR) surface density of $\approx1-20$\MSun yr$^{-1}$ kpc$^{-2}$ \citep{oesch_LF_highz_2014, holwerda_z10_sizes_2015}, contributing to the pre-\jwst characterisation of the UV luminosity function at $z\sim9-10$ \citep[e.g.][]{bouwens_UV_LF_2019}.
Based on its robust photometric redshfit and luminosity, it was included as a bona-fide $z\sim10$ candidate in the HST-selected sample to be followed-up with \jwst/NIRSpec in one of the first observational programmes of the \jwst Advanced Deep Extragalactic Survey (JADES) \citep[PID 1210,][]{Eisenstein_JADES_2023, bunker_hst_deep_DR_2023}, and some of its properties have been already discussed in early, sample-based papers of the collaboration \citep[e.g.][]{cameron_jades_bpt_2023, curti_jades_mzr_2024, laseter_auroral_jades_2023, boyett_EELGs_jades_2024, scholtz_jades_agn_2023}.
It has been then re-observed with NIRSpec in November 2023, in the framework of the JADES Origins Field (JOF) programme \citep[PID 3215,][]{eisenstein_JOF_2023}.

In this paper, we leverage the unprecedented depth provided by the combined programmes 3215 and 1210 in both NIRSpec-MSA PRISM and medium resolution gratings, together with complementary NIRCam imaging, to analyse in more detail its physical properties and chemical enrichment patterns.
We outline the data processing and spectral fitting in Section~\ref{sec:data} and discuss the ionization mechanisms powering \targetshort in Section~\ref{sec:dust_ion}. 
Section ~\ref{sec:abundances} presents the derivation of chemical abundances, and in Section ~\ref{sec:discussion} we discuss possible scenarios of chemical enrichment. Finally, in Section~\ref{sec:ly_break}, we discuss the possible contribution of \targetshort to reionization based on the tentative detection of \Lyalpha in emission and the modelling of the \Lyalpha damping wing.
Our conclusions are summarised in Section~\ref{sec:conclusions}.
Throughout this work, we assume a \cite{planck_2020} cosmology, with $H_0 = 67.4$ km s$^{-1}$ Mpc$^{-1}$, $\Omega_{\mathrm{M}} = 0.315$ and $\Omega_\Lambda = 0.685$. We also assume the Solar abundances of \cite{asplund_solar_2009}. 

% Such emission lines are among the brightest in the UV nebular spectrum, and are now observable using \jwst/NIRSpec up to very-high redshifts (z$\sim$6-10). %moreover, their ratio is only mildly affected by dust reddening due to their proximity in wavelength.
% Even prior to the advent of \jwst, observations of rest-frame UV spectra in z > 6 galaxies have exhibited large-equivalent width (EW), high-ionization emission lines, as seen in none but the most extreme galaxies at z $\sim$ 0, suggesting that extreme radiation fields characterise galaxies in the Epoch of  Reionization (Stark et al., 2015a, 2015b, 2017; Mainali et al., 2017; Senchyna et al., 2017; Hutchison et al., 2019).
% As such, the C/O abundance ratio stands as a crucial tool for understanding the timescales of galaxy formation during the epoch of reionization within the first billion years of cosmic history. 

\section{Data processing and analysis}
\label{sec:data}
\subsection{Observations and data reduction}

\begin{figure}
    \centering
    \includegraphics[width=0.98\columnwidth]{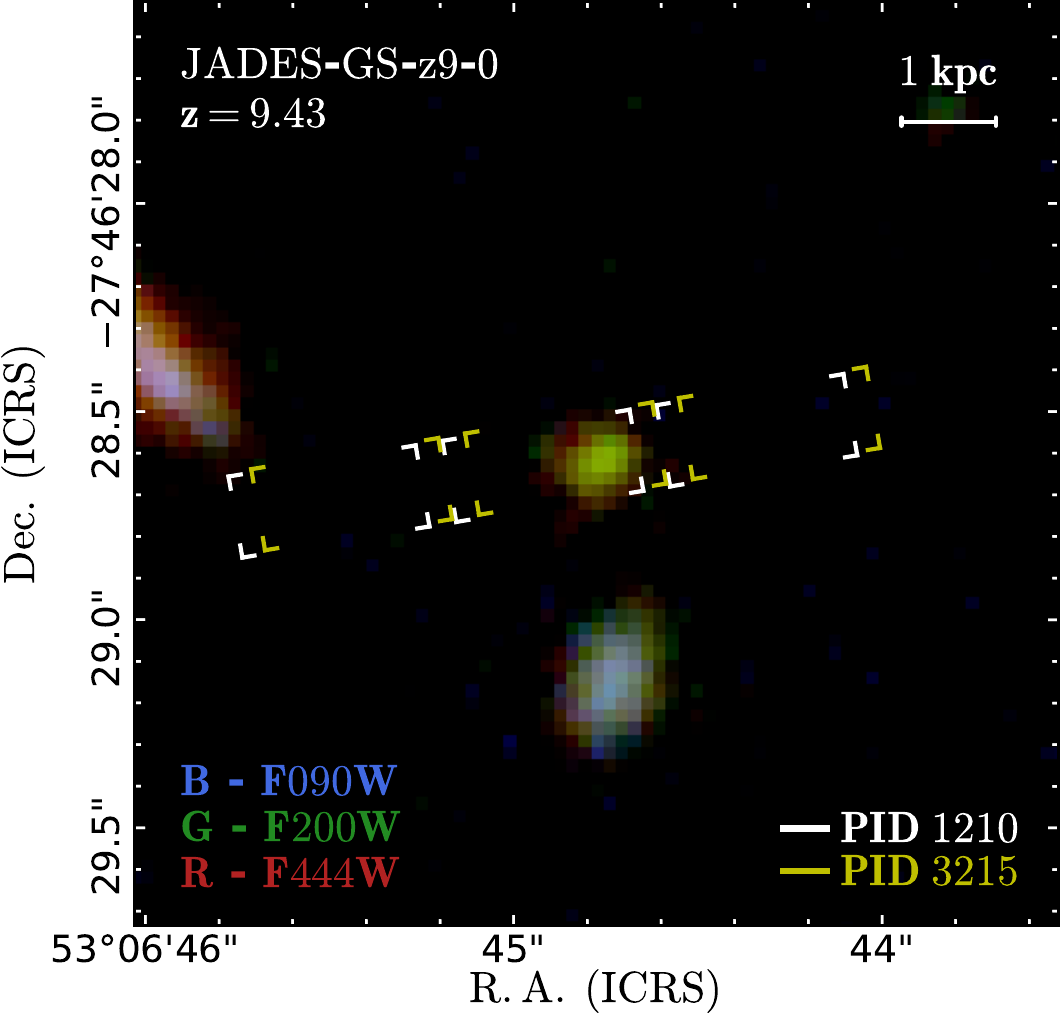}
    \caption{False-colour RGB image of the galaxy \targetshort. The location of the NIRSpec slitlets (first nod) from both the 1210 and 3215 MSA mask designs are overplotted. Only one of the three (for 1210) and five (for 3215) $\sim0.1^{\prime\prime}$ dithered pointings are shown.}
    \label{fig:galaxy_cutout}
\end{figure}

We analyse \jwst/NIRSpec observations carried out in two different programmes, namely PID~1210 \citetext{PI~N.~L\"utzgendorf, \citealp{bunker_hst_deep_DR_2023}} and PID~3215
\citetext{PI~D.~Eisenstein and R.~Maiolino, \citealp{eisenstein_JOF_2023}}, as the target of the present study was included in both NIRSpec/MSA mask configurations, with an almost identical relative slitlet position.
NIRCam imaging covering \targetshort in both wide- and medium-band filters are also available.
In particular, F090W, F115W, F150W, F200W, F277W, F335M, F356W, F410M, and F444W images were taken as part of the medium-depth JADES programme ID 1286, whereas F182M and F210M as part of the FRESCO programme \citetext{PID 1895, PI~Oesch, \citealp{oesch_fresco_2023}}.
Figure~\ref{fig:galaxy_cutout} shows a composite RGB image of \targetshort, with the position of the NIRSpec slitlet from the first visit and first nod of PID 3215 and PID 1210 overplotted.

In virtue of its allocation over multiple programmes, \targetshort has collected a total exposure time with NIRSpec/MSA of $72.3$ hours in PRISM/CLEAR, $44.3$ hours in G395M/F290LP, and $16.3$ hours in G140M/F070LP, as a result of combining 72+114, 18+96, and 18+24 individual integrations from both 1210 and 3215 programmes (19 groups/int, 2 integrations of 1400~s each per exposure, with a three-nodding pattern repeated over three-dithered and five-dithered pointings in 1210 and 3215, respectively).
 We note that the last visit (visit 5) of PID 3215 was affected by short circuits, reducing the exposure time compared to the original allocated time by $8400$~s for PRISM and G140M observations (i.e. 6 integrations lost out of the requested 120), and by $33600$~s for G395M (24 integrations lost, equivalent to the full visit).
The total exposure time in G235M/F170LP is instead of $7$ hours, resulting from solely the 1210 programme as such grating/filter combination has not been repeated in 3215.
In this paper, we leverage primarily the ultra-deep combined 1210+3215 data for PRISM, G140M, and G395M configurations, noting also that the \CIIIL emission line covered by G235M falls unfortunately in the gap between the two NIRSpec detectors.
A summary of the observing modes and total exposure times is provided in Table~\ref{tab:observations}.

\begin{table}
\caption{Summary of NIRSpec observations of \target.}
  \centering
    \begin{adjustbox}{width=0.99\columnwidth}
  \begin{tabular}{l|cccc}
  \hline
  Configuration               & PRISM/      & G140M/ & G235M/ & G395M/ \\
  & CLEAR & F070LP & F170LP & F290LP \\
  Exp. time 1210 [h]                   & 28         & 7    &  7        & 7 \\
  Exp. time 3215$^{\dagger}$ [h]      & 44.3        & 9.3  & -         & 37.3 \\
  Exp. time Total [h]                 & 72.3        & 16.3 & 7         & 44.3 \\
  \hline
  \end{tabular}
    \end{adjustbox}
  \tablefoot{
  \tablefoottext{$\dagger$}{Exposure time reduced from nominal allocated time due to short circuits affecting visit 5 of PID 3215}}
  \label{tab:observations}
\end{table}

The data reduction of the 3215 data follows the same recipe of 1210, as described in other papers of the JADES collaboration \citep[e.g.,][]{bunker_hst_deep_DR_2023, d_eugenio_jades_DR3_2024}.
In brief, we adopt a three-nodding scheme for background subtraction, apply path-loss 
corrections appropriate for point sources (taking into account the intra-shutter position of the source in each nod and dither configuration), and reconstruct 2-D spectra for each individual integration adopting a uniform wavelength sampling for the gratings (with a wavelength bin equal to the average native pixel sampling of the detector), while a highly non-uniform wavelength grid in the case of the PRISM, with the bin width set to account for the largely varying spectral resolution and to avoid oversampling of the line spread function.

From each 2-D spectrum we then extract a 1-D spectrum using a full-shutter window as driven by the light profile inferred from the brightest emission lines (to ensure minimal flux losses and avoid possible biases in the measured line ratios). 
We note that we also repeated the analysis adopting a narrower three-pixel boxcar extraction aimed at possibly maximising the signal-to-noise, noting however a clear flux loss in the rest-optical lines as well as in the UV continuum shortward of $2\mu$m in the PRISM spectrum, as well as no considerable improvement in the significance of the detection of the faintest lines, with the only notable exception of the \NIVL[1483] emission line in the G140M spectrum (see Section~\ref{sec:UV_fitting}).

In order to obtain the final, combined 1-D spectrum, in this work we have co-added the individual reduced 1-D sub-spectra from both 1210 and 3215 for PRISM, G140M, and G395M by inverse-variance-weighted averaging over the surviving entries in each wavelength bin following five passes of iterative three-sigma-clipping aimed at flagging anomalous noise spikes representing outliers to the statistical noise in the data. This approach also allows us to obtain a more robust characterisation of the sources of noise present in the data by leveraging the large number of individual (while nominally identical) exposures in each grating/filter configuration (as discussed in the following Section, see also \citealt{hainline_zgtr10_jades_2024, Witstok_lya_z13_2024}).
Two-dimensional spectra are also reconstructed by the NIRSpec/GTO pipeline, but are generally not considered for extracting the 1-D spectrum to avoid spurious effects and uncertainties possibly introduced by the heavy resampling of the data required to combine the 2-D spectra of sources observed in different intra-shutter positions across different visits (and, in this case, also within multiple programmes).

\begin{figure*}
    \centering
    \includegraphics[width=0.95\textwidth]{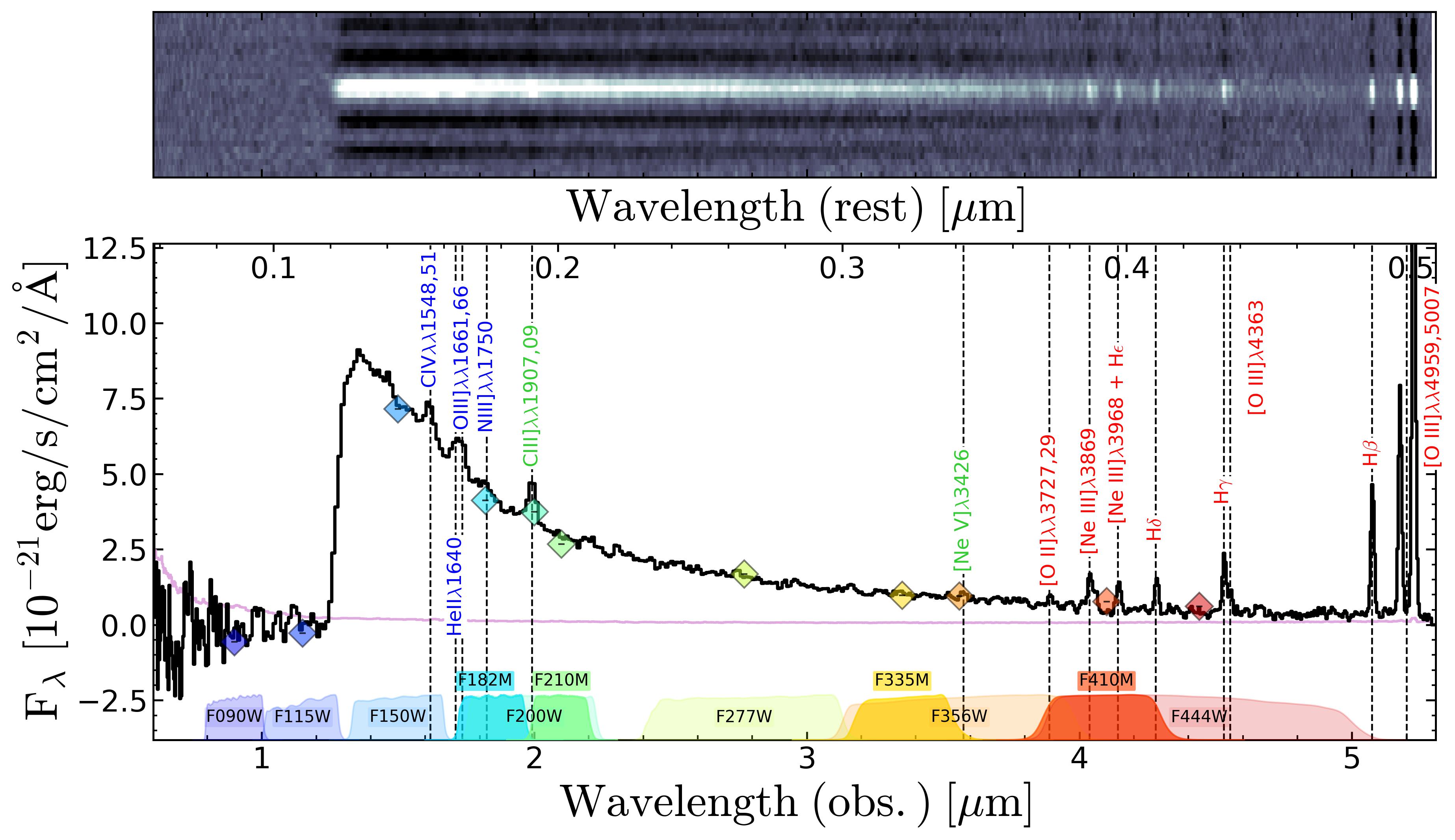}
    
    \caption{PRISM spectrum of \targetshort at z$=9.4327$. The 2-D (top) and 1-D (bottom) spectra are obtained combining observations from both JADES 1210 and 3215 programmes. 
    The pipeline error spectrum is reported in magenta.
    The main emission line features detected in the spectrum are marked in different colours, depending on whether they are also covered by G140M/F070LP (blue), G395M/F290LP (red), or observable only in the PRISM data (green).
    The \textsc{forecepho} photometry extracted from available NIRCam wide- and medium-band imaging is also reported, showing good consistency in the flux calibration with the pathlosses corrected spectrum. 
    }
    \label{fig:plot_spectra}
\end{figure*}

The final, combined 1-D and 2-D PRISM spectra for \targetshort are displayed in Figure~\ref{fig:plot_spectra}.

\subsection{Emission line fitting}
\label{sec:lines_fitting}
We performed emission line fitting separately for low-resolution PRISM and medium-resolution gratings spectra. 
Formally, the error on each parameter of the fit is evaluated exploiting the output pipeline error spectrum. 
%randomly perturbing the flux density spectrum by the error spectrum (assumed to be Gaussian), repeating the fit one hundred times, and taking the standard deviation of the resulting distribution of values.
However, we also perform additional tests to assess the robustness of low-significance detections in both gratings and PRISM spectra by leveraging the large number of individual integrations in G140M (42) G395M (114), and PRISM (186) provided by the combined 1210 and 3215 datasets.
More specifically, we generate 300 bootstrapped spectra by randomly sampling (with replacement) over the set of individual sub-spectra, after five passes of iterative $3\sigma$-clipping have removed strong outliers at each wavelength bin. 
We then repeat the fitting process on each individual bootstrapped combined (averaged) spectrum, and take the square root of the variance of the resulting distribution as the error on the parameter. 
Such empirical bootstrapped uncertainties are expected to be more conservative than those estimated by assuming the output error spectrum from the pipeline, in that they implicitly take into account all possible sources of noise, including the correlated error among wavelength bins present in resampled NIRSpec spectra, which might not be fully accounted for by the pipeline \cite[see also][]{maseda_xmps_lae_2023, hainline_zgtr10_jades_2024}.
We report the extracted emission line fluxes and equivalent widths for both our PRISM and gratings fitting in Table \ref{tab:line_fluxes}, where the signal-to-noise-ratio (S/N) quoted on the emission line fluxes is reported on the basis of both uncertainty estimates.
Overall, we find bootstrapped-based S/N to be generally lower than those based on the pipeline error spectrum, and this is particularly true for medium-resolution grating spectra, whereas the pipeline error for the PRISM spectrum appears intrinsically more conservative in virtue of its attempt to rescale the noise to account for the correlation induced by the spectral resampling.
A more detailed discussion about the noise model in NIRSpec spectra will be presented in a forthcoming paper (Jakobsen et al., in prep.).
We also compare our results with those obtained via the joint continuum and emission lines modelling with \textsc{ppxf} \citep{cappellari_improving_2017} as described in \cite{d_eugenio_jades_DR3_2024}, finding consistent results. 
Further details on the results of our fitting procedure are given below.

\subsubsection{Rest-frame UV spectrum}
\label{sec:UV_fitting}
To begin with, we focus on the rest-frame UV region in the PRISM spectrum. 
We first perform a fit to the \cite{calzetti_dust_2000} region down to $2600\AA$ (rest-frame) but excluding the region blueward of $1450\AA$ (rest-frame) to minimise the impact of the Ly$\alpha$ damping wing (we further discuss the modelling of the Ly$\alpha$ spectral break in Section~\ref{sec:ly_break}).
We model the underlying continuum with a power-law, and include the following emission lines in the fitting procedure: \NIVL[1485] \CIVL, \HeIIL, \OIIIL, \NIIIL, \CIIIL.% \NeIVL. 
% The results are presented in Figure~\ref{fig:prism_fit}.

Each individual line is assumed as spectrally unresolved and is modelled with a single Gaussian whose width is allowed to vary within ten percent of the line spread function (LSF) modelled from 1210 data by \cite{de_graaff_jades_2024}, which takes into account galaxy size and relative NIRSpec intra-shutter position.
Emission line doublets (e.g \CIIIL[1907,1909]   ,  \NIVL[1483,1486]\OIIIL[1661,1666]) or multiplets (e.g. \NIIIall) are assumed unresolved in the PRISM spectra, whereas multiple components are included when fitting the gratings spectra.
To model the \NIIIall multiplet in the PRISM, we set the line centroid to the average between the two brightest transitions in the multiplet, i.e. at 1751.83$\AA$.  
To aid the modelling of the line complex involving \OIIIL and \HeIIL (which are partially blended at the PRISM resolution), their ratio is fixed to that measured from the G140M spectrum (see below).
Furthermore, considering the proximity of the \SiIIIall doublet to \CIIIL, and that such emission line has been observed in $z\sim2.5$ galaxies (with relative intensity of $\sim20-30\%$ that of \CIII, e.g. \citealt{steidel_reconciling_2016}), we also include an additional Gaussian component\footnote{centred at $1885\AA$, noting that \SiIIIL[1883] is expected to be stronger than \SiIIIL[1892]} to account for the \SiIII doublet in our fitting procedure.

\begin{figure*}
    \centering
    \includegraphics[width=0.68\textwidth]{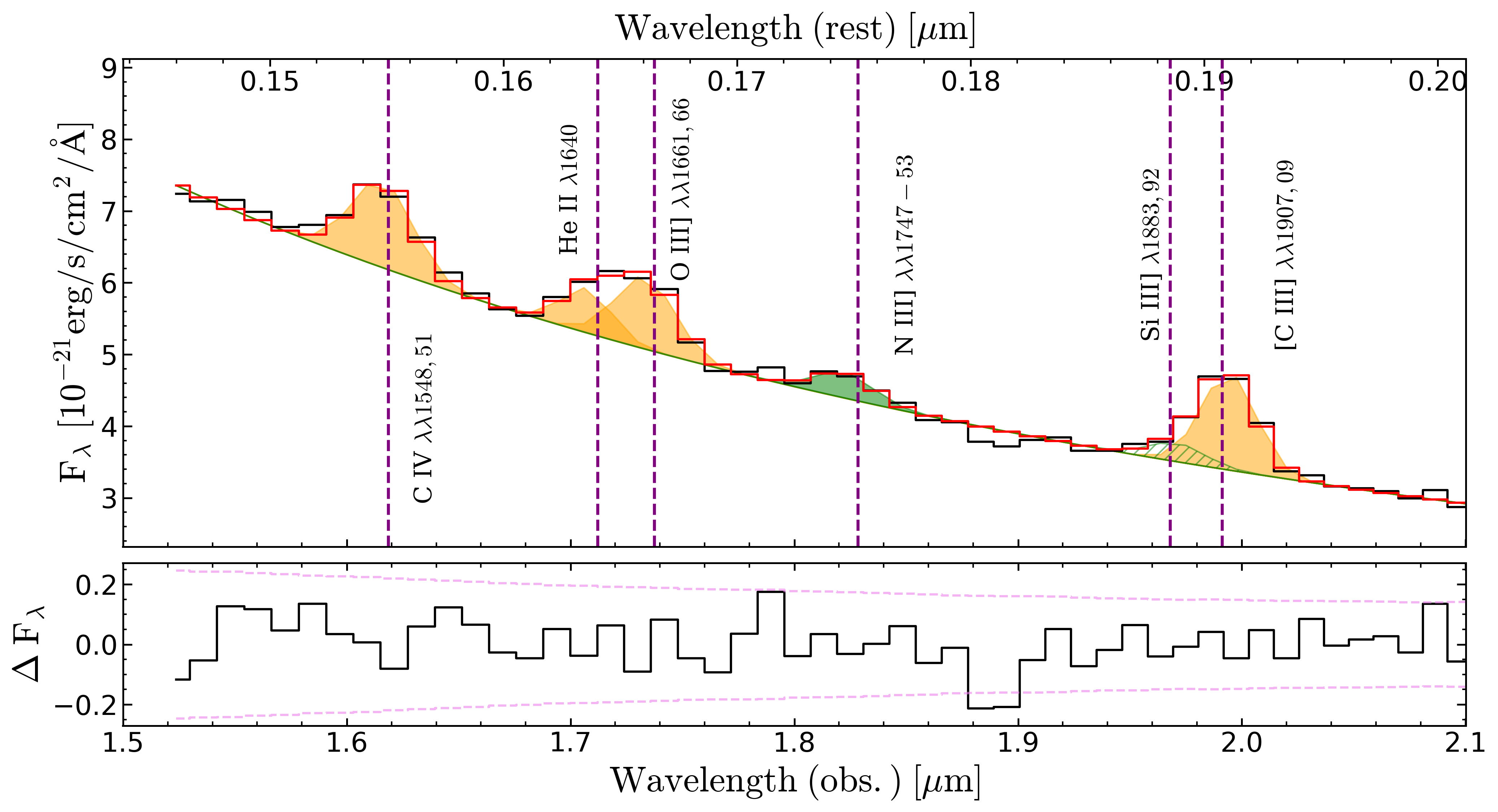}
    \includegraphics[width=0.28\textwidth]{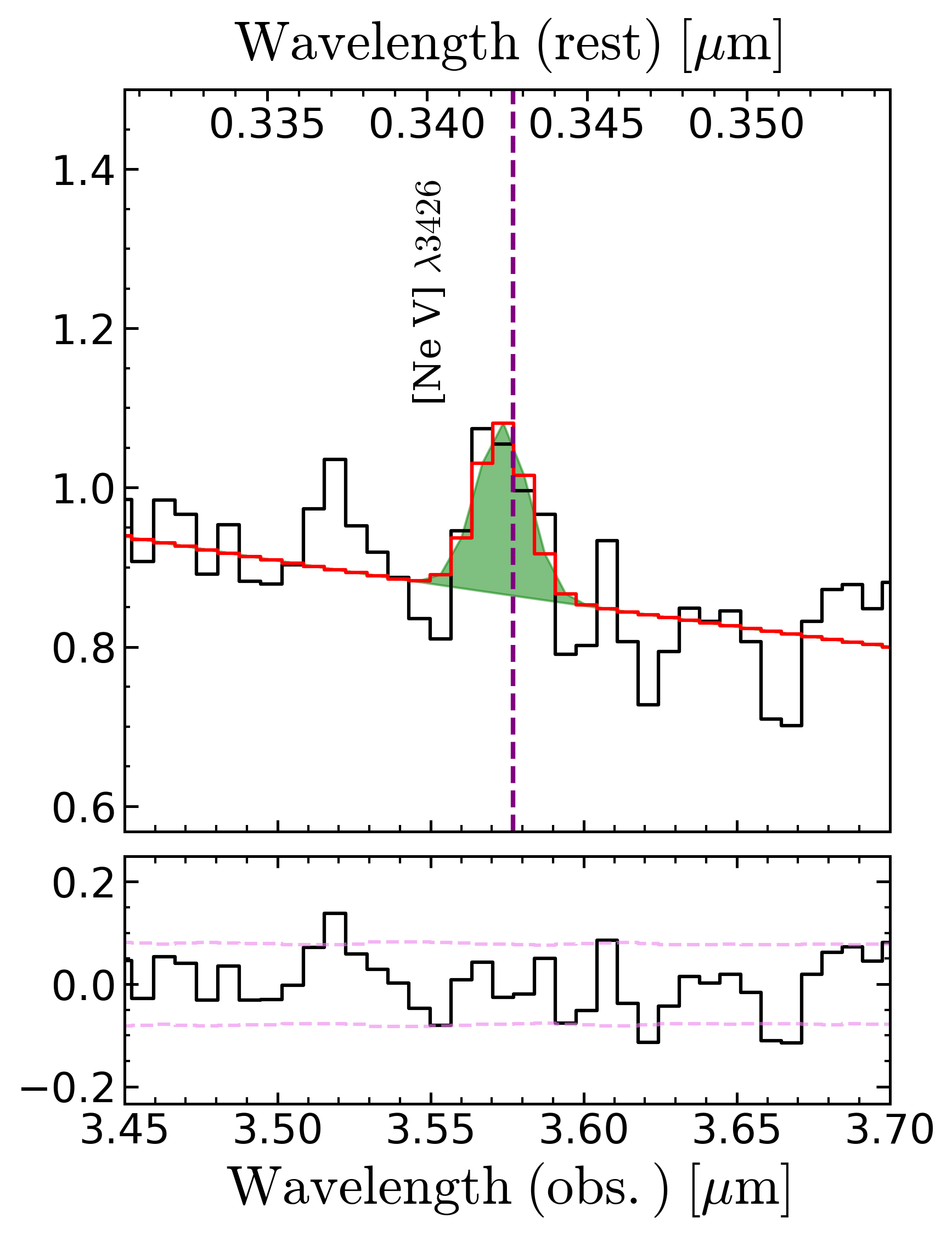}
    \caption{Rest-frame UV emission lines in the PRISM spectrum of \targetshort. 
    The best-fit to the continuum and lines is shown in red.
    Detections at $\geq 4\sigma$ are marked in yellow, while marginal detections ($\sim3\sigma$) are highlighted in green. 
    Vertical purple lines mark the expected location of the emission lines based on the systemic redshift of the source. 
    The bottom panels report the residuals of the fit and the $1\sigma$ uncertainty from the pipeline error spectrum.
    Left panel: zoom-in on the region of the PRISM spectrum
    between $1.5\mu$m and $2.1\mu$m.
     Beside clear detections of \CIV, \HeII, \OIII, and \CIII, marginal detection of \NIIIL is also highlighted. 
    Right panel: Tentative detection of the very high-ionization \NeVL emission line (ionization potential $97.11$~eV). The line is formally detected at $\sim 3\sigma$ from both the pipeline error spectrum and the bootstrapping approach described in Section~\ref{sec:lines_fitting}.
    }
    \label{fig:prism_fit}
\end{figure*}

The results of fitting the rest-frame UV region of the PRISM spectrum are shown in the left-hand panel of Figure~\ref{fig:prism_fit}: \CIVL, \HeIIL, \OIIIL, and \CIIIL are detected above the continuum level at $\gtrsim5\sigma$ significance.  
From the same fit, we constrain the UV slope to $\beta_{\text{UV}}=-2.54\pm0.02$.
Although formally undetected ($\sim1.8\sigma$ significance), and despite its small contribution to the total flux of the complex, we note that including the \SiIIIall component (hatched green) provides a better match to the `blue wing' of the \CIII line profile.
Higher resolution observations of the \CIII complex are needed to assess the real significance of the \SiIII emission\footnote{we recall that the region around \CIII is lost in the detector gap in currently available G235M observations from PID 1210}, and we here note that none of the main results of the analysis depends on the inclusion (or not) of such component in the fitting procedure.

In addition, we report marginal detection (marked in green in Figure~\ref{fig:prism_fit}) of the \NIIIall multiplet. The line is formally detected at $2.7\sigma$ assuming the error spectrum from the pipeline, whereas at 
$3\sigma$ adopting the bootstrapping approach.
The \NIIIall emission arises in a region not contaminated by other emission lines,
while the underlying continuum level is well constrained and anchored by the presence of adjacent high-EW emission lines (\CIII and \OIII), i.e. the best-fit to the continuum is not affected by the inclusion (or not) of this specific line emission component.
% The possible detection of \NIIIall is interesting given the implications of this transition in the determination of the N/O abundance. 
We discuss the implications of the possible detection of \NIIIall for the scenarios of chemical enrichment in \targetshort (and, in particular, for the determination of the N/O abundance) in Section~\ref{sec:discussion}.

Conversely, \NIV emission is difficult to constrain in the PRISM fit. In fact, the fit improves if the \NIV component is not included, whereas forcing an additional Gaussian component at $\sim1485\AA$ impacts the overall level of the continuum as well as the fit of the other emission lines in that region (especially of the adjacent \CIV), decreasing the goodness-of-fit and highlighting the challenges in disentangling faint, low-equivalent width line emission from the continuum at such low spectral resolution.  
When forcing the inclusion of the \NIVL[1483,86] component, we note that the line is formally detected at only $\sim 2\sigma$, though the inferred flux is consistent with the low-significance \NIVL[1483] detection in the G140M spectrum (discussed below). We report the \NIV flux measured in the latter scenario in Table~\ref{tab:line_fluxes}, nonetheless  we ultimately decide to exclude \NIV from our fiducial modelling of the PRISM spectrum (as shown in Figure~\ref{fig:prism_fit}), i.e. all other line fluxes and the spectral slope are measured by fitting a model without such an additional Gaussian component.

Extending our fitting analysis to longer wavelengths, we also report the marginal detection of \NeVL (at $\sim3.3\sigma$ with the pipeline error, $2.9\sigma$ with bootstrapping, right-hand panel of Figure~\ref{fig:prism_fit}).
To produce such very high-ionization ($97.11$ eV) emission line requires an extremely hard photoionization source. The presence of \NeV in galaxy spectra has been attributed to actively accreting black holes in AGN hosts, stellar continuum from an extremely hot ionizing spectrum including Wolf–Rayet stars, or energetic radiative shocks from supernovae \citep{gilli_NeV_2010, izotov_NeV_2012,mignoli_NeV_2013, zeimann_NeV_2015, Backhaus_NeV_2022, cleri_NeV_2023}
We discuss the implications of possible \NeVL detection in the determination of the dominant ionising source of \targetshort in Section~\ref{sec:dust_ion}.
However, we note that we do not detect any significant emission above the continuum level at the location of the \NeIVL emission line in the PRISM spectrum. %at odds with the $4\sigma$ \NeIVL detection reported by \cite{scholtz_jades_agn_2023} in the G235M spectrum. 
One possible explanation for the absence of \NeIVL (in the presence of both \NeIII and \NeV) is that, despite the lower ionization energy of Ne$^{3+}$ compared to Ne$^{4+}$, the strength of the line is hampered on the one hand by its lower emissivity, while on the other by the relatively higher energy required to collisionally excite the ion.

\begin{figure*}
    \centering
    \includegraphics[width=0.315\textwidth]{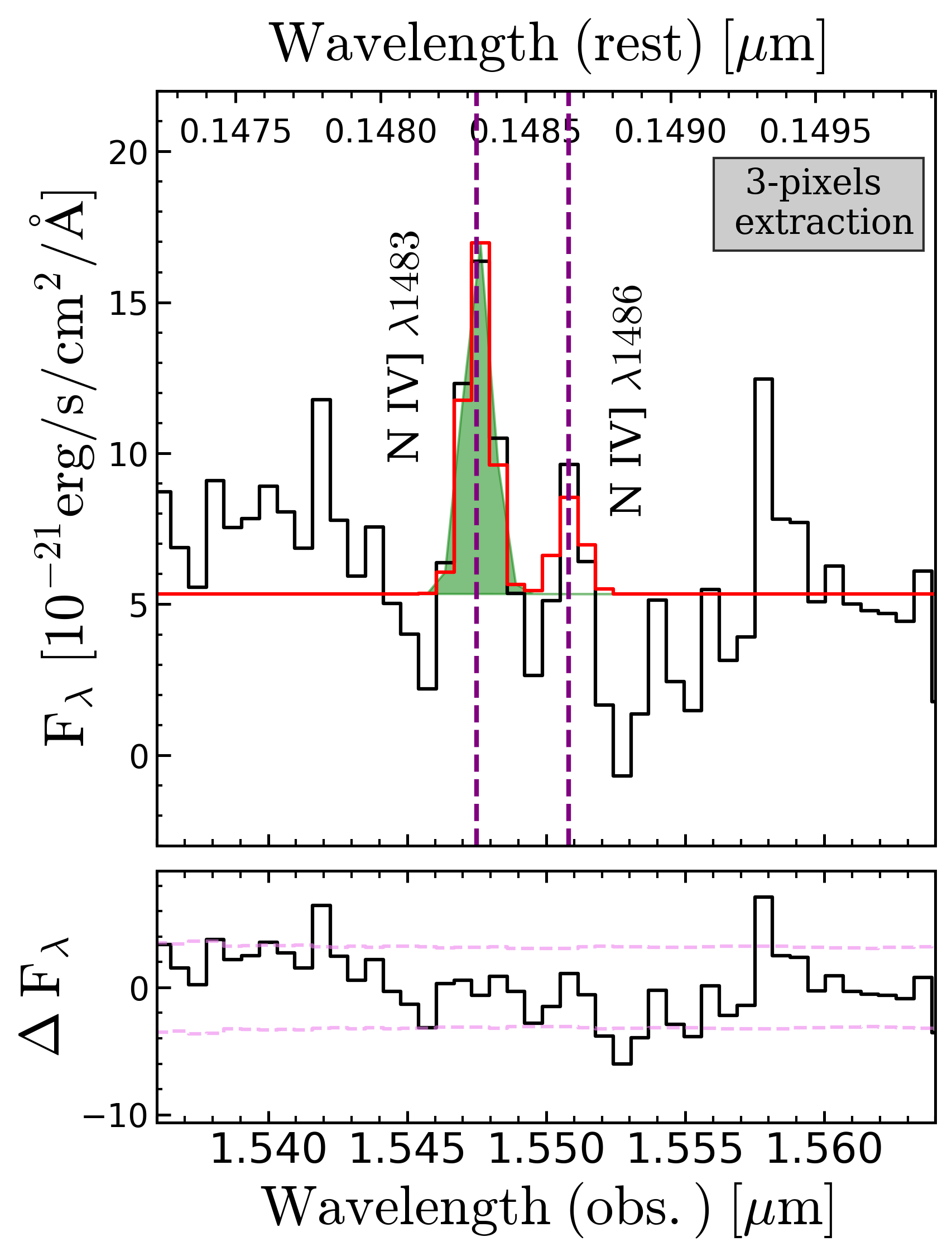}
    \includegraphics[width=0.297\textwidth]{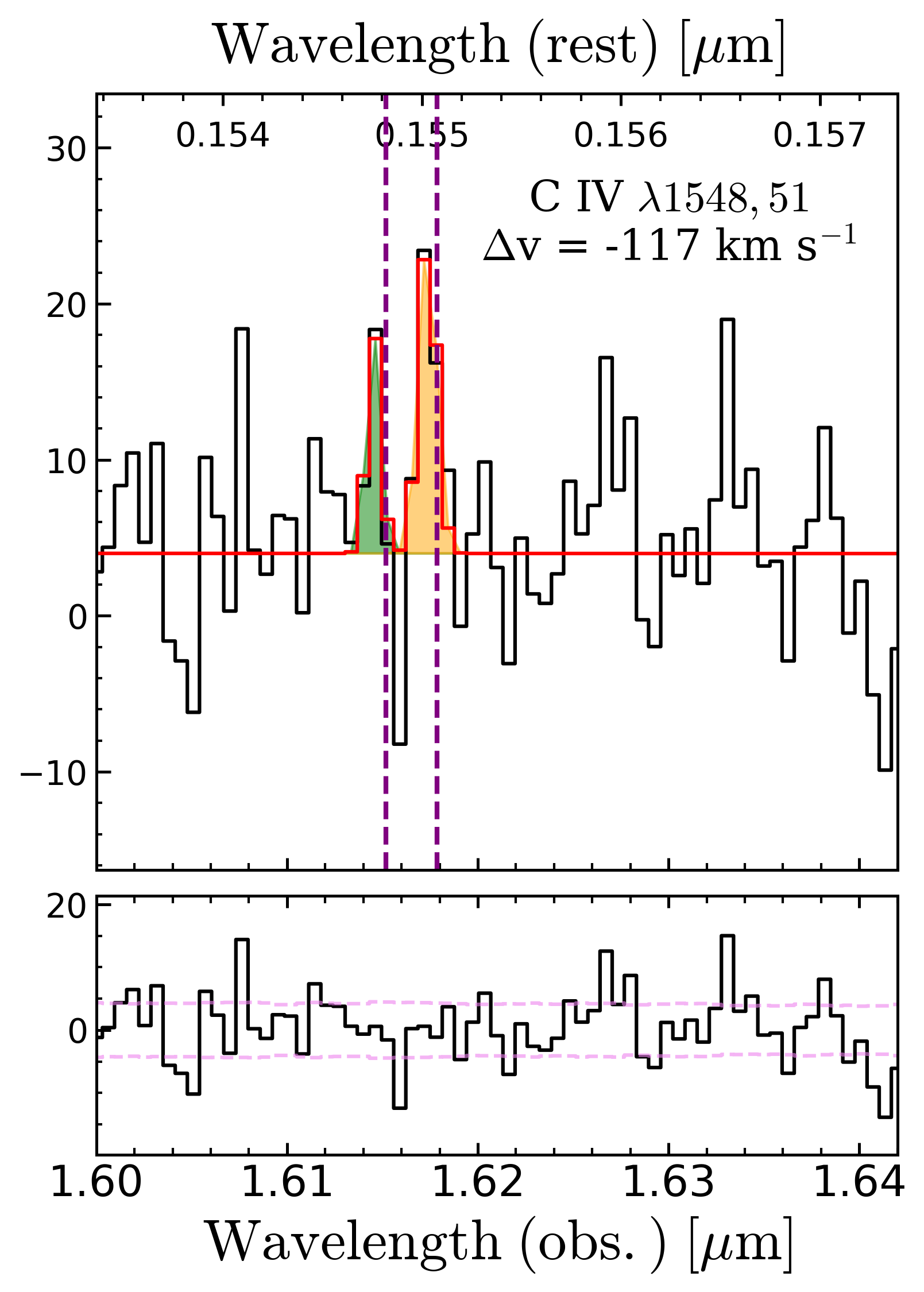}
    \includegraphics[width=0.31\textwidth]{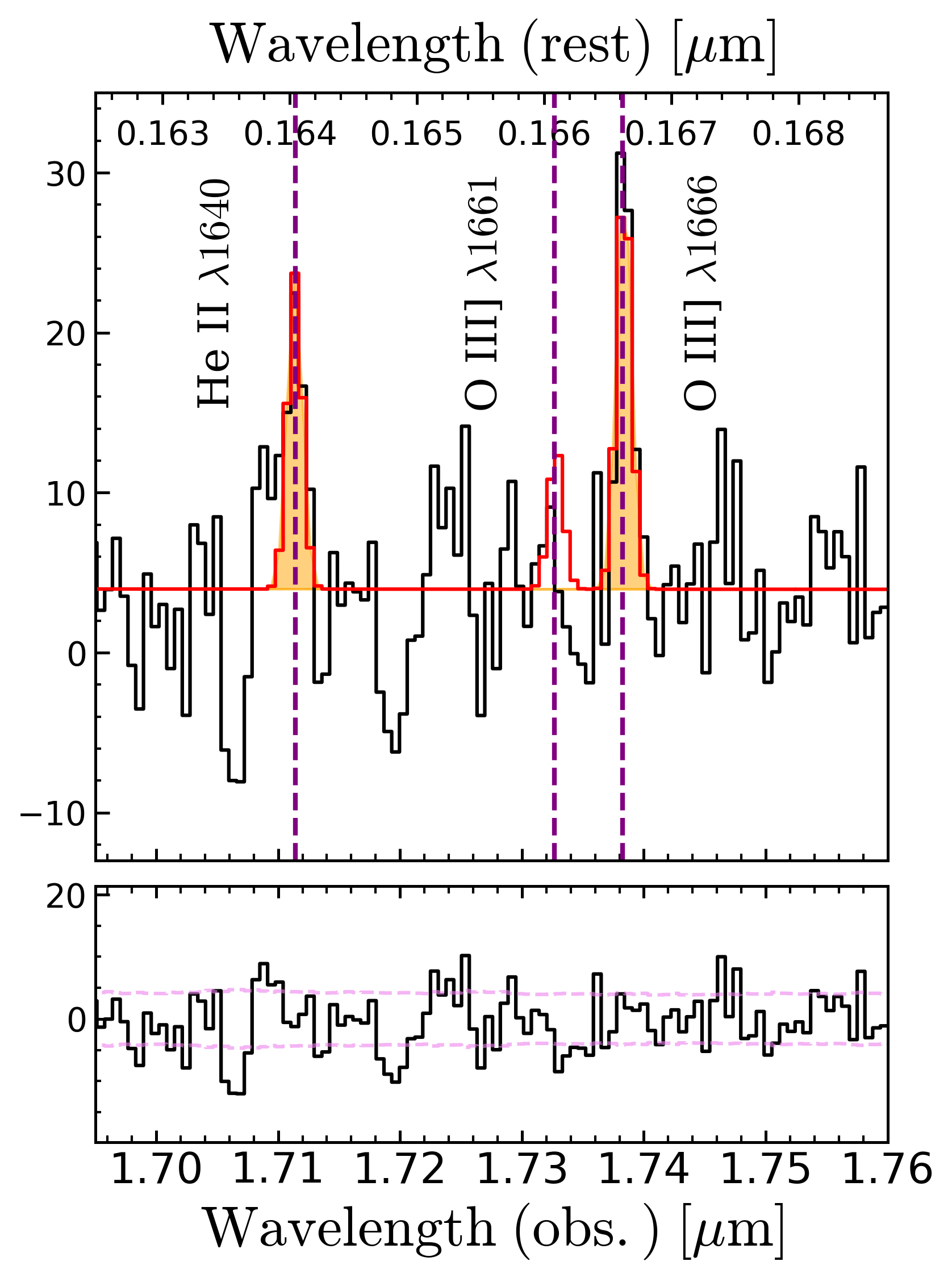}    
    \caption{
    % Tentative detection of the very high-ionization \NeVL emission line (ionization potential $97.11$~eV) in the PRISM spectrum of \targetshort. The line is formally detected at $\sim 3\sigma$ from both the pipeline error spectrum and the bootstrapping approach described in Section~\ref{sec:lines_fitting}.
    Rest-frame UV emission lines in the G140M spectrum of \targetshort.
    In particular, we show a zoom-in on the region of the G140M spectrum around the \NIV doublet (left), \CIV doublet (middle), \HeII, and \OIII emission (right). 
    The \CIV$\lambda1548,51$ doublet is resolved, but only the redder line is detected at $>4\sigma$ significance.
    The \CIV complex is blueshifted by $\sim 117$km s$^{-1}$ compared the systemic redshift, whereas no velocity offset is seen for \OIIIL and \HeIIL. The \NIVL[1483] line is tentatively detected at $\sim3.4\sigma$ only in the 3-pixel extracted spectrum, whereas no significant \NIVL[1486] emission is found.
    }
    \label{fig:uv_g140m}
\end{figure*}

Moving to the fit of the G140M spectrum, we report the detection of both \OIIIL and \HeIIL at $\sim7\sigma$ and $\sim 5\sigma$ significance based on the pipeline errors, respectively; based on the more conservative bootstrapping approach described above, both lines are instead detected at $3.5\sigma$ (right-hand panel of Figure~\ref{fig:uv_g140m}). 
During the fit, we fix the velocity and width of \HeIIL line to that of \OIIIL, in order to restrict the fit only to the \HeII nebular component. 
However, we do not find any
clear evidence for a residual broad component that could be associated with stellar winds features. %or AGN.
The \OIIIL[1661] is not detected, however the ratio of \OIIIL/\OIIIL[1661] is fixed by atomic physics to 2.93. 
The \NIIIL multiplet (whose components are almost fully resolved ) is also undetected, and its $3\sigma$ upper-limit is consistent with the expectations given the lower sensitivity of the G140M observations.

When fitting \CIVall, we leave the velocity and width of the two components of the doublet (spectrally resolved in G140M) free to vary, to account for possible resonant scattering through highly ionised gas that could impact the line profile \citep[see e.g.][]{leitherer_2011, berg_chemical_2019, senchyna_civ_2022,topping_z6_lens_2024}.
In general, the \CIVall spectral feature is challenging to interpret due to its complex composite profiles, with possible contributions from narrow nebular emission, broad stellar emission, stellar photospheric absorption, and interstellar medium absorption and scatter.
We detect (at $4.4\sigma$ assuming the pipeline error, at $3.2\sigma$ with bootstrapping) only one of the two \CIV lines in the doublet, which we interpret as the `red' \CIV$\lambda1551$ component (middle panel of Figure~\ref{fig:prism_fit}). 
% although this is expected to be the weakest of the two (with a relative oscillator strength $\sim$2$\times$ smaller than \CIV$\lambda1548$).
If such interpretation is correct, the line appears shifted by $\sim120$ km s$^{-1}$ with respect to the systemic redshift of the galaxy inferred from strong rest-frame optical lines, possibly indicative of resonant scattering through outflowing gas attenuating the blue component of the doublet, which has a relative oscillator strength $\sim$2$\times$ higher than \CIV$\lambda1551$. 
On the contrary, interpreting such emission line as  \CIV$\lambda1548$ would require a redshift $>400$ km s$^{-1}$. Moreover, this would worsen the tension with the total flux and equivalent width of the total \CIV doublet as inferred from the PRISM.

Finally, we report a tentative detection of \NIVL[1483], whose formal significance is however found to be $>3$ only in the narrower, 3-pixel-boxcar extracted spectrum (left-hand panel of Figure~\ref{fig:uv_g140m}), whereas it is $\sim2.8 \sigma$ in the full-shutter extracted spectrum. Interestingly, no clear evidence of the redder line of the doublet (i.e. \NIVL[1486]) is observed, which seems to exclude extremely-high gas density in the system (see Section~\ref{sec:te_ne}).

\subsubsection{Rest-frame optical spectrum}
\label{sec:optical_fit}

\begin{figure*}
    \centering
    \includegraphics[width=0.46\textwidth]{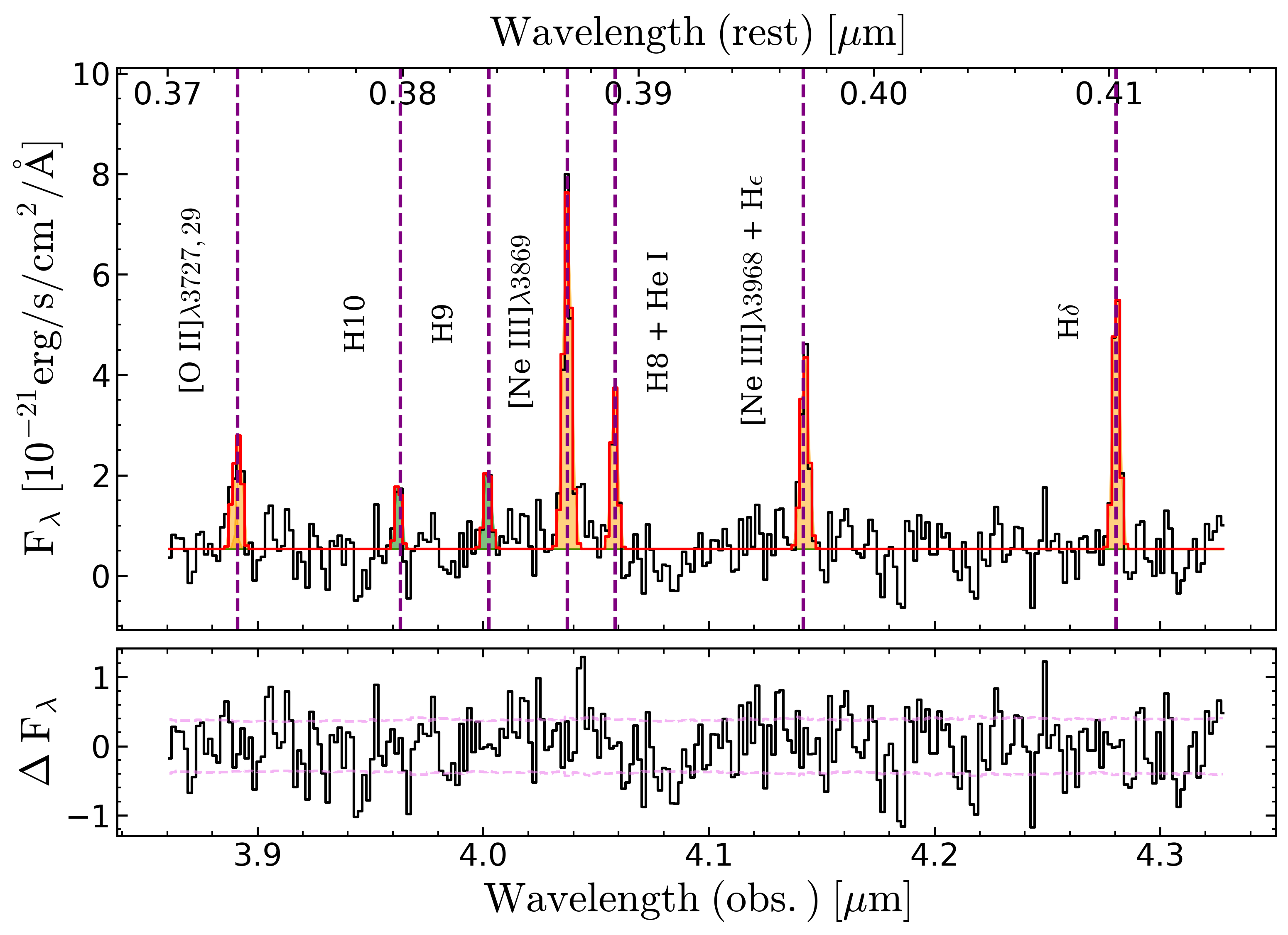}
    \includegraphics[width=0.178\textwidth]{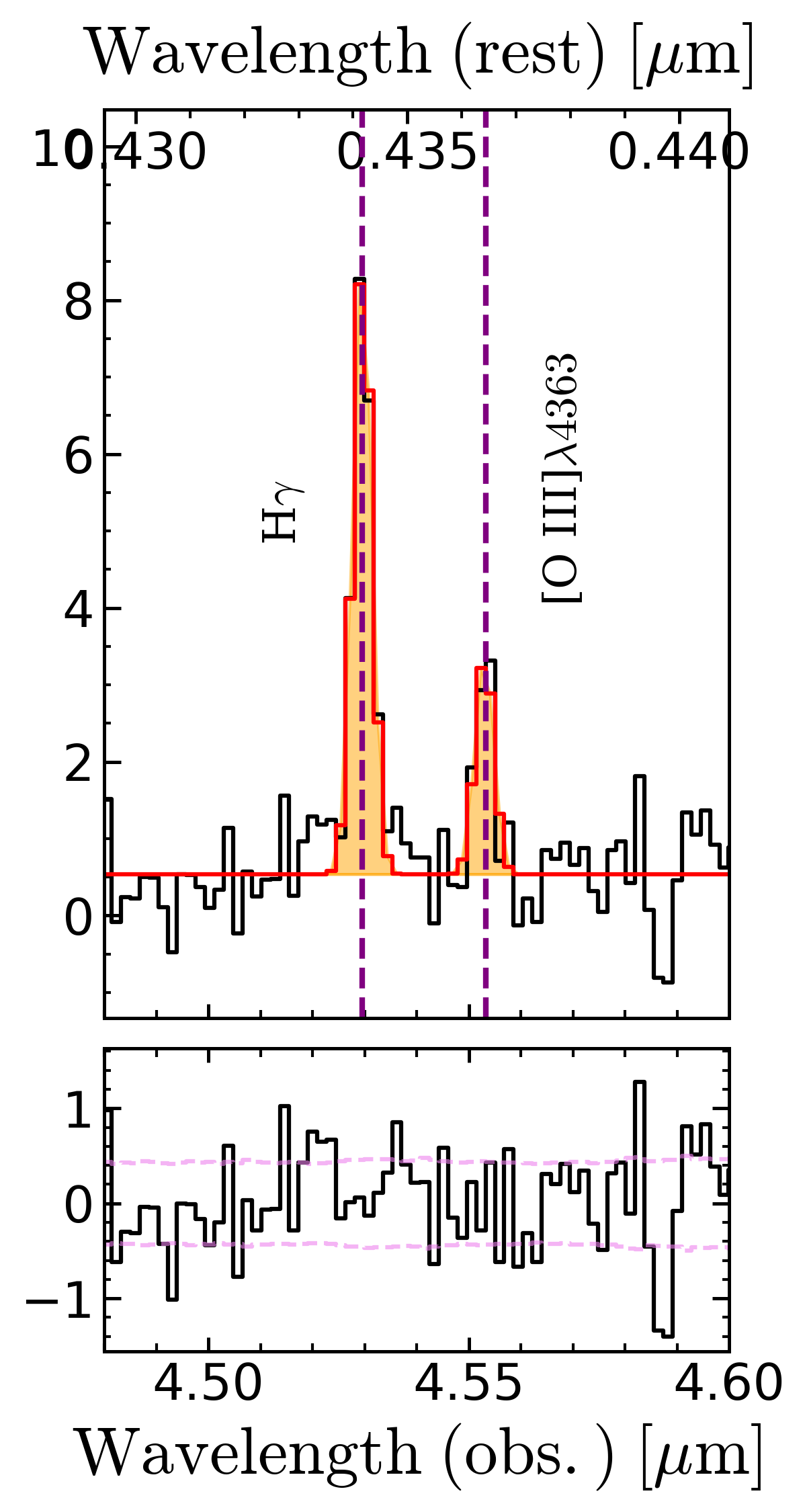}
    \includegraphics[width=0.32\textwidth]{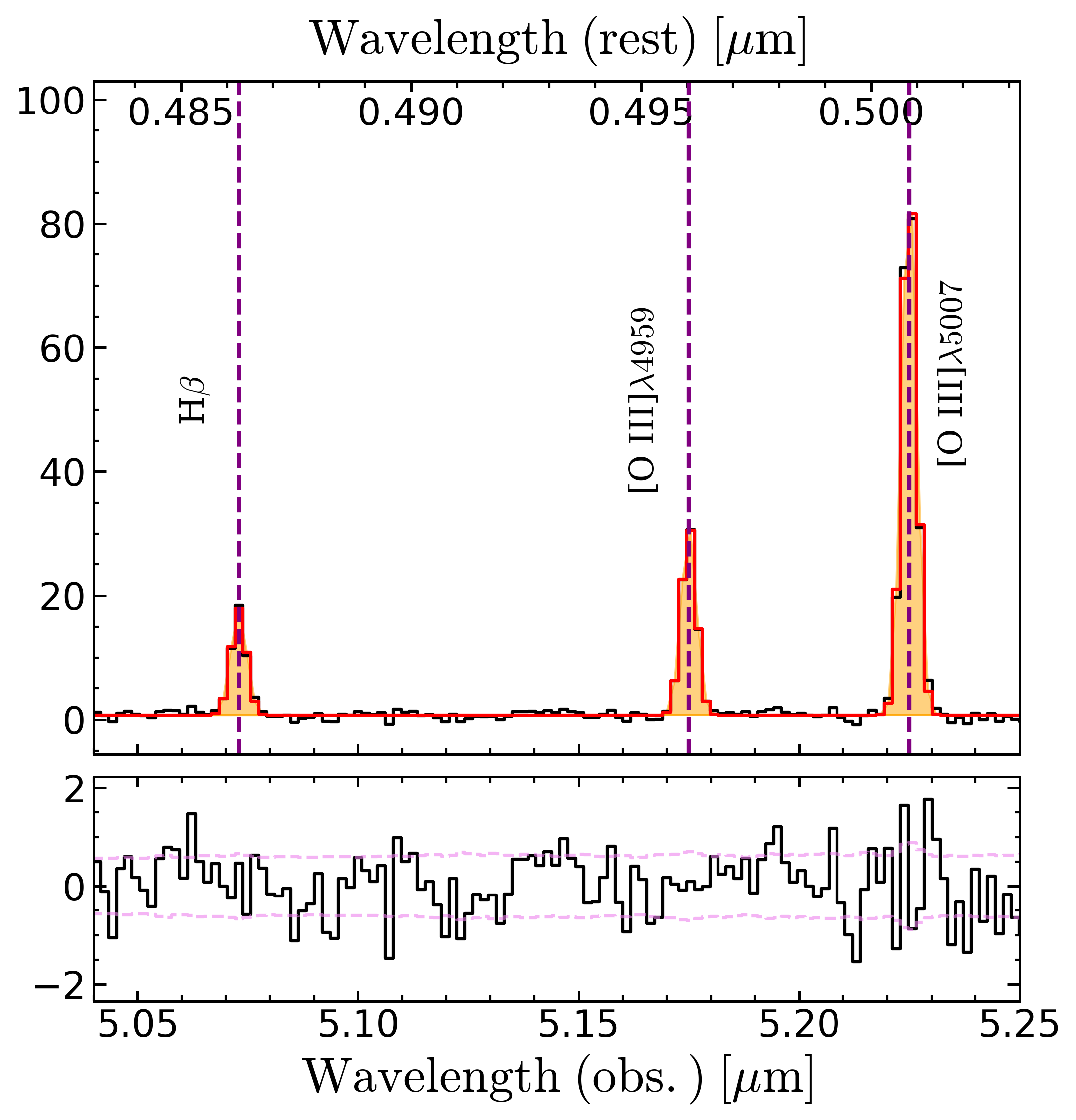}
    \caption{Rest-frame optical emission lines in the G395M medium resolution spectrum of \targetshort.
    The best-fit to continuum and emission lines is overlaid in red, with detected lines highlighted (in green `marginal' $\sim3\sigma$ detections).
    The left-hand panel shows a zoom-in on the region between \OIIall and \Hdelta, highlighting the detection of both strong lines and fainter Balmer lines like H9 and H10.
    The middle panel shows the region around \Hgamma and the \OIIIoptL[4363] auroral line, while the right-hand panel highlights the high S/N detections of the \OIIIoptL[4959,5007] and \Hbeta complex. 
    }    
    \label{fig:G395M_fit}
\end{figure*}

We fit the rest-frame optical lines in both PRISM and G395M spectra modelling the underlying local continuum with a power law. 
Emission lines are assumed as unresolved and modelled with individual Gaussians, the only exception being the \OIIall doublet which is marginally resolved and modelled with two components when fitting the G395M spectrum.
As shown in Figure~\ref{fig:G395M_fit}, rest-frame optical lines are well detected, with high signal-to-noise, as already reported e.g. in \cite{bunker_hst_deep_DR_2023} based on observations from programme PID 1210. 
This includes not only a robust detection of the \OIIIoptL[4363] auroral line, of specific interest for deriving chemical abundances (Section~\ref{sec:abundances}, see also \citealt{laseter_auroral_jades_2023}), but also (more marginal) detections of high-order Balmer lines like H9$\lambda3835$ and H10$\lambda3797$, which, in addition to well detected \Hbeta, \Hgamma, and \Hdelta, provide information to constrain the amount of nebular attenuation over a wide wavelength range (Section~\ref{sec:SED_fitting}).

We derive a fiducial redshift for \targetshort from the combined 1210+3215 spectrum by averaging over the brightest, isolated rest-frame optical lines, finding z$_{\text{G395M}}=9.432681\pm0.000069$ and z$_{\text{prism}}=9.43774\pm0.00020$ (mean and error on the mean), respectively.
These values are consistent with those determined individually from 1210 and 3215 spectra, while revealing a significant discrepancy between G395M and PRISM, possibly caused by offsets in the wavelength calibration introduced by a non proper correction of the relative intra-shutter position in the determination of the wavelength solution \citep{d_eugenio_jades_DR3_2024}.

We also note that existing flux calibration offsets between PRISM and grating spectra, as well as offsets as a function of wavelength and of the location on the MSA detectors, have been reported from the analysis of large galaxy samples observed with NIRSpec \citep[e.g.][]{bunker_hst_deep_DR_2023, d_eugenio_jades_DR3_2024}. 
We test this for \targetshort by comparing the inferred flux for rest-optical lines redward of \NeIIIL (to avoid further uncertainties on the continuum modelling in the PRISM spectrum introduced by the presence of a `Balmer jump'), and find that the gratings' fluxes are higher by $\sim10$--$12$ percent compared to those measured from the PRISM, a value slightly larger than the average reported in JADES data release papers \citep{bunker_hst_deep_DR_2023, d_eugenio_jades_DR3_2024}.
Therefore, to avoid introducing systematics associated with uncertain scaling factors, throughout this paper we consider only line ratios computed within the same spectral configuration (i.e. we do not use gratings-to-prism line ratios), and avoid to adopt UV-to-optical line ratios where possible to infer the physical properties of interest. 
As a general criterion, we primarily adopt line ratios from medium resolution grating spectra where available, especially in the rest-frame optical (while testing the consistency of our results by comparing them with those inferred from PRISM-based fitting), whereas we resort to line ratios measured from the PRISM when no detections are available from the gratings (as in the case of some rest-UV emission lines).
In general, we note that, despite the offset in the absolute flux calibration between PRISM and gratings, we find that line ratios among the same set of lines derived from either configuration generally agree within their respective uncertainties. 
% \CIVL[1548]

% \input{tables/table_fluxes_SN}
\begin{table*}
\caption{Emission line fluxes and rest-frame equivalent widths measured in both PRISM and gratings (G140M and G395M) spectra.}
    \centering
    \setlength{\tabcolsep}{4pt}
    \begin{adjustbox}{width=0.95\textwidth}
    \begin{tabular}{l|cccc|cccc}
    \toprule
    Line & \multicolumn{4}{c}{gratings (R1000)} & \multicolumn{4}{c}{prism (R100)}  \\
    &  Flux & S/N & S/N & EW$_{0}$ &  Flux & S/N & S/N & EW$_{0}$ \\
    & [10$^{-19}$erg s$^{-1}$ cm$^{-2}$] & pipeline & bootstrapping & [$\AA$] & [10$^{-19}$erg s$^{-1}$ cm$^{-2}$] & pipeline & bootstrapping & [$\AA$] \\
    % \vspace{0.2cm}
    \midrule
\Lyalpha & 1.89 & 2.7 & 2.3 & 31$\pm$16 & -- & -- & -- & -- \\
\NIVL[1483]$\dagger$ & 1.49$^{\star}$ & 3.4 & 2.8 & 3$\pm$1 & 1.34$^{\star\star}$ & 2.1 & 1.8 & 2$\pm$1 \\
\NIVL[1486] & < 1.34$^{\star}$ & -- & -- & -- & -- & -- & -- & -- \\
\CIVL[1549] & < 1.85 & -- & -- & -- & -- & -- & -- & -- \\
\CIVL[1551]$\dagger$ & 2.43 & 4.4 & 3.2 & 6$\pm$2 & 4.20 & 6.0 & 6.4 & 7$\pm$1 \\
\HeIIL & 3.07 & 4.9 & 3.5 & 8$\pm$3 & 1.72 & 4.3 & 7.3 & 7$\pm$1 \\
\OIIIL[1661] & < 1.71 & -- & -- & -- & -- & -- & -- & -- \\
\OIIIL[1666]$\dagger$ & 4.40 & 7.0 & 3.5 & 10$\pm$4 & 3.33 & 6.5 & 7.3 & 9$\pm$1 \\
\NIIIall & < 1.73 & -- & -- & -- & 1.16 & 2.7 & 3.0 & 3$\pm$1 \\
\CIIIall & -- & -- & -- & -- & 4.33 & 9.9 & 10.5 & 14$\pm$2 \\
\NeVL & -- & -- & -- & -- & 0.46 & 3.3 & 2.9 & 4$\pm$2 \\
\OIIL[3727] & 0.59 & 3.2 & 5.6 & 19$\pm$5 & -- & -- & -- & -- \\
\OIIL[3729]$\dagger$ & 0.55 & 3.2 & 2.1 & 16$\pm$6 & 0.89 & 6.0 & 5.2 & 15$\pm$4 \\
H10 & 0.43 & 3.0 & 2.6 & 6$\pm$3 & < 0.49 & -- & -- & -- \\
H9 & 0.55 & 3.7 & 3.1 & 9$\pm$4 & < 0.38 & -- & -- & --  \\
\NeIIIL[3869] & 3.04 & 16.7 & 8.6 & 50$\pm$13 & 2.52 & 16.4 & 10.7 & 66$\pm$8 \\
H8 + He I & 1.03 & 5.8 & 5.3 & 15$\pm$5 & 0.87 & 6.9 & 6.3 & 34$\pm$4 \\
\NeIIIL[3968] + H$\epsilon$ & 1.73 & 8.8 & 7.2 & 29$\pm$6 & 1.79 & 12.3 & 10.1 & 37$\pm$5 \\
H$\delta$ & 1.97 & 11.9 & 8.1 & 31$\pm$7 & 1.88 & 13.3 & 11.7 & 39$\pm$4 \\
H$\gamma$ & 4.03 & 16.3 & 12.3 & 66$\pm$16 & 3.18 & 20.3 & 18.8 & 79$\pm$7 \\
\OIIIoptL[4363] & 1.35 & 6.4 & 5.7 & 22$\pm$7 & 1.09 & 7.0 & 7.0 & 49$\pm$8 \\
H$\beta$ & 7.54 & 24.3 & 16.4 & 92$\pm$29 & 6.76 & 36.5 & 35.9 & 182$\pm$19 \\
\OIIIoptL[4959] & 13.24 & 40.7 & 32.5 & 161$\pm$42 & 11.07 & 58.0 & 48.0 & 249$\pm$18 \\
\OIIIoptL[5007] & 37.36 & 102.5 & 74.3 & 452$\pm$113 & 29.22 & 149.2 & 114.7 & 600$\pm$43 \\
\bottomrule
\end{tabular}
\end{adjustbox}
\tablefoot{The signal-to-noise on the lines is estimated either from the pipeline error spectrum, and by repeating the fit on 300 realisations of the spectra generated via randomly bootstrapping (with replacement) over the 186, 42, and 114 individual integrations available for the PRISM, G140M, and G395M configurations, respectively. $3\sigma$ upper limits are reported for non-detections.
\tablefoottext{$\dagger$}{\NIVL[1483,1486], \CIVall, \OIIIall, \OIIall are modelled as unresolved doublets in the PRISM spectrum.} 
\tablefoottext{$\star$}{From 3-pixel boxcar extracted G140M spectrum} 
\tablefoottext{$\star\star$}{When including the \NIVL Gaussian component in the PRISM fitting (see Section~\ref{sec:UV_fitting} for details)} }
    \label{tab:line_fluxes}
\end{table*}

\subsection{Photometry and full spectral fitting}
\label{sec:SED_fitting}
We performed \textsc{forcepho} (Johnson et al. in prep.) fitting of the \targetshort system to forward model the light distribution and extract the photometry and morphological parameters from all the available NIRCam images in individual medium- and wide-band filters, namely F090W, F115W, F150W, F182M, F200W, F210M, F277W, F335M, F356W, F410M, F444W.
The \textsc{forcepho} setup follows that adopted in \cite{tacchella_gnz11_2023} and \cite{baker_core_bulge_2025}, and we model the galaxy with a single Sérsic component.
The extracted \textsc{forcepho} photometry is reported in Table~\ref{tab:fpho_phot}.
Based on the morphological parameters derived from the \textsc{forcepho} fitting, we infer a high compactness for this source, with an effective radius as small as $\text{R}{_e} = 110\pm9$ pc. %and a high S\'ersic index $n=5.50$.  

\begin{table*}
\caption{\textsc{forcepho} photometry from available NIRCam images in both wide and medium band filters.}
\centering
\setlength{\tabcolsep}{4pt}
\begin{adjustbox}{width=0.99\textwidth}
\begin{tabular}{ccccccccccc}
\toprule
F090W & F115W & F150W & F182M & F200W & F210M & F277W & F335M & F356W & F410M & F444W \\
\midrule
$<$0.42$^\dagger$ & $<$0.88$^\dagger$ & 53.67$\pm$0.30 & 45.59$\pm$0.66 & 49.99$\pm$0.34 & 39.31$\pm$0.64 & 43.21$\pm$0.42 & 36.66$\pm$0.25 & 40.31$\pm$0.39 & 42.90$\pm$0.30 & 40.25$\pm$0.43 \\
\bottomrule
\end{tabular}
\end{adjustbox}
 \tablefoot{Units are in nJy, and quoted errors only report formal statistical uncertainties.
 \tablefoottext{$\dagger$}{~3$\sigma$ upper limit.} }
\label{tab:fpho_phot}
\end{table*}

\begin{figure*}
    \centering
    \includegraphics[width=0.48\textwidth]{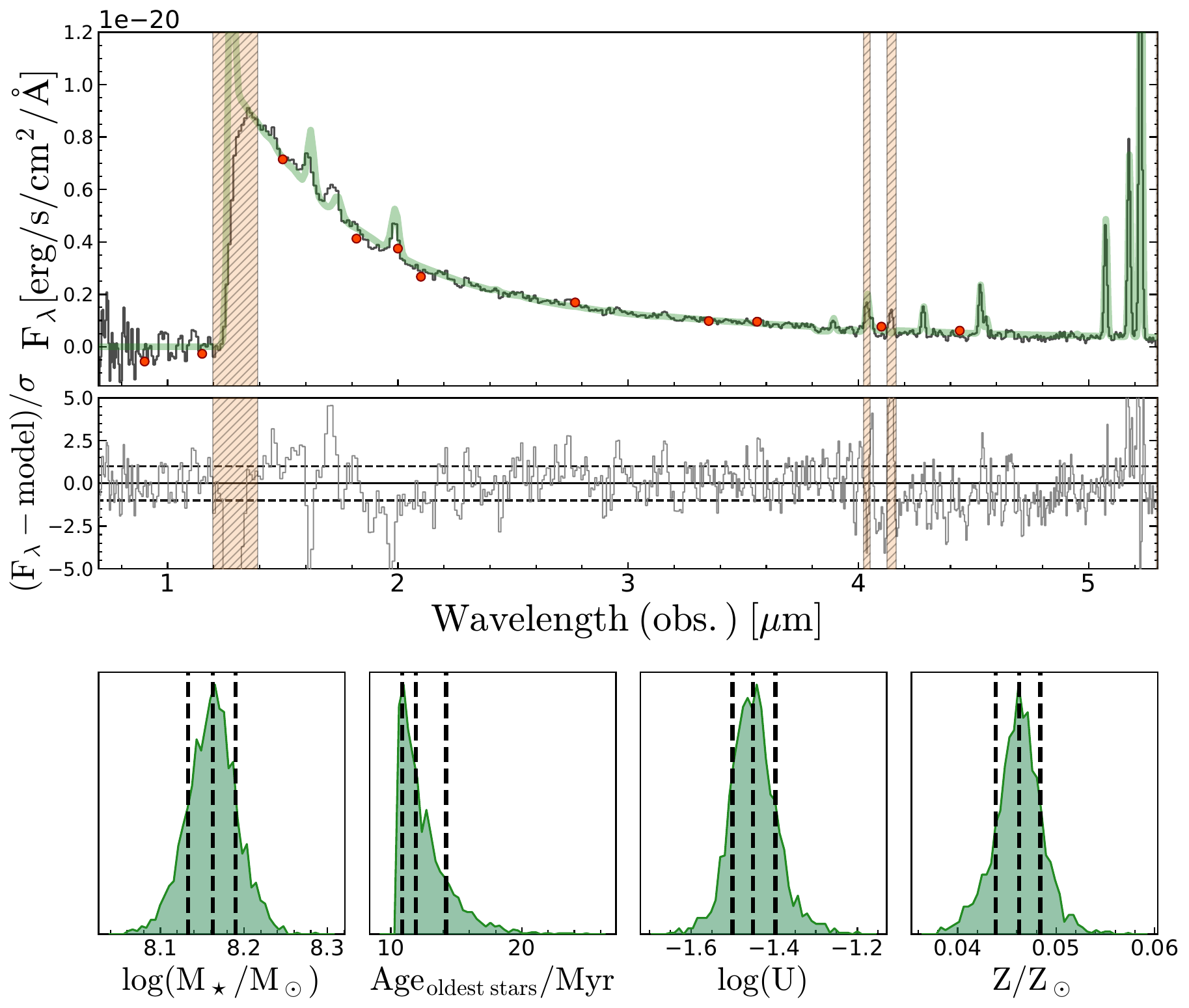}
    \includegraphics[width=0.48\textwidth]{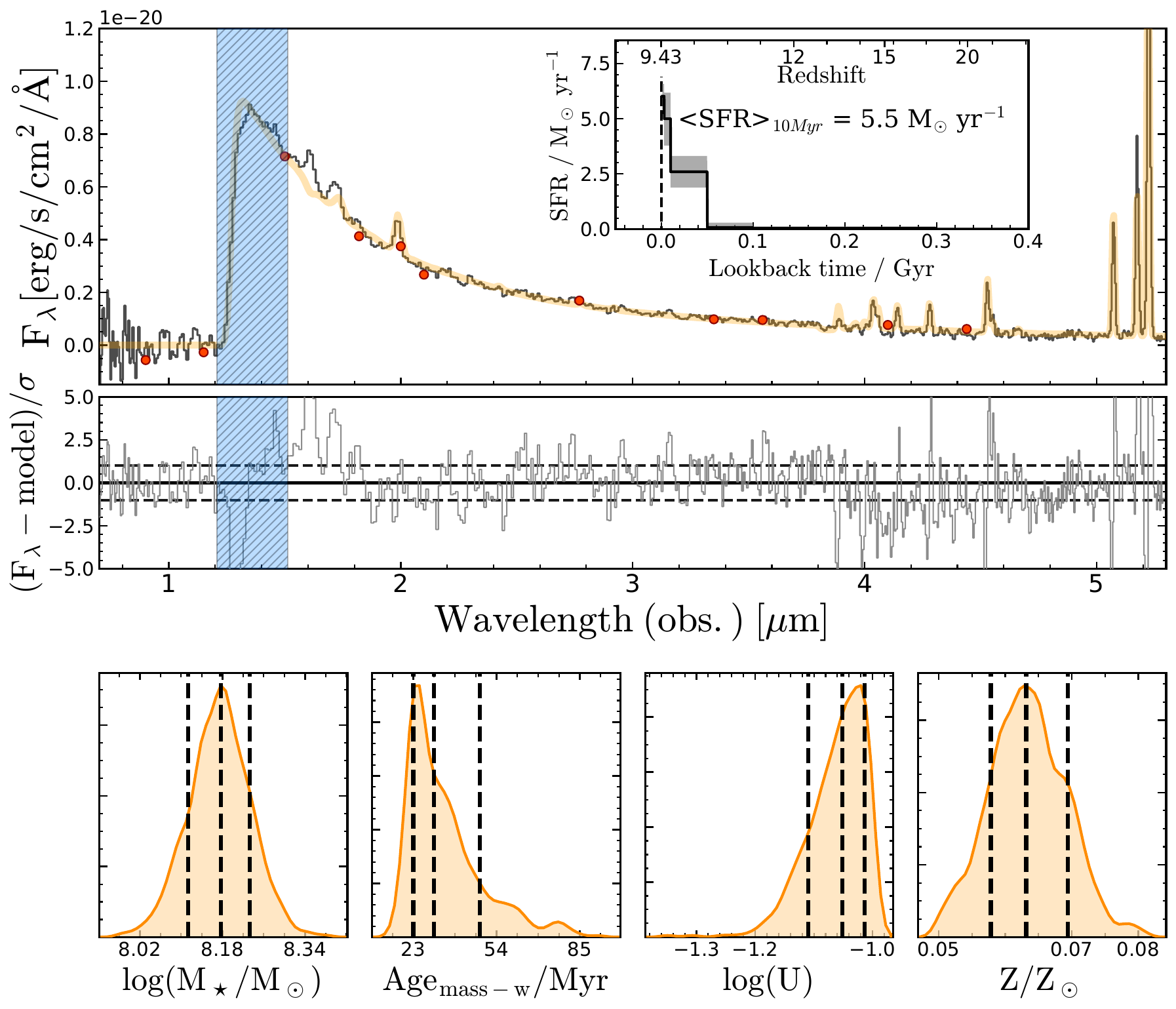}
    \caption{SED fitting to the PRISM spectrum of \targetshort.   
    The left-hand panel shows the fit performed with \textsc{beagle}, while the right-hand panel shows the fit adopting \textsc{bagpipes} with a non-parametric star-formation history.
    In both cases, the upper panel shows the best-fit (minimum chi-square) model spectrum superimposed on the observed spectrum (in grey) and the \textsc{forcepho} photometry (red points). 
    The shaded areas mark the wavelength intervals masked during the fit (in particular, the region around the \Lyalpha break).
    The inferred non-parametric star-formation history is depicted in the inset panel for the \textsc{bagpipes} fit.
    The lower panels report the marginalised posterior PDFs from the two fits for stellar mass, stellar age, ionization parameter, and metallicity (relative to solar), respectively.}
    \label{fig:bagpipes}
\end{figure*}

We then model the full SED of the galaxy by simultaneously fitting PRISM spectrum and photometry with different codes, namely \textsc{beagle} \citep{chevallard_beagle_2016} and \textsc{bagpipes} \citep{carnall_bagpipes_2018}.
We note that, given the excellent agreement between the pathlosses corrected PRISM spectrum and the extracted \textsc{forcepho} photometry (see Figure~\ref{fig:plot_spectra}), no significant re-scaling of the spectrum is required in the procedure.
We summarise the results of SED fitting from both codes in Table~\ref{tab:properties} and Figure~\ref{fig:bagpipes}.
The resulting best-fit spectra (corresponding to the minimum chi-square model) are shown in the top panels of Figure~\ref{fig:bagpipes}, whereas in the bottom panels we report the posterior PDFs for stellar mass, stellar age, ionization parameter, and metallicity (median, 16th, and 84th percentiles are marked with dashed black lines).

The \textsc{beagle} setup mimics that employed in previous studies \citep[e.g.][]{d_eugenio_gsz12_2023, hainline_zgtr10_jades_2024}: in brief, we set an upper-mass cut-off for a \cite{chabrier_galactic_2003} IMF to 300 \MSun, and model the SFH as a delayed-exponential with a burst occurring in the last $10$~Myr. The metallicity of stars is tied to that of the nebular gas, and we implement dust attenuation following the prescriptions of \cite{charlot_simple_2000}. 
Overall, the best-fit spectrum provides a good match to the observed continuum and spectral shape, while struggling to match the intensity of some of the rest-frame UV emission lines such as \HeII and \CIV.
We discuss further on the possible mechanisms powering nebular emission lines in Section~\ref{sec:dust_ion}. 
The \textsc{beagle} fit favours a recent history of star-formation for \targetshort, 
(log(sSFR/yr$^{-1}$)$=-7.52$, with an inferred light-weighted age of $\sim3$~Myr and an age of the oldest stars $\sim12$~Myr), a scenario consistent with the chemical abundance patterns discussed in Section~\ref{sec:discussion}.
We infer a low metallicity ($0.046 \pm 0.002$~Z$_{\odot}$), in good agreement with that derived from the \Te-method (Section~\ref{sec:abundances}), and a relatively high ionization parameter, log(U)$=-1.46 \substack{+0.06\\-0.04}$. 
For comparison, adopting for instance the photoionization models presented in \cite{berg_chemical_2019} we would derive a relatively lower log(U)=$-1.78 \pm 0.10$ based on the \OIIIoptL/\OIIall ratio and given the measured \Te-metallicity.

% Results from \textsc{beagle} generally agree well with \textsc{bagpipes} estimates, modelling GS-z9 as a galaxy with a recent history of star formation (log(sSFR)$=-7.52$, max stellar age of $12 \substack{+2\\-1}$ Myr), a scenario consistent with the independent constraints set by chemical abundance patterns, as discussed later. We note \textsc{beagle} infers a relatively lower (though still `high' in an absolute sense) ionization parameter log(U)$=-1.41$ compared to \textsc{bagpipes} (log(U)$=-1.06$).
% We summarise the results of SED fitting from both codes in Table~\ref{tab:properties}. 
% \textsc{bagpipes} \citep{carnall_bagpipes_2018}, by fixing the source redshift to the value inferred from emission lines.

We explored the impact of systematics associated with the adoption of different stellar population synthesis models and star-formation histories on our inferred physical properties by fitting the data also with \textsc{bagpipes}.
In this setup, we employ the Binary Population and Stellar Synthesis (BPASS) v2.2.1 \citep{Stanway_Eldrige_BPASS_2018} templates including the evolution of binary systems, an upper mass cut-off of 300 \MSun for a Kroupa IMF, dust extinction modelled by a \cite{calzetti_dust_2000} law, and IGM attenuation models from \cite{inoue_igm_2014}.
Nebular emission (in form of lines and continuum) is included by processing the BPASS stellar templates through the CLOUDY photoionization code \citep{ferland_cloudy_2017}.
We leave the gas metallicity and ionization parameter as free parameters in the fit, but inform our priors (especially for metallicity) by exploiting the information empirically derived from emission lines (see Section~\ref{sec:abundances}); in particular, we adopt uniform priors on log(Z/Z$_{\odot}$) $\in$ [0.01, 0.25] and log(U) $\in$ [-3, -1]. 
Finally, we assume a so-called non-parametric SFH following the recipe of \cite{leja_sfh_2019}, defining six age bins counted backwards from the epoch of observations (the first two spanning $0$--$3$ and $3$--$10$ Myr, respectively).
The priors on $\Delta$log(SFR) values among adjacent bins are modelled as a Student's-\textit{t} distribution with scaling factor $\sigma=0.3$ (`continuity' model).
The top-right inset panel of Figure~\ref{fig:bagpipes} shows the inferred SFH, reporting the SFR in each of the temporal bins considered, as a function of lookback time. 
The \textsc{bagpipes} fit confirms the very recent history of mass assembly for \targetshort, predicting the majority of star-formation to have occurred within the last two bins\footnote{We note however that younger stars generally outshine older populations, whose contribution might be difficult to infer from SED fitting}, with a total stellar mass of log(M$_{\star}) = 8.18\substack{+0.06\\-0.06}$ \MSun, and a mass-weighted age of $32\substack{+20\\-9}$ Myr, while the star-formation rate averaged over the past 10 Myr is
SFR $= 5.5\substack{+0.2\\-0.2}$ \MSun yr$^{-1}$, in agreement with the values inferred both by \textsc{beagle} ($= 4.34\substack{+0.10\\-0.08}$ \MSun yr$^{-1}$), and by 
applying the calibration for low metallicity systems from \cite{reddy_lyA_2022, shapley_balmer_2023} to the measured \Hbeta flux ($= 5.46\pm1.04$ \MSun yr$^{-1}$).
Assuming the latter as fiducial value, and given the compactness of the system ($\text{R}_{e}\sim110$pc), this translates into a high star-formation rate surface density of $\Sigma_{\text{SFR}}=72 \pm 14$~\MSun yr$^{-1}$ kpc$^{-2}$, and into a stellar mass surface density of $\Sigma_{\text{M}_{\star}}=1.85\pm 0.3 \times 10^3$~\MSun pc$^{-2}$.
We note \textsc{bagpipes} prefers an even more extreme ionization parameter log(U)$=-1.06$ compared to \textsc{beagle} (log(U)$=-1.46$), whereas the derived metallicity is fully consistent.
A possible source of the discrepancy between the two inferred estimates of log(U) stems from its definition: in \textsc{beagle}, the ionization parameter is defined at the Str{\"o}mgren radius (Equation~7 in \citealt{gutkin_modelling_2016}), whereas \textsc{bagpipes} follows the prescription of \cite{byler_nebular_2017} and defines log(U) at the inner radius of the illuminated cloud, hence its value is expected to be systematically higher.

\begin{figure}
    \centering
    \includegraphics[width=0.48\textwidth]{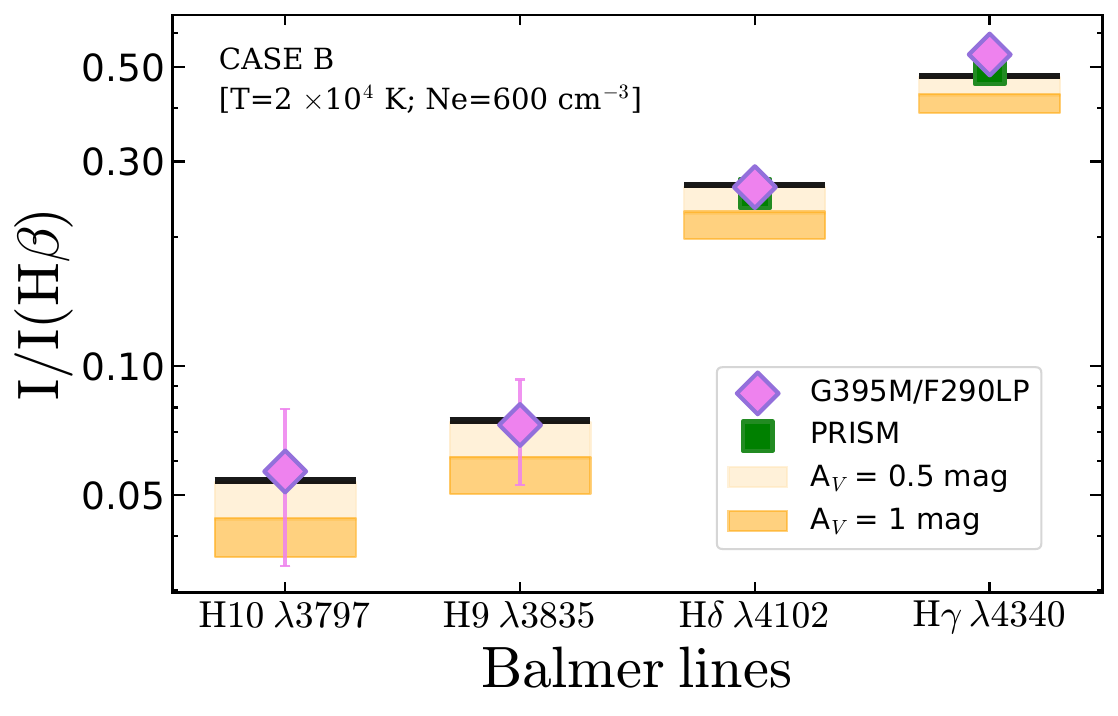}
    \caption{Nebular attenuation inferred from Balmer decrements.
    The ratios between different Balmer lines and \Hbeta as measured from both G395M and PRISM spectra are compared with the theoretical values set by Case B recombination (solid black bars) for T$_{e}$=20,000 K and n$_{e}$=600 cm$^{-3}$ (consistent with electron temperature and density derived in Section~\ref{sec:abundances}). The ratios measured from both PRISM and gratings suggest negligible dust attenuation, with best-fit \AV values consistent with zero within $1\sigma$ uncertainty.}
    \label{fig:BD}
\end{figure}

Both SED fitting codes infer very low dust attenuation (\AV$=0.050\substack{+0.008\\-0.006}$ and \AV$=0.004\substack{+0.005\\-0.002}$ from \textsc{beagle} and \textsc{bagpipes}, respectively), in agreement with the measured \lq decrement\rq among Balmer lines ratios.
Having access to spectrally resolved \Hbeta, \Hgamma, and \Hdelta in both G395M and PRISM observations (with additional detections of higher order H9 and H10 lines in G395M), we can in fact obtain an independent constraint on the amount of nebular attenuation.
In Figure~\ref{fig:BD}, we compare the observed \Hgamma/\Hbeta, \Hdelta/\Hbeta, H9/\Hbeta, and H10/\Hbeta ratios with the theoretical values expected for case B recombination, assuming a temperature of $20000$~K and density n$_{e}$=600 cm$^{-3}$, consistent with the values measured directly from emission lines as detailed in Section~\ref{sec:abundances}.
The ratios measured independently from G395M and PRISM spectra are in excellent agreement, and are consistent within their uncertainties with negligible-to-no dust attenuation. 
In fact, the best-fit \AV derived from simultaneously fitting all the available Balmer line ratios with a \cite{gordon_LMC_attenuation_2003} SMC attenuation curve and R$_{\text{V}}=2.505$ are \AV$_{\text{G395M}}=-0.07\pm0.07$ and  \AV$_{\text{PRISM}}=0.01\pm0.07$, respectively.   

Finally, by converting the UV luminosity density at $1500\AA$ rest-frame as measured in the PRISM spectrum, we derive a UV magnitude of M$_{\text{UV}}$=--20.43, consistent with previous determinations based on the 1210 spectrum \citep{boyett_EELGs_jades_2024}. The inferred SFH, UV brightness, UV $\beta$ slope, and negligible dust attenuation in \targetshort aligns with observations of several other $z>10$ galaxies \citep[e.g.][]{bunker_gnz11_2023, Arrabal_Haro_nature_2023, curtis-lake_2023, castellano_ghz12_2024, carniani_z14_2024}, in agreement with  theoretical scenarios invoking rapid star-formation and dust-free environments (with dust possibly expelled by fast outflows, e.g. \citealt{ferrara_BM_2024a, ferrara_BM_2024b}) as the responsible for the overabundance of luminous systems at high-z observed by the \jwst \citep[e.g.][]{Ferrara_UV_bright_JWST_2023, mason_UVLF_2023}.

\section{Ionization source}
\label{sec:dust_ion}

The ratios among different emission lines (either collisionally excited or produced by recombination) can be modelled to constrain the nature of the dominant source of photoionization in galaxies. 
In the case of \targetshort, we can leverage high S/N detections of several emission lines in both rest-frame optical and rest-frame UV regimes to explore a variety of different diagnostic diagrams. %with the aim of constraining the nature of its ionization source. 
Rest-frame optical diagrams for this source (e.g. \OIIIopt/\OII vs (\OIIIopt+\OII)/\Hbeta) have been explored already in \cite{cameron_jades_bpt_2023} as based on the data from the 1210 programme. In those diagrams, \targetshort is observed to occupy the region belonging to local analogues of high-z galaxies characterised by low-metallicity and high ionization parameter.

More recently, \cite{scholtz_jades_agn_2023} explored some rest-frame UV diagnostics with the aim of selecting robust type-2 AGN candidates within JADES.
According to the criteria outlined in \cite{scholtz_jades_agn_2023}, and following, in particular, a possible ($\sim4\sigma$) detection of \NeIVL in the G235M spectrum (a high-ionization emission line which requires very hard ionising continua to be powered, e.g. \citealt{brinchmann_ERO_2023}), \targetshort is classified as a type-2 AGN. As mentioned already in Section~\ref{sec:lines_fitting}, this emission line, however, is formally undetected in our PRISM spectrum (we quote a $2.4\sigma$ significance based on the pipeline error spectrum, and only $1.6\sigma$ based on bootstrapping).
% In addition, they reported the detection of Ne IV$\lambda2424$ in the G235M spectrum, a high-ionization emission line (ionization energy = 65 eV \todo{TBC}) which requires very hard ionising continua (typically associated to AGN) to be powered \citep[see e.g.][]{brinchmann_2022}.
%and which represents currently the best evidence in support of the AGN scenario. 

\begin{figure*}
    \centering
    \includegraphics[width=0.47\textwidth]{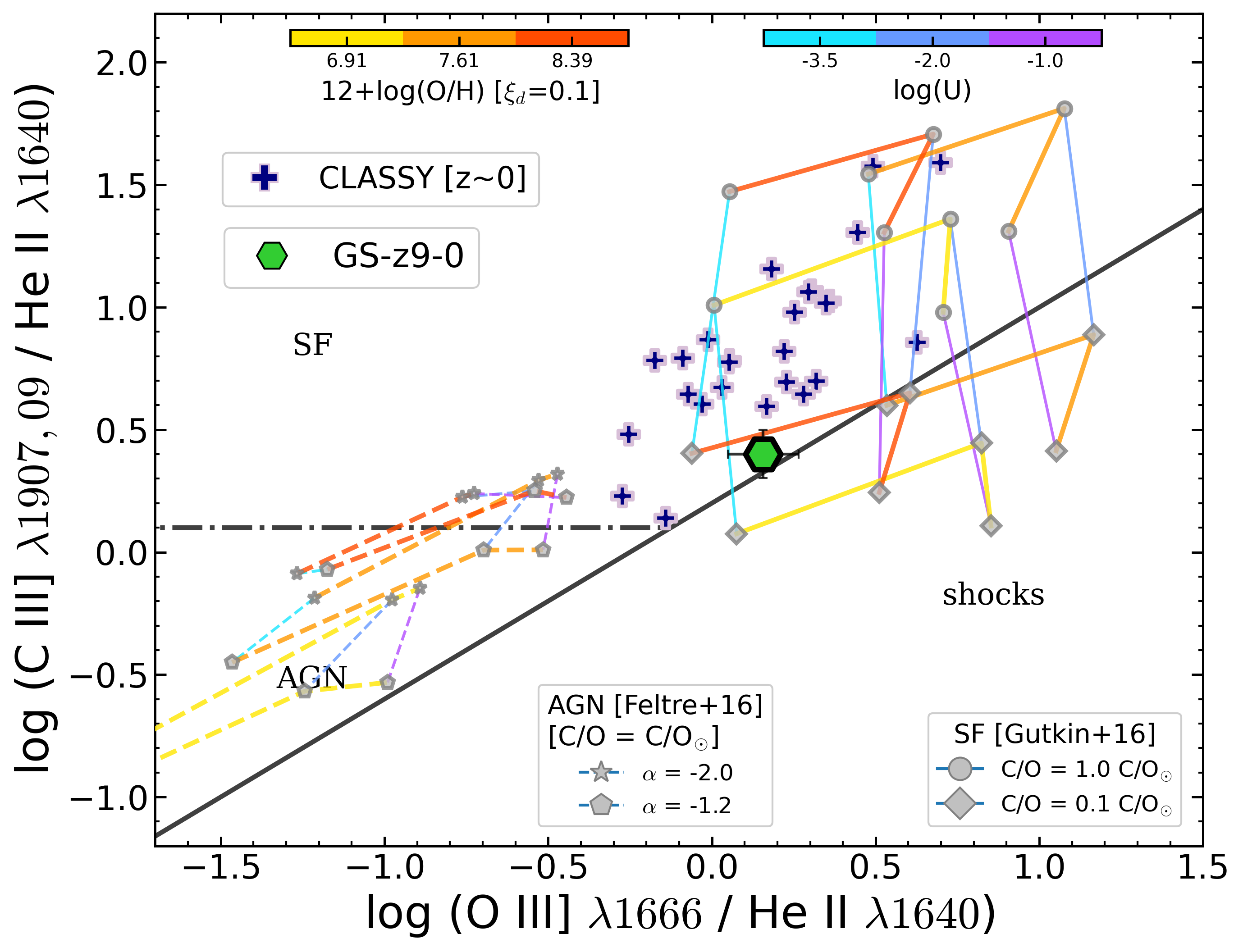}
    \includegraphics[width=0.47\textwidth]{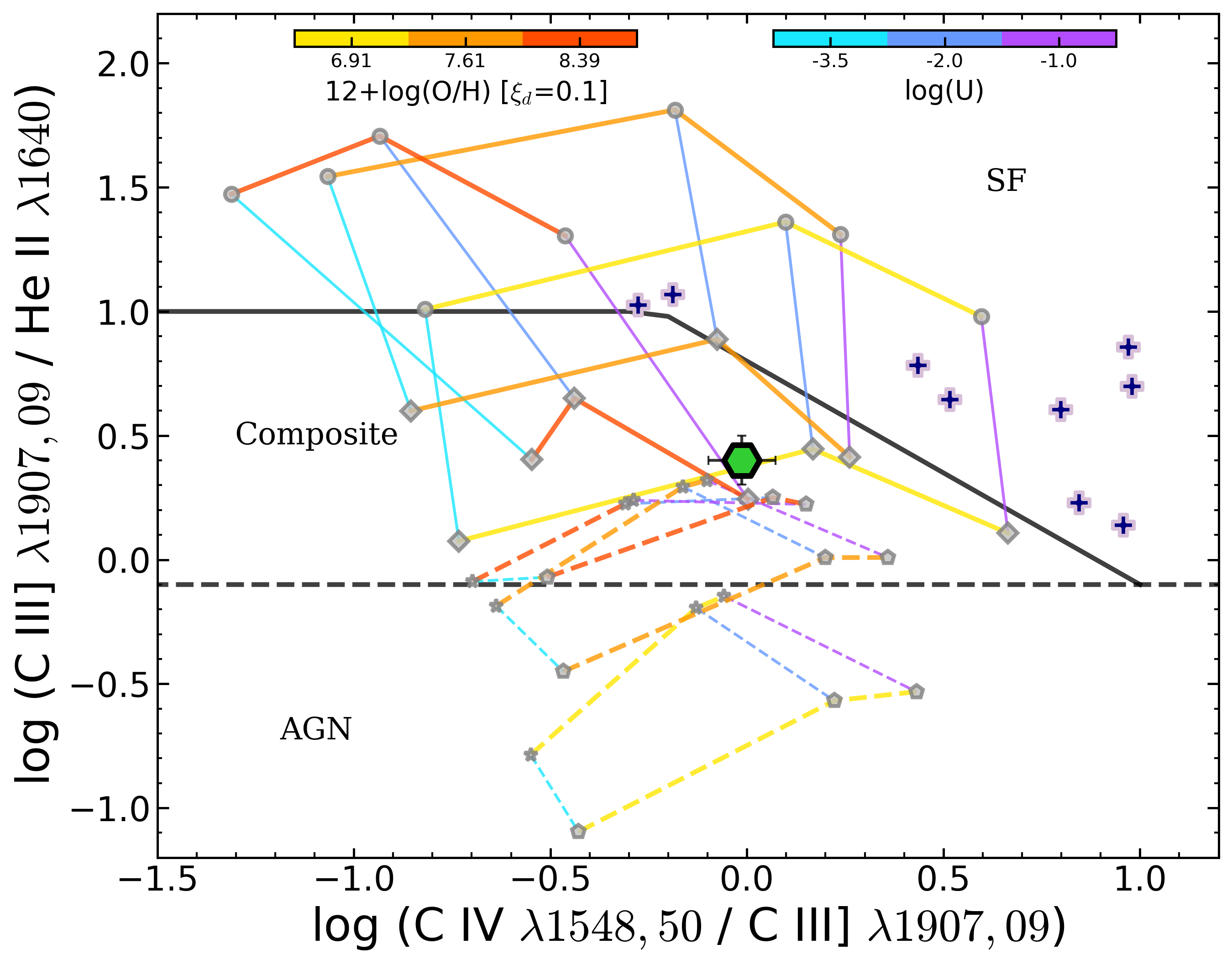} \\

     \centering
   \includegraphics[width=0.47\textwidth]{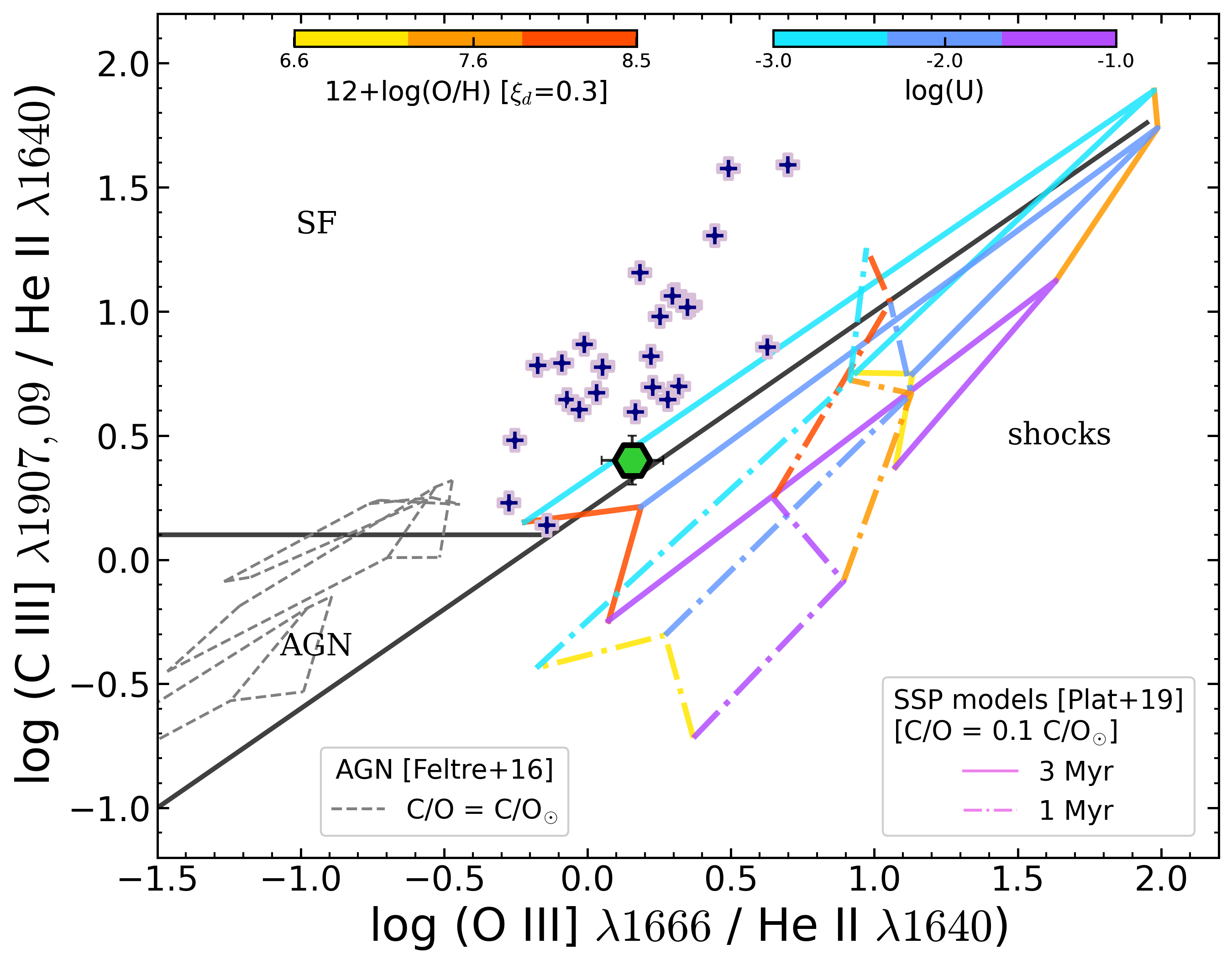}
    \includegraphics[width=0.47\textwidth]{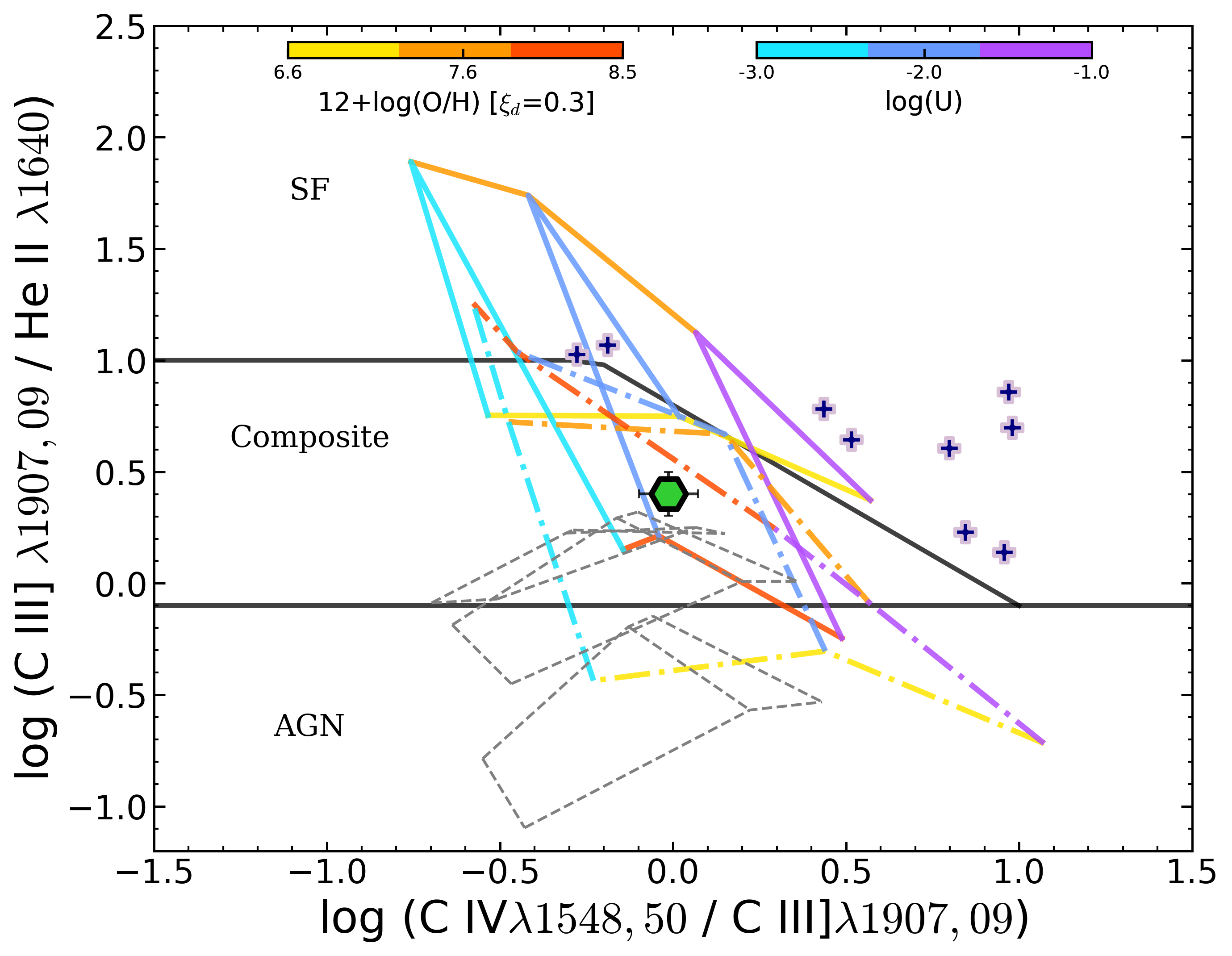} \\    
    
    \caption{Rest-frame UV diagnostic diagrams. The location of \targetshort is reported and compared to a sample of local, high-z analogues from the \textsc{CLASSY} survey \citep{berg_classy_2022, mingozzi_uv_2024}.
    Top panels: \CIIIall/\HeIIL vs \OIIIL/\HeIIL (C3He2-O3He2, left-hand panel) and \CIIIall/\HeIIL vs \CIVL/\CIIIall (C3He2-C43, right-hand panel) line ratios are compared with photoionization models by \cite{gutkin_modelling_2016} and \cite{feltre_uv_2016} for constant star-formation (solid lines), and NLR of AGNs (dashed lines), respectively.
    We fix the dust-to-metal ratio to $\xi_{d}$=0.1, while varying metallicity and ionization parameter (colour-coded), C/O abundance and slope $\alpha$ of the AGN continuum (different symbols). 
    \targetshort occupies a region fully consistent with star-formation driven ionization in the C3He2-O3He2 diagram, whereas it sits in a region of overlapping star-formation and AGN grids in the C3He2-C43 diagram.
    In the C3He2-O3He2 diagram, the solid and dot-dashed black lines represent the empirical demarcation between SF-, AGN-, and shock-driven ionization as proposed by \cite{mingozzi_uv_2024}, whereas in the C3He2-C43 diagram we report the demarcation lines between SF, AGN, and composite proposed by \cite{Hirschmann_UV_diags_2019}.
    Bottom panels: Same as upper panels, but for single stellar population models from \cite{plat_2019} (3 Myr and 1 Myr old for solid and dashed lines, respectively, with fixed $\xi_{d}$=0.3, and $\text{C/O}=0.1\text{C/O}_{\odot}$). The grids from 1 Myr stellar populations at low metallicity fully overlap with those of AGN-like ionization of higher metallicity and solar C/O ($\xi_{d}$=0.1) in the C3He2-C43 diagram.
    }
    \label{fig:diagnostic_diagrams}
\end{figure*}

Here, we compare several UV-based diagnostics with predictions from suites of photoionization models assuming different input ionising continua.
These are presented in Figure~\ref{fig:diagnostic_diagrams}.
More specifically, we focus our model predictions by exploring range of values in the physical properties matching (or in broad agreement with) those independently inferred for \targetshort from the emission lines (e.g. for O/H, C/O). 
This is done in an attempt to limit the degeneracies induced by comparing model grids from different ionising sources under very different physical conditions, even when they do not match those (empirically and independently) estimated for an individual galaxy.

More specifically, to model emission line ratios as predicted by ionization from star-formation, we adopt an updated version of the model grids described in \cite{gutkin_modelling_2016, plat_2019} for the two cases of constant star-formation history (upper panels of Figure~\ref{fig:diagnostic_diagrams}) and single burst of different ages (lower panels of \ref{fig:diagnostic_diagrams}), respectively.
We assume an upper-mass cut-off of the IMF = $300$ \MSun, a dust-to-metal mass
ratio $\xi_{d}$ = 0.1, and the gas density $n_{e}$ = 10$^{2}$ cm$^{-3}$ as fiducial values.
In the first scenario (constant star-formation) we assume a maximum stellar age of $100$ Myr, whereas for the single stellar population scenario, we explore two bursts of 1 and 3 Myr age, respectively (more reflective of the SFH inferred in Section~\ref{sec:SED_fitting}).
We then explore a range of values in ionization parameter (spanning between $-3<\text{log(U)}<-1$) and gas-phase metallicity ($0.015<\text{Z/Z}_{\sun}<0.7$).
In the upper panels, we explore also the variation in C/O abundance relative to solar ([C/O] $\in$ [-1, 0], where different [C/O] values are marked by different symbols), whereas in the middle panels [C/O] is fixed to $-1$  (in better agreement with that inferred from direct measurements as described in Section~\ref{sec:abundances}).
To model line ratios produced by the narrow-line-region (NLR) of AGN instead, we adopt the models from \cite{feltre_uv_2016}, fixing $\xi_{d}$ = 0.3 and 
$n_{e}$ = 10$^{3}$ cm$^{-3}$ (more consistent with the typical densities of the NLR of AGN), while varying ionization parameter, metallicity, and slope of the ionising continuum ($\alpha$ $\in$ [-2, -1.2]). For AGN grids, the C/O abundance is fixed to the solar value.

In Figure~\ref{fig:diagnostic_diagrams} we show the location of \targetshort on two different diagrams based on rest-frame UV lines, namely \CIIIall/\HeIIL vs \OIIIL/\HeIIL (left-hand panel, hereafter C3He2-O3He2) and \CIIIall/\HeIIL vs \CIVL/\CIIIall (right-hand panel, hereafter C3He2-C43). 
These diagrams have been recently suggested as some of the most reliable in discriminating the dominant ionising source in galaxy spectra \citep{mingozzi_uv_2024}, on the basis of observations of a sample of local, high-z analogues with full coverage of rest-UV spectrum from the CLASSY survey \citep{berg_classy_2022}: these objects are included in our plots for comparison.
In the C3He2-O3He2 diagram, \targetshort occupies the region probed by low-C/O grids at low-to-intermediate metallicity from the SF-models, and appears inconsistent with the model tracks produced by AGN-NLR ionization (while lying at the boundary of the SF-shocks demarcation line from \citealt{mingozzi_uv_2024}).
In the C3He2-C43 diagram instead, \targetshort falls at the intersection between the SF and AGN model grids. Notably, 1 Myr SSP model grids (dot-dashed lines in the bottom panels) span a region of the diagram that overlaps with the AGN/NLR grids, making it harder to disentangle the dominant contribution to ionization. 
However, we note that, because AGN grids are computed assuming a solar C/O abundance ratio, one could expect line-ratio grids for AGN-like ionization of lower C/O (e.g. $0.1\times$(C/O)$_{\odot}$) to be shifted from those shown in the panel by a similar amount as observed between [C/O]=-1 and [C/O]=0 SF-grids (at fixed other parameters): this in general would apply to all diagrams that intrinsically involve a dependence on the C/O abundance, and in the case of the C3He2-C43 diagram would make the agreement between \targetshort and AGN-like grids worse.

\begin{figure*}
    \centering
    \includegraphics[width=0.33\textwidth]{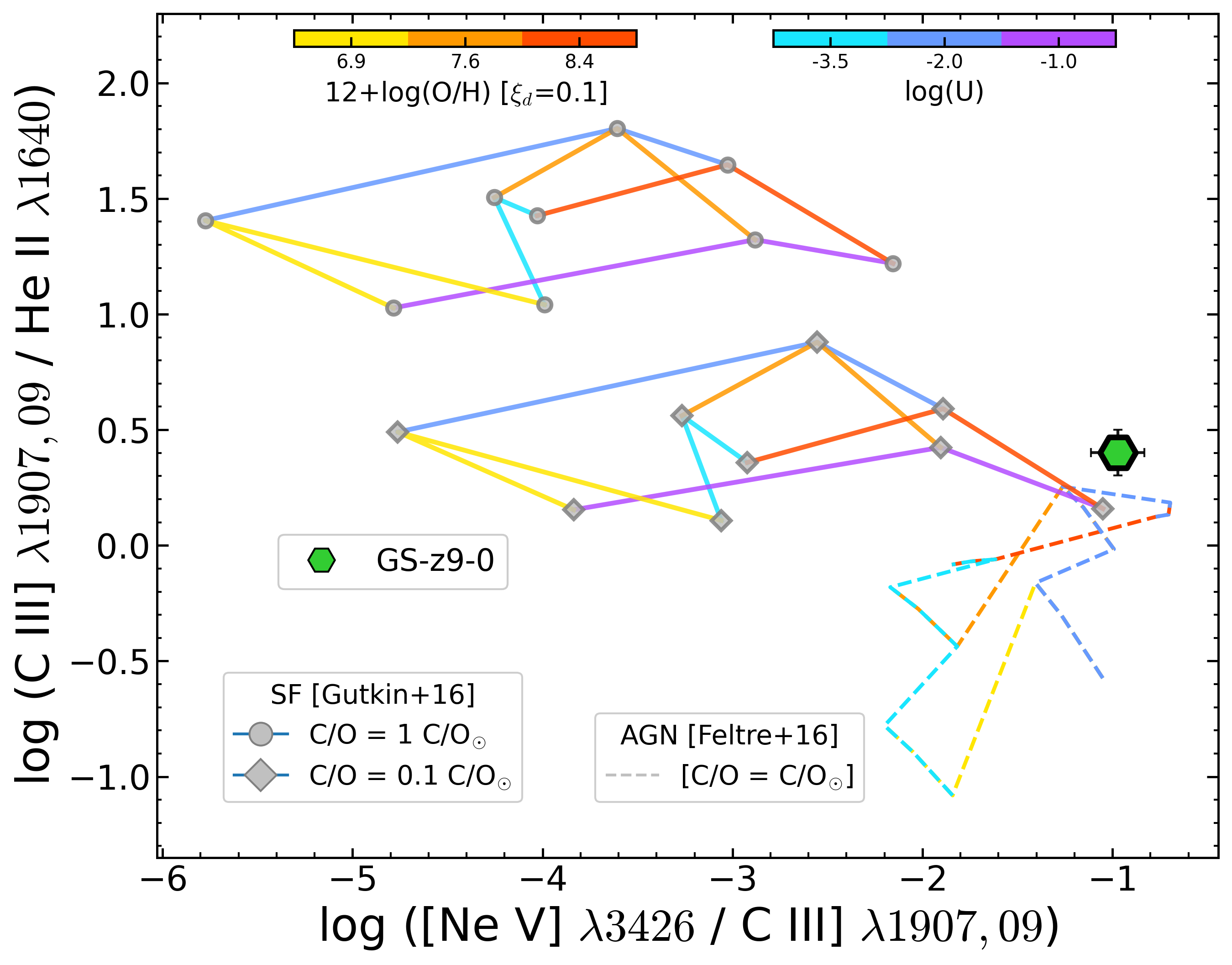}
    \includegraphics[width=0.33\textwidth]{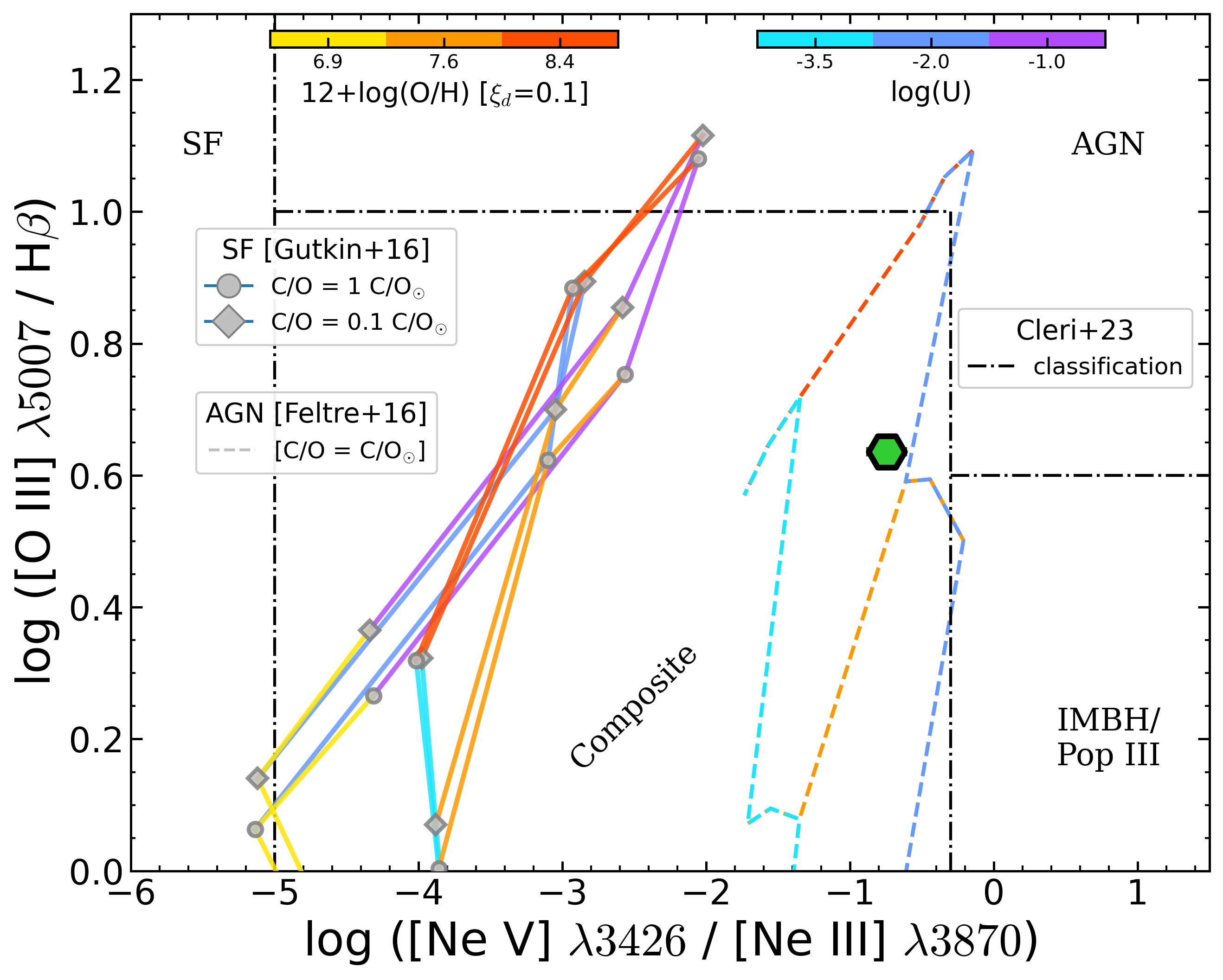}
    \includegraphics[width=0.33\textwidth]{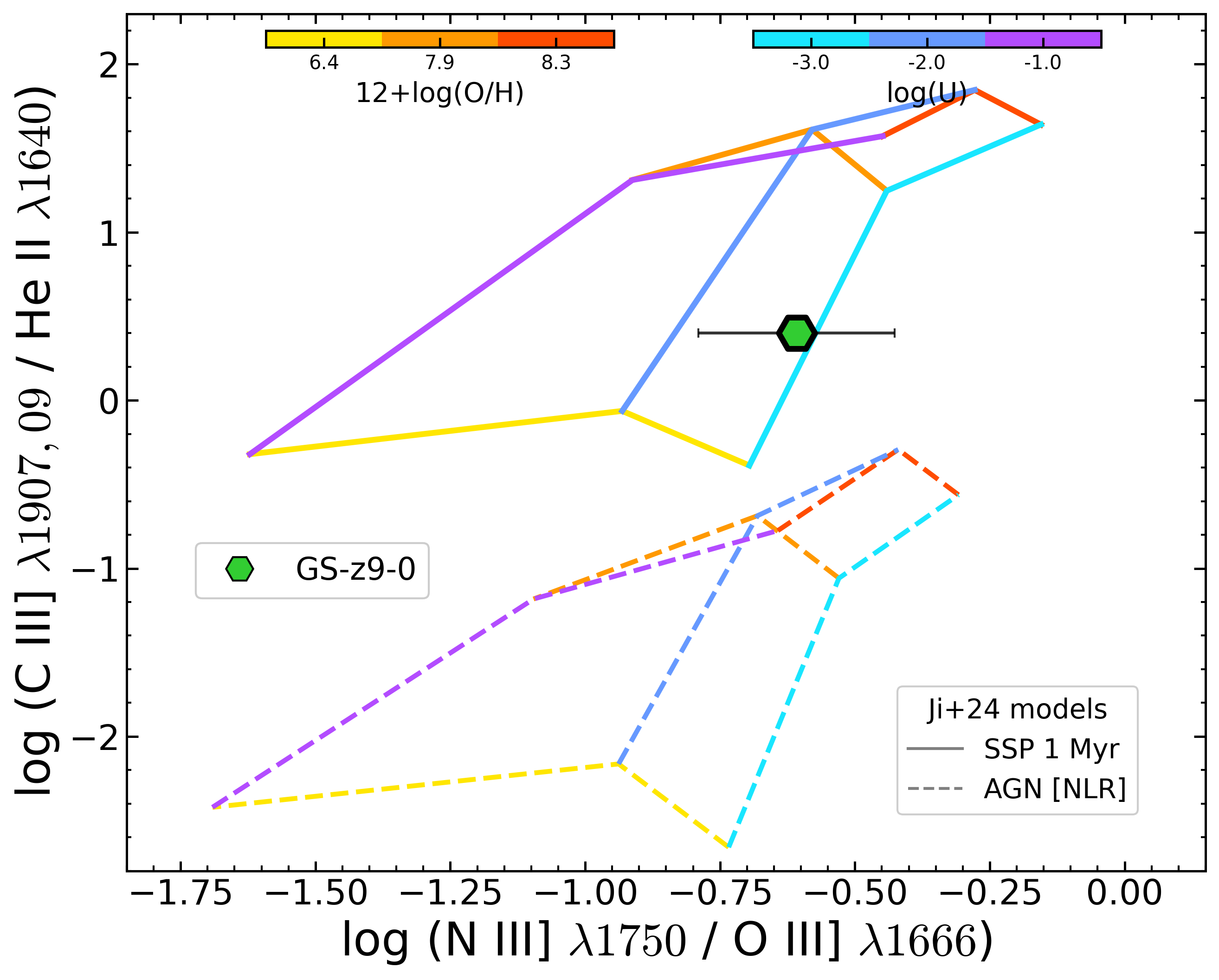}
    \caption{Alternative diagnostic diagrams for \targetshort. 
    %The green arrows mark the $3\sigma$ upper limits on line ratios based on assuming \NIIIL and \NeVL as non-detections, as discussed in Section~\ref{sec:lines_fitting}.
    The left-hand and middle panels leverage the very high ionization ($\sim97$eV) [Ne V]$\lambda3426$ emission line, and the same set of models from \cite{gutkin_modelling_2016} and \cite{feltre_uv_2016} as in Figure~\ref{fig:diagnostic_diagrams}. 
    The left-hand diagram (C3He2-Ne5C3) is sensitive to the C/O abundance, while the diagnostic considered in the middle panel (O3HB-Ne53) is not. Here, the location of \targetshort is more in line with AGN-powered line ratios, especially in the \OIIIoptL/\Hbeta vs \NeV/\NeIII diagram, where we show also the empirical demarcation lines between SF, AGN, and IMBH/PopIII models from \cite{cleri_NeV_2023}. 
    %although we note the model grids from \citealt{feltre_uv_2016} are computed only for solar C/O abundance.
    Right-hand panel: \CIIIL/\HeIIL vs \NIIIL/\OIIIL diagram. The location of \targetshort is compared with the set of photoionization models for star-formation (solid lines) and AGN/NLR (dashed lines) ionization from \cite{ji_nitrogen_AGN_z5_2024}. 
    }
    \label{fig:diagnostics_N3}
\end{figure*}

However, the spectrum of \targetshort reveals also the possible presence of a very high ionization emission line, which is challenging to explain by standard stellar population models, namely \NeVL (Figure~\ref{fig:prism_fit}). 
% We note that \cite{scholtz_jades_agn_2023} reported a $4\sigma$ detection of \NeIV$\lambda2424$ in G235M, however we do not detect this line in the combined 1210+3215 PRISM spectrum (we report a formal $2.4\sigma$ significance based on the pipeline error spectrum, and only $1.6\sigma$ based on bootstrapping). 
Therefore, we can exploit the detection of such emission line in our PRISM spectrum to explore different diagnostic diagrams: these are shown in the left and middle panels of Figure~\ref{fig:diagnostics_N3} for 
\CIIIL/\HeIIL vs \NeVL/\CIIIL (C3He2-Ne5C3) and \OIIIoptL/\Hbeta vs \NeVL/\NeIIIL (O3HB-Ne53), respectively.
\targetshort is located in-between SF and AGN grids in the C3He2-Ne5C3 diagram, but appears in slightly better agreement with AGN-ionization (though still broadly consistent with SF models of very high ionization parameter log(U)=-1), regardless of the dependence on C/O of the different grids (we recall that AGN/NLR models from \citealt{feltre_uv_2016} are computed assuming solar C/O). 
However, based on the O3HB-Ne53 diagram, \targetshort appears totally consistent with AGN ionization grids, with pure stellar population models struggling to produce significant \NeV.
The advantage of the O3HB-Ne53 diagram is that it exploits the large difference in minimum energy required to produce \NeV and \NeIII emission lines \citep[which trace different ionization zones][]{berg_4zones_2021}; moreover, it is based on lines closely spaced in wavelength (hence avoiding potential issues associated with wavelength-dependent slit loss correction and potential dust reddening), and it is also not affected by degeneracies in chemical abundances introduced by the use of emission lines of different elements.
Such diagram has also been recently proposed as a possible way to discriminate between the ionization produced by AGN with accreting supermassive black holes (M$_{\text{BH}} \geq 10^6$\MSun), AGN with intermediate-mass black holes (IMBH, M$_{\text{BH}} \lesssim 10^5$\MSun), and extreme stellar populations or even Population III stars \citep{cleri_NeV_2023}.
In the middle panel of Figure~\ref{fig:diagnostics_N3}, we  report the empirical demarcation lines for SF, AGN, and IMBH/Pop III stars based on the set of photoionization models presented in \cite{cleri_NeV_2023} (dashed black lines).
Interestingly, we note that although \targetshort resides in the `composite' region, its \OIIIoptL/\Hbeta ratio (which is not as high as in local AGN given the low metallicity of the system) places it not far from the region dominated by IMBH/Pop III models.
We note that in \cite{cleri_NeV_2023}, the latter models assume zero metallicity for the stellar population but a slightly pre-enriched gas-phase metallicity of Z=0.05Z$_{\odot}$ (similar to what inferred for \targetshort based on \Te\ measurements, Section~\ref{sec:abundances}) from primordial supernovae or stellar mass-loss events.

Finally, in the right-hand panel of Figure~\ref{fig:diagnostics_N3}, we plot a different diagnostic diagram, now leveraging the tentative detection of \NIIIL in the PRISM spectrum.
In particular, we explore 
%\NIIIL/\HeIIL vs \CIIIL/\HeIIL, which is primarily sensitive to the spectral hardness and ionization parameter \citep[see also][]{hirschmann_simul_jwst_2023}, and 
\CIIIL/\HeIIL vs \NIIIL/\OIIIL, which introduces explicitly the dependence on the N/O abundance.
For this purpose, we employ a set of grids generated with \textsc{cloudy}, with 1 Myr old SSP templates from BPASS as the input spectra to model star-formation, and a canonical SED with an effective big blue bump temperature of TBB = $10^{6}$ K, a UV-to-X-ray slope of $-1.4$, a UV slope of $-0.5$, and an X-ray slope of $-1.0$ to model AGN continuum.
We vary the metallicity Z/Z$_{\odot}$ between 5$\times10^{-4}$ and $2$, and the ionization parameter log(U) between $-3.5$ and $-1$. 
We adopt the prescriptions from \cite{groves_2004} to compute the initial N/O (before dust depletion) at every given O/H, while C/O is scaled accordingly assuming solar abundance patterns. More details are given in \cite{ji_nitrogen_AGN_z5_2024}.
In this diagram, \targetshort occupies a region consistent with stellar ionization. 

Summarising, on the basis of rest-UV diagnostics explored here, the line ratios observed in \targetshort are consistent with ionization from a population of massive and metal-poor stars, although the marginal \NeVL detection reported in the present work bring some evidence in support of the AGN scenario \citep[see e.g. the recent observations of such transition in the galaxy GN42437 at $z\sim6$, ][]{chisholm_NeV_2024}. 
Nonetheless, radiative shocks driven by stellar winds and SNe explosions have been also proposed as physical mechanisms capable of boosting the \NeVL emission in metal-poor star-forming galaxies
\citep{izotov_NeV_BCD_2012, lecroq_models_2024}.
In general, one should be probably careful about interpreting these diagnostic diagrams too rigidly. 
%Indeed, previous studies have shown that adopting different models, with different densities and ionization shape can result in broader (and overlapping) distributions of predicted line ratios \citep[e.g.][]{maiolino_gnz11_2023,castellano_ghz12_2024}. 
%For instance, the AGN models developed by \cite{nakajima_maiolino_2022} would reproduce the ratios observed in \targetshort.
In fact, it is not unlikely that the nebular spectrum of \targetshort is the result of mixing between different sources of ionization, as recently suggested by the analysis of similar spectra of high-redshift galaxies (e.g. GS-9422 at $z\sim6$, \citealt{tacchella_GS9422_2024}, GN-z11 at $z\sim10.6$, \citealt{bunker_gnz11_2023, maiolino_gnz11_2023}, GHz2 at $z\sim12.3$, \citealt{castellano_ghz12_2024}), 
and decoupling their relative contribution would require more detailed modelling and a better understanding of the shape of the ionising continua of young, massive stellar populations.

\section{Chemical abundances}
\label{sec:abundances}

\begin{table}
\caption{Derived physical properties for \targetshort.}
    \centering
    \setlength{\tabcolsep}{4pt}
    \begin{tabular}{l|c}
    \toprule
    Property & Value \\
    % \vspace{0.2cm}
    \midrule
 JADES ID & \target \\
 NIRSpec ID [3215] & 265801 \\
 NIRSpec ID [1210] & 10058975 \\
 RA & 53.1124351 \\
 Dec & -27.7746258 \\
 z$_{\text{G395M}}$ & 9.432681 $\pm$ 0.000069 \\
 z$_{\text{PRISM}}$ & 9.43774 $\pm$ 0.00020 \\
 M$_{\text{UV}}$ & -20.43 \\
$\beta_{\text{UV}}$ & -2.54$\pm$0.02 \\
log($\xi_{\text{ion}}$/erg$^{-1}$Hz) & 25.64 \\
\midrule
\multicolumn{2}{c}{Morphology (\textsc{forcepho} fitting)} \\
R$_{e}$ [arcsec] & 0.025$\pm$0.002 \\
R$_{e}$ [kpc] & 0.110$\pm$0.009 \\
PA [degree] & -64$\pm$3 \\
axis ratio $q$ & 0.44$\pm$0.04\\
% Sérsic $n$ & 5.50$\pm$0.02\\
\midrule
\multicolumn{2}{c}{SED modelling} \\
\multicolumn{2}{c}{\textsc{beagle}} \\
log(M$_{\star}$/M$_{\odot}$) & 8.17$\substack{+0.02\\ -0.04}$ \\
SFR [M$_{\odot}$yr$^{-1}$] & 4.34$\substack{+0.1\\ -0.08}$ \\
log(sSFR/yr$^{-1}$) & -7.52$\substack{+0.02\\ -0.04}$ \\
A$_{\text{V}}$ & 0.050$\substack{+0.008 \\ -0.006}$ \\
Age $_{\text{(light-weighted)}}$ [Myr] & 2.80$\substack{+0.06\\ -0.03}$ \\
Age $_{\text{(oldest stars)}}$ [Myr] & 12$\substack{+2\\ -1}$ \\
log(U) & -1.46$\substack{+0.06\\ -0.04}$ \\
Z/Z$_{\odot}$ & 0.046$\substack{+0.002\\ -0.002}$ \\
\multicolumn{2}{c}{\textsc{bagpipes}} \\
log(M$_{\star}$/M$_{\odot}$) & 8.18$\substack{+0.06 \\ -0.06}$ \\
SFR [M$_{\odot}$yr$^{-1}$] & 5.5$\substack{+0.2 \\ -0.2}$ \\
log(sSFR/yr$^{-1}$) & -7.44$\substack{+0.07 \\ -0.07}$ \\
A$_{\text{V}}$ & 0.004$\substack{+0.005 \\ -0.002}$ \\
Age $_{\text{(mass-weighted)}}$ [Myr] & 32$\substack{+20 \\ -9}$ \\
log(U) & -1.06$\substack{+0.03 \\ -0.06}$ \\
Z/Z$_{\odot}$ & 0.06$\substack{+0.06 \\ -0.06}$ \\\midrule
\multicolumn{2}{c}{ISM properties (emission lines)} \\
A$_{\text{V}}$ & 0.00$\pm$0.07\\
SFR (\Hbeta) [M$_{\odot}$yr$^{-1}$] & 5.46$\pm$1.04\\
$\Sigma_{\text{SFR}}$ [M$_{\odot}$ yr$^{-1}$ kpc$^{-2}$] & 72 $\pm$ 14 \\
n$_{e}$ [cm$^{-3}$] & 650$\pm$430 \\
 \Tiii (\OIIIopt[4363]) \ [K] & 20137$\pm$1940 \\
 \Tiii (\OIII[1666]) \ [K] & 24405$\pm$1700 \\
 12+log(O$^{+}$/H) & 6.00$\pm$0.14 \\
 12+log(O$^{++}$/H) & 7.38$\pm$0.09 \\
 12+log(O/H) (no ICF) & 7.40$\pm$0.09 \\
 ICF(O) & 1.23$\pm$0.20 \\
 12+log(O/H) (fiducial) & 7.49$\pm$0.11 ($\substack{+0.11 \\ -0.15}$ syst.)\\
 log(C$^{++}$/O$^{++}$) & -0.95$\pm$0.12 \\
 log(C$^{3+}$/C$^{++}$) & -0.36$\pm$0.10 \\
 log(C/O) (fiducial) & -0.90$\pm$0.12 ($\substack{+0.27 \\ -0.19}$ syst.) \\
 $[$C/O$]$ & -0.64 \\
 log(N/O) (fiducial) & -0.93 $\pm$ 0.24 ($\substack{+0.28 \\ -0.28}$ syst.) \\
 $[$N/O$]$ & -0.07 \\ 
 log(N/O)$^{\dagger}$ & -0.77 $\pm$ 0.18  \\
 $[$N/O$]$ $^{\dagger}$ & 0.09 \\ 
 % log(N/O)$^\dagger$ & < -0.66 \\
 log(C/N) (fiducial) & 0.03 $\pm$ 0.19 ($\substack{+0.24 \\ -0.24}$ syst.) \\
 % log(C/N)$^\dagger$ & >-0.24 \\
 log(Ne/O) & -0.68$\pm$0.06 \\
 $[$Ne/O$]$ & -0.05 \\  
    \bottomrule
    \end{tabular}
    \tablefoot{
    Fiducial values on chemical abundances are quoted with both their statistical uncertainties and including (co-added in quadrature) systematic uncertainties as discussed in Section~\ref{sec:systematics}.     
    \tablefoottext{$\dagger$}{Based on marginal \NIVL[1483] detection in the 3-pixel extracted G140M spectrum.}}
    \label{tab:properties}
\end{table}

The simultaneous detection of both nebular and auroral lines in the spectrum of \targetshort enables us to perform a detailed study of chemical abundance patterns in this galaxy, employing the `direct', \Te-method. 
Throughout this Section, we assume that ionisation comes primarily from star-formation (see Section~\ref{sec:dust_ion}).
However, we note that even in the case of possible AGN contribution to ionization as discussed in Section~\ref{sec:dust_ion}, this does not prevent a direct measurement of the abundances via the auroral lines, provided that the proper ionization correction factors (ICF) are included, which in the case of \targetshort we expect to be similar between AGN and stellar spectrum with hard ionizing continuum and/or high ionization parameter.

Throughout this Section, unless stated otherwise, we adopt \textsc{pyneb} for chemical abundances derivation, with atomic data and collision strengths tabulated from the \textsc{chianti} database.
%as reported in Table XY.
The errors on all the derived quantities are estimated by randomly perturbing the input emission line fluxes by their uncertainties (assumed Gaussian) and repeating the full procedure 100 times, taking the standard deviation of the distribution of values for each inferred parameter at each step of the procedure as our estimate of the (statistical) uncertainty associated to the fiducial value.
Additional systematic uncertainties in the derivation of chemical abundances are discussed throughout individual subsections for each given element, and more generally in Section~\ref{sec:systematics}.

\subsection{Gas temperature and density}
\label{sec:te_ne}

As a first step, we derive the temperature associated with the emitting region of O$^{++}$ (hereafter \Tiii) exploiting the high S/N detection of both \OIIIoptL[4363] and \OIIIoptL[5007] in the G395M grating spectrum.
The gas density is simultaneously derived exploiting the \OII doublet ratio, which is marginally resolved in G395M observations, while the temperature of the O$^{+}$ emitting region (hereafter \Tii) is 
assumed in the process to follow the temperature-temperature relation from \cite{izotov_low_2006}, i.e. $t_{2} = 0.693\xspace t_{3} + 2810$.
% The temperature of the O$^{+}$ emitting region is instead estimated adopting the temperature-temperature relation from \cite{izotov_low_2006}, $t_{2} = 0.693\xspace t_{3} + 2810$, which provides  
We infer electron temperatures of $t_{3}=20137\pm1940$ K and $t_{2}=16765\pm1345$ K, respectively. 
The gas density is not well constrained ($n_{e}=650\pm430$ cm$^{-3}$), given the \OIIall doublet is only marginally resolved in the G395M spectrum,
however its best-fit value is consistent with typical densities measured in high-redshift galaxies \citep[e.g.][]{isobe_density_jwst_2023}.
We assume, therefore, $n_{e}=650$ cm$^{-3}$ in our abundance calculations, noting that varying density between $100$ and $1000$ cm$^{-3}$ would produce a difference in the inferred oxygen abundance of only $\approx 0.01$~dex.
Nonetheless, the tentative detection of \NIVL[1483] in the absence of \NIVL[1486] (Figure~\ref{fig:uv_g140m}) provides complementary information on the density of the emitting gas.
In fact, this UV transition is another density-sensitive doublet which, in contrast to low-ionisation optical lines such as \OIIall, traces much higher density regimes.
The $3\sigma$ lower limit that can be placed on the \NIVL[1483/1486] ratio rules out extremely-high gas densities ($n_{e}>10^4$ cm$^{-3}$), being in broad agreement with the density regime probed by the \OII doublet. We can therefore reasonably exclude any significant contribution from high-density regions to the global emission line spectrum (whereas densities of the order of $n_{e}\approx 10^5-10^6$ cm$^{-3}$ have been measured in other bright UV galaxies at high-z, e.g. \citealt{topping_deep_UV_2024)}, which might hamper the simultaneous interpretation of rest-UV and rest-optical features and possibly bias also the metallicity determined with the \Te-method due to their unknown impact on emission lines of very different critical densities \citep{mendez_delgado_density_biases_2023, marconi_homerun_2024}.

\begin{figure}
    \centering
    \includegraphics[width=0.95\columnwidth]{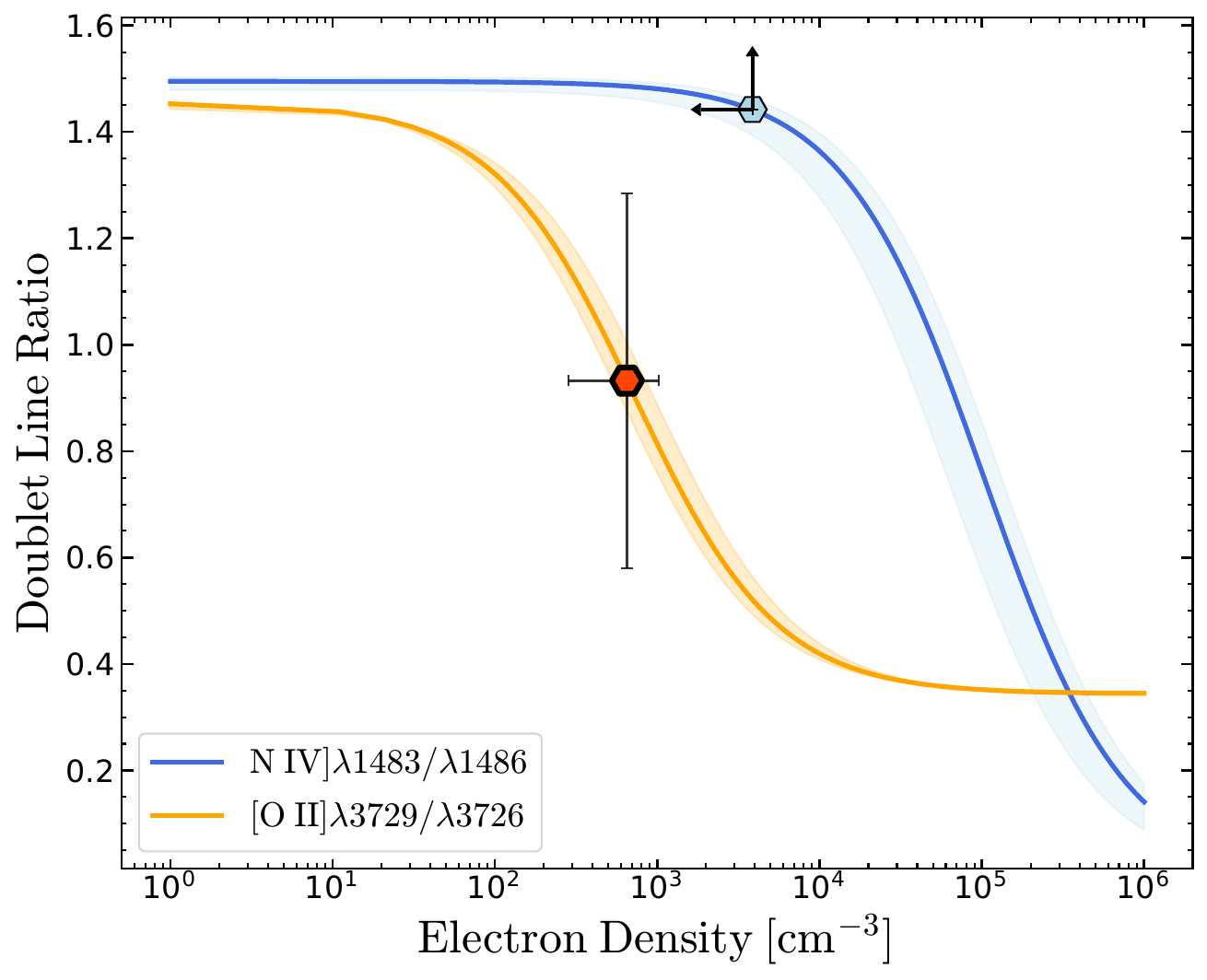}
    \caption{
    Electron density diagnostics.
    The \OIIL[3729/3727] ratio and the lower limit on the \NIVL[1483/1486] ratio, as well as the inferred electron densities, are reported as measured for \targetshort.  
    The expected behaviour of both line ratios as a function of density is depicted by the solid lines for \Te$=2\cdot10^4$~K (with shaded regions spanning \Te$=1-3\cdot10^4$~K). The lower limit on \NIVL[1483/1486] suggests no significant contribution from very-high ($n_e \gtrsim 10^4$~cm$^{-3}$) density regions, in broad agreement with the value inferred from the optical \OIIL[3729/3727] diagnostics.
    }
    \label{fig:density_comparison}
\end{figure}

\subsection{Derivation of oxygen abundance}
\label{sec:OH}

We can then derive the relative ionic abundance of two elements comparing the intensity $I(\lambda)$ in the emission lines of each species while taking into account the different temperature- and density-dependent volumetric emissivity of the transitions $J$, i.e.
\begin{equation}
    \frac{N(X^l)}{N(Y^m)} = \frac{I_{(\lambda)l}}{I_{(\lambda)m}} \frac{J_{(\lambda)m} (T,n)}{J_{(\lambda)l} (T,n)} \ .
\end{equation}
We compute the abundance of O$^{++}$/H and O$^{+}$/H from the \OIIIoptL/\Hbeta (assuming t=\Tiii) and \OIIall/\Hbeta (assuming t=\Tii) ratios, respectively, and derive log(O$^{++}$/H)$=-4.62\pm0.09$ and log(O$^{+}$/H)$=-6.00\pm0.14$.

One question is whether a significant fraction of oxygen could be in the triple-ionised state as possibly suggested by the detection of \HeIIL, since He$^{2+}$ shares the same ionization potential as O$^{3+}$ ($54.9$eV).
We do not detect any significant \OIVL emission in the G140M spectrum, and photoionization modelling from \cite{berg_chemical_2019} suggests a fractional contribution of O$^{3+}$/O $<0.01$ given the measured ionization parameter and the relative C$^{3+}$/C$^{++}$ ratio (see below).
The measured 3-$\sigma$ upper limit on the \OIVL flux $< 1.82 \times 10^{-19}$ erg s$^{-1}$ cm$^{-2}$ only provides an upper limit on O$^{3+}$/O$^{++}$ $\lesssim 0.4$.
We note that a small peak possibly associated with \OIVL is seen in the PRISM spectrum at $\approx1400\AA$, but the low spectral resolution makes it impossible to properly disentangle the \OIV contribution from other nearby features like \permittedEL[Si][iv]$\lambda1394$, considering also the uncertainty on the intrinsic shape of the underlying continuum and the presence of the damping wing of the nearby Ly$\alpha$ break.

Alternatively, we can exploit an ICF for oxygen based on the relative abundance of single- and double-ionised Helium, following
\cite{torres-peimbert_PNe_1977} and leveraging the similar ionization potential of He$^{++}$ and O$^{3+}$ \citep[see also][]{izotov_chemical_2006,valerdi_He_2021, dors_2020b, dors_seyfert_2022}. 
First, we estimate the intensity of He I$\lambda3889$ emission by correcting the flux measured in the G395M spectrum by the contribution of the blended H8 Balmer line, which we infer from the expected theoretical ratio to the nearby \Hdelta (i.e. H8/\Hdelta=0.406) assuming \Tii and density as measured above, and under the assumption of no dust attenuation as suggested by the analysis in Section~\ref{sec:SED_fitting} (see also Figure~\ref{fig:BD}).
Then, we compute the relative He$^{+}$/H$^{+}$ and He$^{++}$/H$^{+}$ abundances from the He I$\lambda3889$/\Hbeta and \HeIIL/\Hbeta flux ratios, assuming \Tii and \Tiii respectively.
The ICF(O) is finally given by the (He$^{+}$+He$^{++}$)/He$^{+}$ ratio, which we measure as ICF(O)$=1.23$, corresponding to a fractional contribution of O$^{3+}$/O $\sim 19$ per-cent.
We note here that such an ICF is prone to large uncertainties due to the significant impact of radiative transfer effects on the He I$\lambda3889$ line; while adopting the simplest optically thin case as fiducial, we explore variations in the assumed optical depth between $\tau \in [0,5]$ implementing the correction coefficients from \cite{Benjamin_helium_1999}, and estimate an uncertainty on the ICF(O) of $\sim10$ percent.
To account for additional sources of uncertainties (e.g. He I$\lambda3889$-H8 deblending, underlying line absorption) we conservatively include an additional $10$ percent error budget on the ICF(O).
% we are neglecting possible additional uncertainties on the ICF(O) determination associated with radiative transfer effects of He I lines (as we assume the simple case B); therefore, we assume a conservative 20 percent error on ICF(O).
The total oxygen abundance is hence O/H = ICF(O) $\times$ (O$^{++}$/H + O$^{+}$/H), and for \targetshort this corresponds to 12+log(O/H)=7.49$\pm$0.11, which we assume as our fiducial value in the following analysis.
The fraction of doubly-ionised oxygen over the total is O$^{++}$/O = 0.84$\pm$0.01.
Neglecting instead any O$^{3+}$ contribution (i.e. assuming O/H=O$^{++}$/H + O$^{+}$/H) would turn into 12+log(O/H)=7.40$\pm$0.09.
Finally, we note that repeating the procedure assuming emission line fluxes measured from the PRISM spectrum delivers a total (ICF-corrected) 12+log(O/H)=7.41$\pm$0.13, lower but consistent with our fiducial value based on G395M within statistical uncertainties.

\subsection{C/O abundance}
\label{sec:CO}
We derive C$^{++}$/O$^{++}$ from the \CIIIL/\OIIIL ratio, assuming the same electron temperature \Tiii when calculating the emissivity of the two ions.
Because \CIIIL is observed only in the PRISM spectrum (while falling in the detector gap in G235M observations), here we adopt the \CIIIL/\OIIIL ratio as derived from the PRISM, to avoid introducing uncertainties associated with flux calibration differences between PRISM and grating spectra.
We also note that the relative temperature associated with \CIII and \OIII emission might be different, as \CIII is possibly associated with an `intermediate-ionization zone', which could translate into a higher (lower) C$^{++}$/O$^{++}$ abundance by $0.15$~dex \citep{garnett_electron_1992, croxall_chaos_2016, rogers_chaos_2021, jones_CO_z6_2023}
in case of lower (higher) C$^{++}$ temperature, respectively.
In principle, it would be possible to derive C$^{++}$/O$^{++}$ also from the \CIIIL/\OIIIoptL ratio, however, we prefer to adopt \CIIIL/\OIIIL to minimise uncertainties on the reddening correction and the choice of the attenuation curve given the short involved wavelength separation, as well as for consistency with the vast majority of literature studies.

Since the ionization potential of O$^{2+}$ is higher than that of C$^{++}$ ($54.9$~eV versus $47.9$~eV, respectively), systems subject to hard ionising spectra such as \targetshort may have a significant amount of carbon in the C$^{3+}$ state.
Therefore, we correct the inferred C$^{++}$/O$^{++}$ abundance applying an ICF. 
The \CIVL emission is detected in both the PRISM spectrum (though possibly blended with other spectral features) and in the G140M grating spectrum (though at lower significance), therefore we can exploit the \CIV/\CIII ratio to estimate the C$^{3+}$/C$^{++}$ relative abundance and derive the ICF for C/O or, alternatively,
we can compare its strength to that of \OIII to directly measure a C$^{3+}$/O$^{++}$ abundance; the two approaches give consistent results.
We note that we have to assume that the \CIV emission is purely nebular in origin (as indeed suggested by its narrow line profile in G140M), although the full spectral profile of the \CIVL doublet can be further complicated by resonant scattering through highly ionised gas, as well as interstellar absorption or contribution from stellar emission, that can either under- or over-estimate the total \CIV flux \citep{berg_window_2018, berg_chemical_2019, senchyna_civ_2022}.
% (Berg+2018,2019, Senchyna+2017,2022).
% whereas either interstellar absorption or contribution from stellar emission can under- or over-estimate the C$^{3+}$ abundance, respectively. 
% Moreover, the full spectral profile of the \CIVL doublet is further complicated by resonant scattering through highly ionised gas, hence determining the real intrinsic flux of \CIVL is challenging.
Given the likely non-negligible contribution of O$^{3+}$ to the total oxygen abundance, we can not simply assume that C/O = C$^{3+}$/O$^{++}$ + C$^{++}$/O$^{++}$.
Therefore, we first compute C$^{++}$/H$^{+}$ and C$^{3+}$/H$^{+}$ by multiplying  C$^{++}$/O$^{++}$ and C$^{3+}$/O$^{++}$ by O$^{++}$/H$^{+}$ as derived in Section~\ref{sec:OH}, and then we assume that the contribution from single-ionised carbon is negligible, so that the total C/H abundance is = C$^{++}$/H$^{+}$ + C$^{3+}$/H$^{+}$.
Finally, we divide C/H by O/H to infer a total log(C/O) = -0.90 $\pm$ 0.12 dex. % which we assume as our fiducial estimate.
We note that if we assign the flux of the emission line detected in G140M to the red component of the \CIV doublet (i.e., \CIV$\lambda1551$), we can compare its strength to that of \OIIIL to obtain a grating-based measurement of the C$^{3+}$/O$^{++}$ abundance. This ultimately translates into a total log(C/O) abundance of $-0.91\pm0.13$, fully consistent with the previous estimate.

These values are also consistent, within their uncertainties, with the C/O inferred by assuming an ICF based on the photoionization models presented by \cite{berg_chemical_2019}, assuming photoionization from star-formation and the ionization parameter self-consistently inferred for \targetshort from the same models (log(U)$=-1.78$), in which case ICF(C$^{++}$/O$^{++}$)=$1.20\pm0.05$\footnote{Such ICF is larger than what inferred from the relations presented in \cite{amayo_ICFs_2020}, i.e., ICF = 1.05. We note that feeding instead the ICF equations from \cite{berg_chemical_2019} with the higher log(U) values as derived from SED fitting from both \textsc{beagle} (log(U)$=-1.46$) and \textsc{bagpipes} (log(U)$=-1.05$), although not based on self-consistently generated \textsc{cloudy} grids, would translate into ICF = 1.45 and ICF = 2, hence in higher log(C/O)=$-0.80 \pm 0.11$ and log(C/O)=$-0.65 \pm 0.11$, respectively.} and C/O = ICF $\times$ C$^{++}$/O$^{++}$, providing log(C/O)=$-0.86 \pm 0.11$.
% We note that the steep rise in predicted ICF(C$^{++}$/O$^{++}$) at high O$^{++}$/O ($\gtrsim 0.9$) in the Berg2019 models translates into an uncertainty on the derived ICF of $\substack{+0.11 \\ -0.04}$ when propagating the formal uncertainty on the oxygen ionization fraction, but would translate into a much larger uncertainty of $\substack{+0.23 \\ -0.06}$ if assuming a slightly more conservative $2\%$ uncertainty on O$^{++}$/O.
% Assuming instead the total \CIVL doublet as measured from the prism spectra provides (by directly comparing \CIVL and \CIIIL prism fluxes) C$^{3+}$/C$^{++}$ $=1.015$, which we can then convert to C$^{3+}$/H$^{+}$ and use to derive C/H and C/O; in this way, we find a slightly higher total log(C/O)=$-0.92\pm0.15$, which is nonetheless still consistent with our fiducial value.
Finally, we estimate the C/O abundance via the equations outlined in \cite{perez-montero_using_2017}, which deliver log(C/O)=$-0.80 \pm 0.12$.

Throughout the rest of the paper, we assume the C/O derived including both \CIIIall and \CIVL line fluxes as our fiducial estimate, i.e. log(C/O)$=-0.90\pm0.11$ (statistical uncertainty), corresponding to $\sim23$ per-cent the solar C/O abundance, or [C/O]$=-0.64$.
However, we note that C/O estimates based on the \CIIIall/\OIIIL ratio and ICFs from photoionization models generally provide C/O abundances up to $\sim0.1$~dex higher; therefore, we include an additional $0.1$~dex uncertainty (co-added in quadrature) on the upper value to take into account these systematics. 

% Finally, we note that assuming the log(O/H) as derived from the prism fluxes, we obtain log(C/O)=$-0.82 \pm 0.15$. 

% Finally, we test the impact of assuming the temperature derived from \OIIIL instead of the optical temperature in our C/O derivation, finding a slightly higher but consistent log(C/O)=$-0.95\pm 0.10$.

% We here assume the \Tiii temperature to compute the relative  C$^{3+}$/O$^{++}$ abundance, and derive the total C/O abundance as (C$^{3+}$+C$^{++}$)/O$^{++}$, ultimately finding log$_{10}$(C/O)=-0.89$\pm$XX, which we adopt as our fiducial value.
% \textbf{TBC}
% % The relative contribution of C$^{3+}$/O$^{++}$ to the total C/O abundance is therefore 0.21~dex, corresponding to an ICF(C$^{++}$/O$^{++}$) of $\sim$1.6.
% This value is much larger than both the ICF derived from the relations presented in Amayo+2021 (ICF = 1.05) and that predicted by the grids of photoionization models of Berg+2019 (ICF = 1.23) %,especially at the low metallicity of GS-z9
% ; in both cases, the ICF(C$^{++}$/O$^{++}$) is estimated on the basis of the ionization fraction of oxygen (O$^{++}$/O=0.96 for GS-z9, see above). 
% We note that the steep rise in predicted ICF(C$^{++}$/O$^{++}$) at high O$^{++}$/O ($\gtrsim 0.9$) in the Berg2019 models translates into an uncertainty on the derived ICF of $\substack{+0.11 \\ -0.04}$ when propagating the formal uncertainty on the oxygen ionization fraction, but would translate into a much larger uncertainty of $\substack{+0.23 \\ -0.06}$ if assuming a slightly more conservative $2\%$ uncertainty on O$^{++}$/O.

\subsection{N/O abundance}
\label{sec:N_O}
We estimate the N/O ratio by exploiting the detection of \NIIIL in emission.
Leveraging the fact that the ionization potentials of N$^{++}$ and C$^{++}$ are basically identical ($47.448$~eV and $47.887$~eV, respectively), we assume the C$^{++}$/N$^{++}$ abundance calculated from the \CIIIL/\NIIIL ratio (accounting for the emissivity of all five lines of the \NIIIL multiplet) as a proxy of the relative C/N enrichment, finding log(C/N)$=0.03\pm0.19$. 
Then, we combine this ratio with our fiducial C/O ratio, measured as described in the Section~\ref{sec:CO}, to derive a total N/O of log(N/O)$=-0.93\pm0.24$. 
% Such N/O abundance is higher than typical values found in local, metal-poor dwarf galaxies with similar oxygen abundance as \targetshort \citep{perez-montero_impact_2009, berg_direct_2012, vale_asari_bond_2016, vincenzo_extragalactic_2018}.

Alternatively, we can exploit the tentative detection of \NIVL[1483] in the 3-pixel extracted G140M spectrum for an alternative derivation of N/O.
From the \NIVL[1483]/\OIIIL[1666] ratio\footnote{in this case we adopt both fluxes as measured from the 3-pixel extracted spectrum to avoid relative flux calibration issues; F(\OIIIL[1666])~[3-pix]=$(3.29\pm0.52) \cdot 10^{-19}$~erg/s/cm$^{2}$} we derive the N$^{3+}$/O$^{++}$ abundance (assuming t=$t_3$), which we then multiply by O$^{++}$/H to infer N$^{3+}$/H; the same procedure is applied to N$^{++}$/O$^{++}$ (as measured from the \NIIIL in the prism spectrum) to infer N$^{++}$/H, which we then co-add with N$^{3+}$/H to derive the total N/H abundance (assuming negligible contribution from the singly-ionised N$^+$ state).
Finally, we divide N/H by O/H to obtain a total N/O of log(N/O)$=-0.77\pm0.18$, higher than though consistent with our fiducial estimate.

Finally, based on the $3\sigma$ upper limit on the total flux of the \NIIIall multiplet from the G140M spectrum, we obtain a lower limit on log(C/N)$>-0.24$, and an upper limit on log(N/O)$<-0.66$.
% \todo{report N/O as derived from N IV detection}

\subsection{Ne/O abundance}

We observe intense emission from Ne in its Ne$^{++}$ form and the \NeIIIL emission line.
We derive the Ne$^{++}$/O$^{++}$ abundance ratio assuming the \Tiii temperature for both ions, and the correct to the total Ne/O abundance exploiting the ICF presented in \cite{amayo_ICFs_2020}.
We note that, given the different ionization potentials of Ne$^{+}$ and O$^{+}$, coupled with the charge-transfer
recombination rate of the two ions, such ICF is quite uncertain in low-ionization, high-metallicity systems. However, in the case of \targetshort, the fractional contribution of unseen ionization states to the total Ne/O is expected to be small, with ICF(Ne$^{++}$/O$^{++}$)=1.02, nonetheless we conservatively assume a 10 per-cent uncertainty on such ICF.
We therefore infer a total log(Ne/O) = -0.68$\pm$0.06. 
We note that this value already `includes' the possible contribution from Ne$^{4+}$: in fact, accounting separately for the marginal detection of \NeVL (but without applying any ICF) would deliver log(Ne/O) = -0.73$\pm$0.07.

\subsection{Systematics in the abundance measurements}
\label{sec:systematics}

Despite the high S/N of most of the emission lines detected in the \targetshort spectrum, a number of additional systematics uncertainties affect our chemical abundance measurements beyond those already mentioned in the previous sections.
Some of the most relevant are associated with electron temperatures, as briefly discussed here.

In Section~\ref{sec:abundances}, we have adopted the temperature inferred from the \OIIIopt4363/\OIIIoptL ratio as our fiducial estimate for \Tiii. However, the \OIIIL/\OIIIoptL is another temperature diagnostics that can be adopted in studies of high-z galaxies where the \OIIIopt4363 emission line is not detected \citep[e.g.][]{Revalski_mzr_z1_2024}.
Exploiting the \OIIIL flux measured from G140M (where it is spectrally resolved from \HeIIL), we infer \Tiii=24405$\pm$1490 K, higher (significant at $1.6\sigma$) than that inferred from \OIIIoptL[4363].
This propagates into a lower inferred O/H by $0.15$~dex, whereas only in a difference of $0.05$~dex in C/O, as the \CIIIL/\OIIIL ratio is only mildly sensitive to temperature.
We include further $0.1$~dex systematic uncertainty on the lower value of log(O/H) to reflect this difference.

The origin of such discrepancy is uncertain.
Although similar differences in \Tiii as inferred from either \OIIIL or \OIIIopt4363 have been reported before \citep[e.g.][]{berg_carbon_2016}, these were generally attributed to offsets in the flux calibration between \OIIIL and \OIIIoptL (which, before the advent of \jwst, were usually observed by instruments on different facilities), or to the large uncertainty on the relative reddening correction between UV and optical wavelength regimes.
In our case, assuming no attenuation correction as suggested by the measured Balmer decrements (see Section~\ref{sec:dust_ion}), this would imply an unrealistic (and never observed in NIRSpec to our knowledge) factor $\times 1.9$ offset in the flux calibration between G140M and G395M (in the sense that \OIIIL is almost a factor of two brighter than it should be) in order to match the \Tiii estimates from optical and UV lines. %\todo{any reference on known inter-calibration issues between \jwst gratings ?}
% Conversely, assuming negligible flux calibration offset between different NIRSpec gratings/filters,  \todo{TBC}

More physically motivated explanation involves the fact that \OIIIL and \OIIIoptL[4363] come from different energy levels, with different critical densities and different collisional excitation rates; hence, their ratio also depends on the gas density, and can therefore be impacted by the presence of density inhomogeneities within galaxies \citep[e.g.][]{marconi_homerun_2024}.
Moreover, when different temperature estimators rely on collisionally excited lines originating from well-spaced energy levels, the inferred temperatures can also differ by an amount that depends on the variance of the temperature distribution within the nebula, even if the diagnostics pertain to the same ionic species (O$^{++}$ in this case). 
This possibly suggests that rest-frame UV and rest-optical emission lines (even of the same ionic species) could originate from different regions in the galaxy, 
and that this effect could be particularly relevant in high-redshift systems like \targetshort characterised by highly ionised gas, where temperature inhomogeneities are expected to be enhanced.
This effect has been historically described with the t$^2$ parameter (see \citealt{peimbert_chemical_1969}, and more recently \citealt{mendez_delgado_t_inhomogeneities_2023}), which represents the root mean square deviation from the average temperature.
In the case of \targetshort, the difference measured between \OIIIL-based and \OIIIoptL[4363]-based temperatures would translate into t$^{2} \sim 0.24$, a value higher than measured in local HII regions. However, given such a large t$^2$ one would expect a much lower temperature inferred from the nebular continuum ($\sim 11,000$K) than observed. In fact, fitting a broken power-law continuum to the `Balmer jump' region of \targetshort one infers $T = 2.0\pm 0.5\times 10^4$ K, consistent with the \Te-based measurements presented above \citep[see also][]{Ji_GNz11_2024}.

In a similar manner, another source of systematic uncertainty is associated with the adoption of the same temperature for the emitting region of \OIIIoptL and \CIIIL.
As C$^{++}$ has a lower ionization energy than O$^{++}$, it might be better associated with an `intermediate' ionization region, probed by ions like S$^{++}$ \citep{berg_4zones_2021}. 
Since we do not have access to transitions suitable for directly inferring the temperature of such intermediate zone, we can employ Equation (3) from \cite{berg_chaos_2020}, which provides T$_{\text{intermediate}}= 23150$~K and consequently a C$^{++}$/O$^{++}$ ratio lower by $0.20$~dex (which propagates into a total C/O lower by $0.13$~dex, if still assuming the same \Tiii for O$^{++}$ and C$^{3+}$).
We note, however, that such temperature-temperature relation can be quite scattered and uncertain, especially at high temperatures and ionization \citep{hagele_2006, binette_discrepancies_2012, berg_chaos_2015, jones_CO_z6_2023}; moreover it is mostly calibrated on samples of local HII regions.
Therefore, we choose to not adopt it in our fiducial analysis, nonetheless we note that the amount of systematic uncertainty it carries can be significant.

A similar argument can also be made for the temperature-temperature relation employed to infer the abundance of singly-ionised oxygen.
In Section~\ref{sec:abundances} we have adopted the \Tii-\Tiii relation from \cite{izotov_chemical_2006}
; adopting instead the relation calibrated by e.g. \citet{pilyugin_electron_2009}, or the theoretical one by \cite{campbell_stellar_1986}, provides higher \Tii by $\sim2,500$~K and $\sim322$~K, respectively. However, this translates to a total $\Delta$log(O/H) of $\lesssim 0.005$~dex, since O$^{+}$ is contributing only $\sim 2.5$ percent to the total oxygen abundance.

Finally, to take into account the abundance of highly ionised ionic species like C$^{3+}$ and N$^{3+}$, \cite{berg_4zones_2021} proposed a four-zone ionization model to provide a more accurate description of the regions of line emission in galaxies characterised by `extreme' conditions. 
In such framework, the temperature of the `very-high' ionization zone is better probed by the \NeIII $\lambda$3342/$\lambda$3868 ratio than the classical \Tiii (as the ionization potential of Ne$^{2+}$ is higher than that of O$^{2+}$). 
However, we do not detect \NeIII $\lambda$3342 in our prism spectra, and therefore we have to assume \Tiii when computing the emissivity of C$^{3+}$ and N$^{3+}$ in our carbon and nitrogen abundance calculations. The extent of the associated systematic uncertainty is difficult to assess. 
\cite{berg_4zones_2021} find only a mild (mostly within 1-$\sigma$ uncertainty) impact on the total inferred O/H, C/O, and N/O when adopting either the three- or four-zone ionization model for the galaxies J104457 and J141851, though we note that the fraction of highly ionised species of oxygen, carbon, and nitrogen are minimal in those systems.
On the other hand, we note that our derivation of the N/O abundance in \targetshort follows from the measurement of C/N, which is based on emission line ratios with relative weak temperature dependence, whereas for what concerns the C/O derivation, conservatively assuming a temperature much higher than \Tiii (e.g., $25,000$ K) as representative of the \CIV emitting region (as it would be in case C$^{3+}$ preferentially occupies a region closer to the ionising source) would translate into a lower total inferred C/O by $\sim 0.24$~dex compared to our fiducial value.
This effect might be even larger for higher states of neon like Ne$^{4+}$ that could originate from extremely high-ionization regions (especially in the case of the NLR of AGN); however, given the small relative Ne$^{4+}$/Ne$^{++}$ abundance, we verified that even assuming a temperature as high as $25,000$ K it would not significantly impact the final inferred Ne/O abundance (which would be lower by 0.02~dex compared to our fiducial estimate).
% \todo{TBC}

To account for the hereby discussed effects, we include an additional $0.15$~dex of systematic uncertainty also on the C/O and N/O\footnote{which is derived by combining C/N and C/O, Section~\ref{sec:N_O}} abundances beyond the formal statistical uncertainties propagated from the error on the measured line fluxes.
This, together with the uncertainties on C/O from photionisation-based ICFs discussed already in Section~\ref{sec:CO}, translates into the final uncertainties quoted in parenthesis in Table~\ref{tab:properties}.

\section{Possible scenarios for chemical enrichment}
\label{sec:discussion}

\begin{figure*}
 %   \centering
    \includegraphics[width=0.48\textwidth]{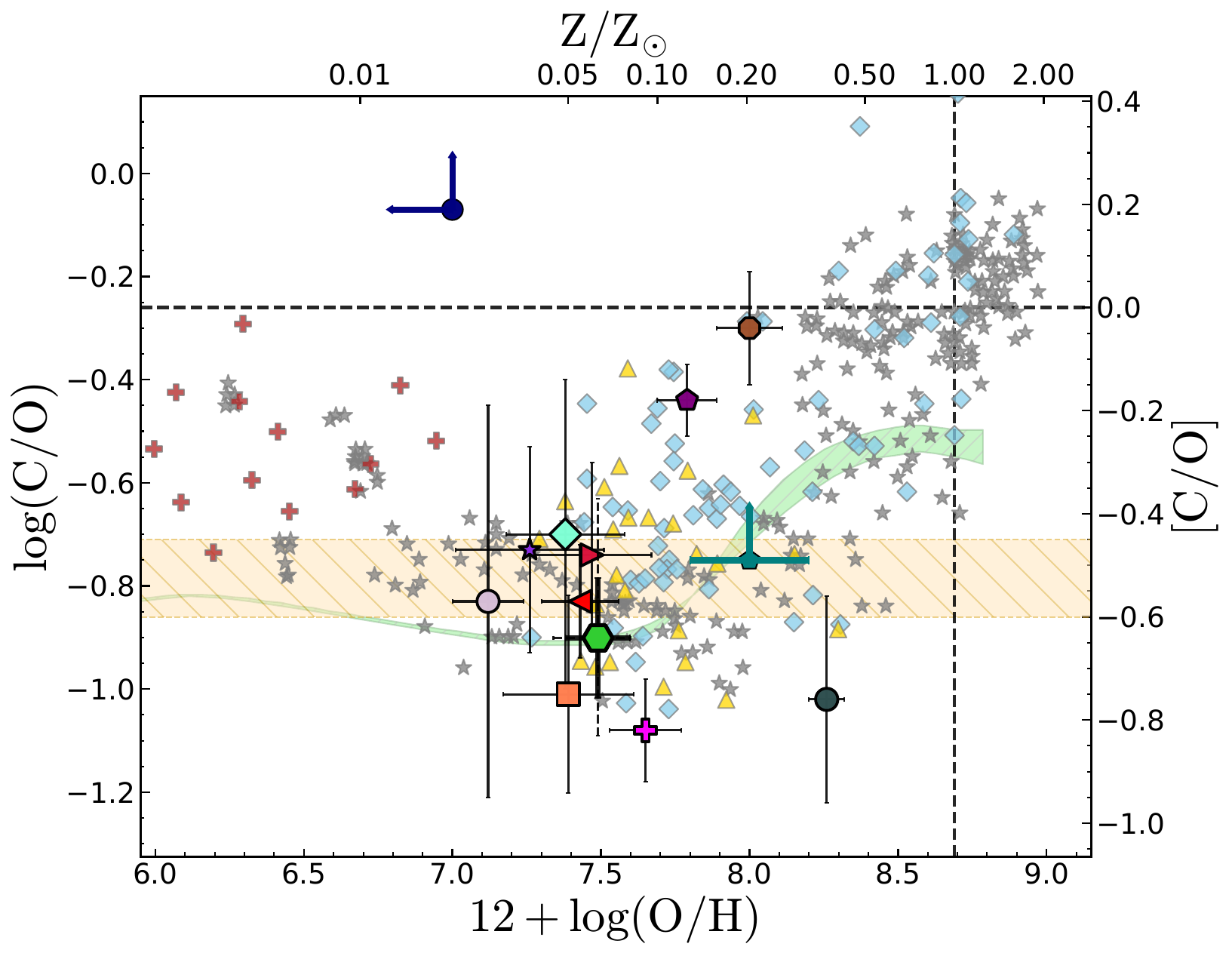}
    \hspace{1.5cm}
    \includegraphics[width=0.3\textwidth]{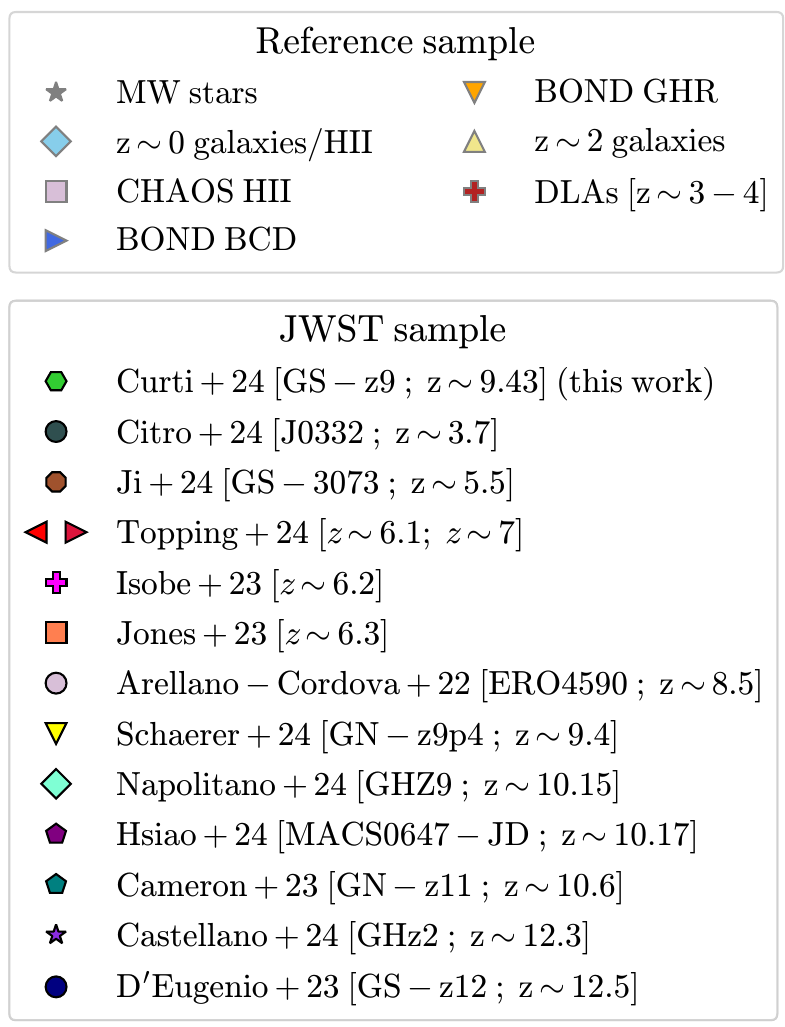}\\
    \includegraphics[width=0.48\textwidth]{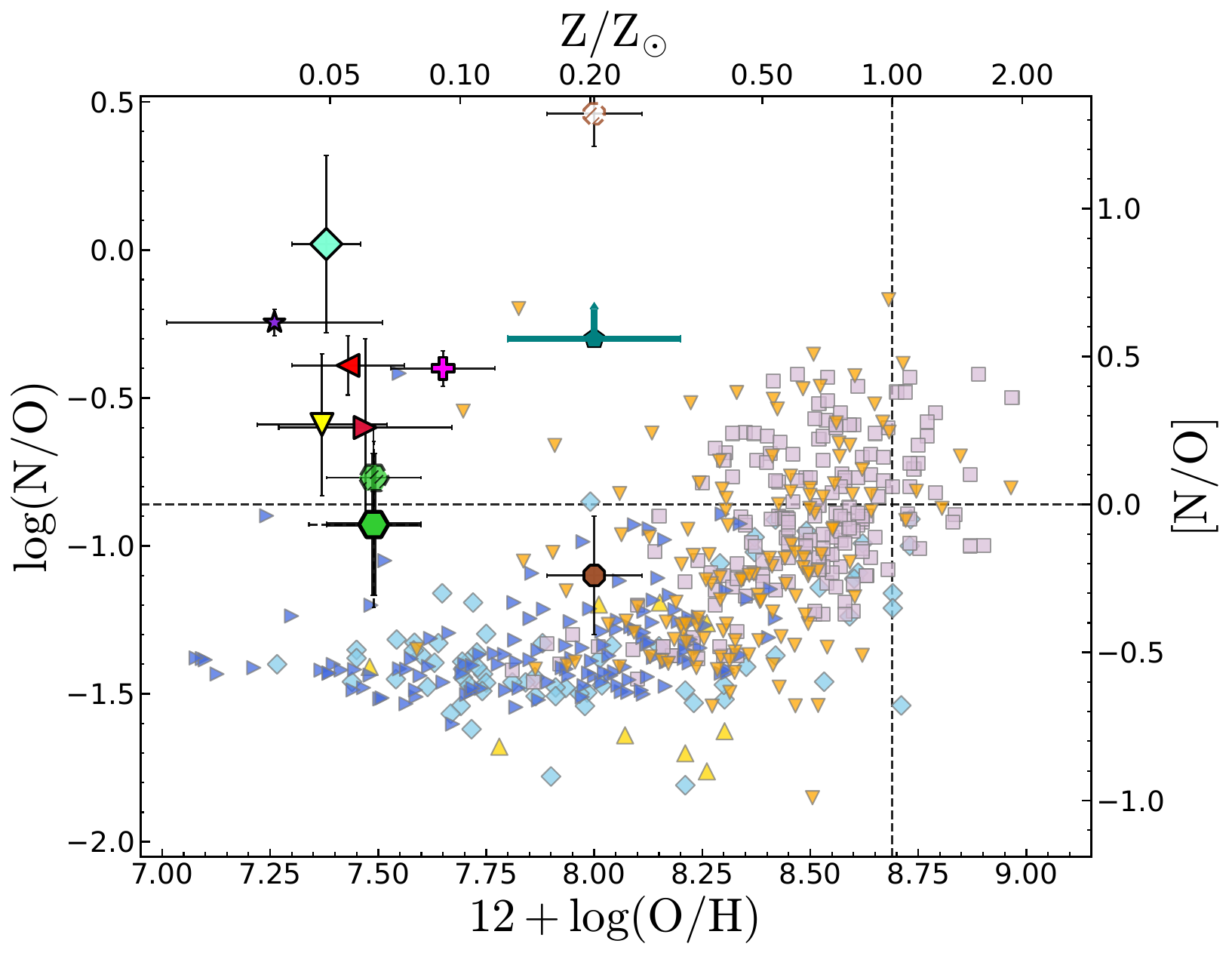}
    \includegraphics[width=0.48\textwidth]{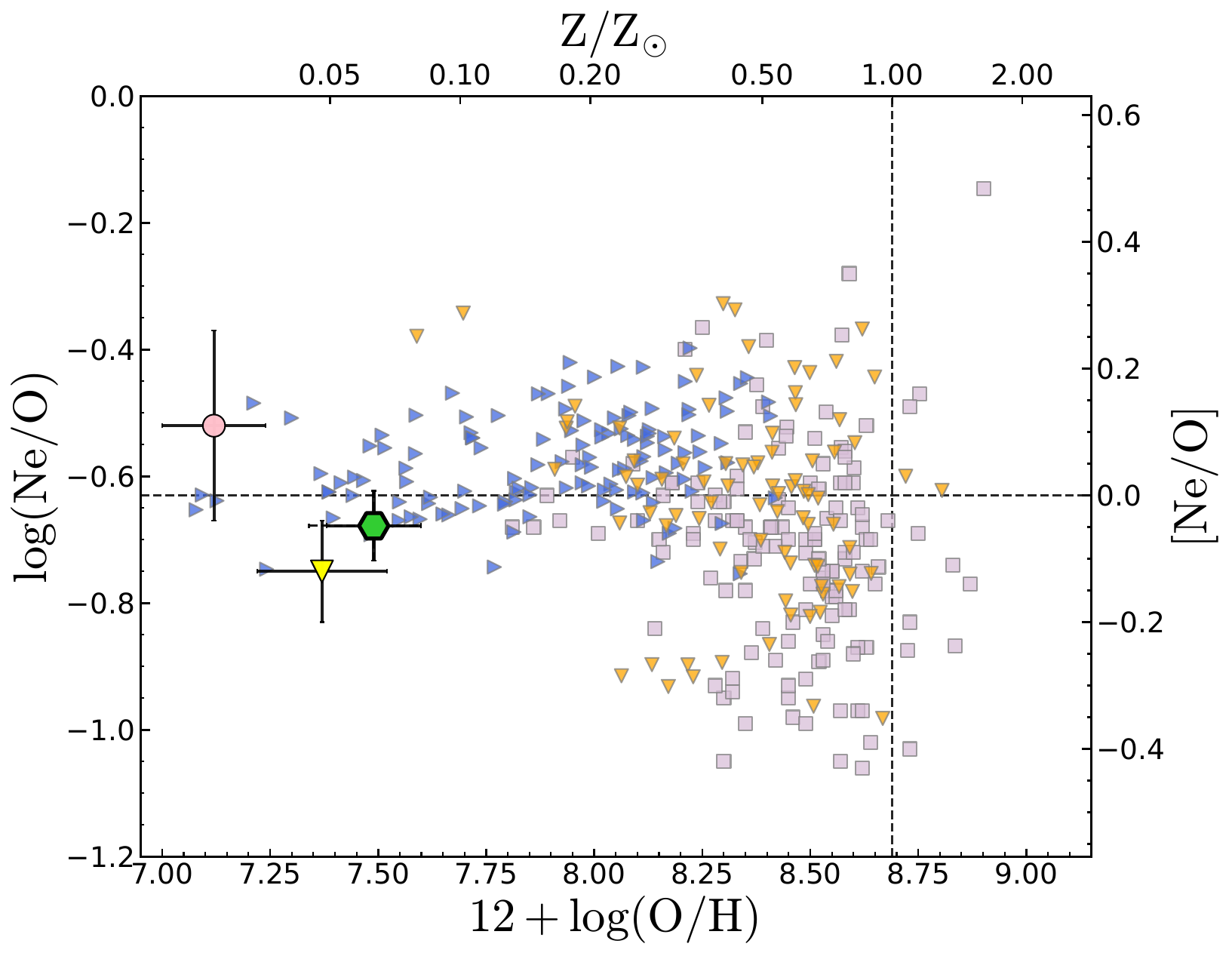}
    \caption{Chemical abundance patterns in \targetshort.  The location of \targetshort is marked by the green hexagon, with solid errorbars reporting formal statistical uncertainties on our fiducial values and dashed errorbars accounting for additional systematic uncertainties as discussed in \ref{sec:systematics}.
    Dashed black lines in the diagrams mark the solar abundance values.
    We compare \targetshort with a compilation of recent abundance measurements in high-z galaxies from \jwst, i.e. from \citealt{ji_nitrogen_AGN_z5_2024} (plain marker for optical-based and hatched marker for UV-based abundances, respectively), \citealt{topping_z6_lens_2024}, \citealt{isobe_CNO_2023} and \citealt{jones_CO_z6_2023} at z$\sim 6$, 
    \citealt{topping_deep_UV_2024} at at z$\sim 7$,
    \citealt{arellano-corodva_2023} at z$\sim8.5$, 
    \citealt{Napolitano_GHZ9_2024} at z$\sim10$,    \citealt{castellano_ghz12_2024} at z$\sim12$, as well as upper limits on GN-z11 and GS-z12 from \citealt{cameron_gnz11_2023} and \citealt{d_eugenio_gsz12_2023}, respectively.
    We also include measurements in Milky Way disk and halo stars, local HII regions, local dwarf galaxies, z$\sim2$ galaxies, and Damped Lyman-Alpha systems (DLA). 
    Upper-left panel: C/O versus O/H diagram. The C/O abundance range predicted from type II SNe yields from \citealt{tominaga_2007} is marked by the golden hatched region, whereas the C/O vs O/H pattern from \citealt{kobayashi_origin_2020} galactic chemical evolution (GCE) models is shown by the green sequence.
    The C/O level observed in \targetshort is consistent with the enrichment expected from core-collapse SNe for a recently assembled, low-metallicity system.
    Bottom-left panel: N/O versus O/H diagram, reporting for \targetshort both the N/O as inferred from the detection of \NIIIL in the PRISM spectrum and the C/N ratio (fiducial value, plain green hexagon), as well as that based on the marginal detection of \NIVL[1483] in the G140M spectrum (hatched symbol, see Section~\ref{sec:N_O}).    
    The N/O measured in \targetshort is higher than the plateau occupied by local galaxies and HII regions at low metallicity, although the nitrogen enhancement observed in this galaxy appears less extreme compared to other high-z systems recently observed by \emph{\jwst}.
    % The green dashed line marks the upper limit on N/O from the $3\sigma$ upper limit \NIIIL flux in the G140M grating.
   Bottom-right panel: Ne/O abundance for \targetshort, which we observe consistent with the abundance pattern expected at low-metallicity for $\alpha$-elements. \\
   Additional references for literature points: compilation of MW stars \citep{gustafsson_MW_1999, akerman_CO_2004, fabbian_CO_2009, Nissen_CO_2014}; compilation of z~$\sim0$ galaxies and HII regions \citep{Tsamis_HII_2003, esteban_reappraisal_2004, esteban_keck_2009, esteban_carbon_2014, esteban_abundance_2017, garcia_rojas_2004, garcia_rojas_2005, garcia_rojas_2007, peimbert_2005, garcia_rojas_esteban_2007, lopez_sanchez_2007, toribio_san_cipriano_2016, toribio_san_cipriano_carbon_2017, senchyna_2017}; HII regions from CHAOS \citep{berg_carbon_2016, berg_chemical_2019}; blue compact dwarfs (BCD) and giant HII regions (GHR) from BOND \citep{ vale_asari_bond_2016}; z~$\sim2$ galaxies \citep{fosbury_2003, erb_2010, christensen_gravitationally_2012, bayliss_2014, james_testing_2014, stark_ultraviolet_2014, steidel_reconciling_2016, vanzella_2016, amorin_analogues_2017, berg_window_2018, mainali_2020, matthee_2021, rigby_2021, Iani_2023}; DLAs at z~$\sim3-4$ \citep{cooke_dla_2017, saccardi_dla_2023}.
    }
    \label{fig:CO_OH}
\end{figure*}

% \begin{figure}
%     \centering
%     \includegraphics[width=0.98\columnwidth]{figs/Ne_O_plots_z94_te_opt.pdf}
%     \caption{Ne/O abundance for GS-z9, compared to other \Te-based measurements in both local and high redshift galaxies.}
%     \label{fig:N_NE_O}
% \end{figure}

In Figure~\ref{fig:CO_OH} we report \targetshort (green hexagon marker) on different diagrams which explore trends between various chemical abundance patterns (i.e. C/O, N/O, and Ne/O) and metallicity (probed by oxygen abundance). The solid errorbars report the formal statistical uncertainties on our fiducial estimates of chemical abundances, whereas dashed errorbars account for the systematics discussed in Section~\ref{sec:systematics}.

The top-left panel shows the C/O versus O/H diagram, with \targetshort compared with previous abundance determinations in both local and high-redshift systems, as well as those measured in Milky Way (MW) disk and halo stars (see caption of Fig.~\ref{fig:CO_OH} for details and references). 
The low C/O abundance in \targetshort is comparable to that observed in metal-poor systems at lower redshift, being in line with some of the most recent determinations in $z\gtrsim6$ galaxies observed with \jwst from \citet[][$z\sim8.5$]{arellano-corodva_2023}, \citet[][$z\sim6.23$]{jones_CO_z6_2023}, \citet[][$z\sim6.2$]{isobe_CNO_2023}, \citet[][$z\sim6.3$]{topping_z6_lens_2024}, \citet[][$z\sim12.3$]{castellano_ghz12_2024} (whereas in contrast, the upper limits on GS-z12 from \citealt{d_eugenio_gsz12_2023} suggest significantly super-solar C/O in that system at $z\sim12.5$). 

Given the very young age of the system (mass-weighted age $\sim30$~Myr as inferred from the SED fitting), we expect chemical yields dominated by core-collapse SNe (CCSNe), whereas low- and intermediate-mass stars in their AGB phase (with lifetimes $\approx 100$ Myr) would not have had enough time to significantly contribute to the overall enrichment of the galaxy.
Indeed, the datapoint for \targetshort on the C/O versus O/H plane is consistent with the range in C/O predicted by integrating the yields of type II SNe from very-metal-poor stars over a Salpeter IMF \citep[][golden shaded region]{tominaga_2007}.
In the same panel, we also compare our observations with the C/O enrichment pattern predicted by galactic chemical evolution (GCE) model of the solar neighbourhood presented in \cite{kobayashi_origin_2020} (shaded green curve), which follows the time evolution of elemental abundances in a one-zone model with ISM instantaneously mixing but no instantaneous recycling approximation (i.e, the delay of chemical enrichment from different sources is taken into account), assuming a \cite{kroupa_IMF_2008} IMF and solar abundances from \cite{asplund_solar_2009}.
% \todo{describe SFH in K20}
% On top of yields from core-collapse supernovae (type II, Ib, Ic), GCE models from Kobayashi2020 include contribution from AGB and super-AGB stars (with yields from Lugaro et al. (2012), Fishlock et al. (2014), Karakas et al. (2018), and Karakas \& Lugaro (2016)) and `failed' supernovae, i.e. very massive ($\gtrsim 30$\MSun), non-exploding stars whose CO core fall onto the newly formed black hole and hence is not ejected into the ISM; conversely, it does not include `faint' supernovae from carbon enhanced metal-poor stars (CEMP, e.g., Nomoto 2013) or pair-instability supernovae (Woosley 2002; Umeda \& Nomoto 2002).
Our observations agree well with the sequence in C/O of \cite{kobayashi_origin_2020} at the measured oxygen abundance for \targetshort, as this system resembles the C/O enrichment pattern of metal-poor stars in the MW.   

% The golden shaded region in Figure~\ref{fig:CO_OH} encompass instead the range in C/O predicted by integrating over a Salpeter \todo{check} IMF the yields of type II SNe from very-metal-poor stars by Tominaga+2007 \textbf{TBC}.
The position of galaxies on the C/O vs O/H diagram is also sensitive to the amount of metals lost due to outflows.
Possible indications of outflowing gas in \targetshort comes from the observed blueshifted \CIVall doublet in G140M (Figure~\ref{fig:prism_fit}). 
The spread observed in C/O at fixed O/H in samples of metal-poor local dwarfs and z$\sim2$ galaxies can be reproduced in fact not only in terms of different SFH and IMF slopes, but even accounting for selective oxygen removal in outflows driven by CCSNe 
\citep{yin_CO_2011, berg_chemical_2019}
According to the formalism from \cite{berg_chemical_2019}, developed to model the C/O vs O/H pattern observed in local, metal-poor dwarf galaxies, the low O/H and low C/O measured in \targetshort would indicate negligible oxygen-enhancement in SN-driven outflows. 
However, we note that in such a framework low C/O abundances are associated with single, long bursts of star-formation characterised by relatively small star-formation efficiency, whose timescales ($\sim300-400$Myr) can be hardly reconciled with the inferred age and star-formation history of \targetshort.

In the bottom-left panel of Figure~\ref{fig:CO_OH}, we 
report instead the location of \targetshort on the N/O vs O/H diagram. 
Based on the N/O derived from the \NIIIL detection in the PRISM spectrum as detailed in Section~\ref{sec:N_O}, \targetshort appears nitrogen-enriched compared to the typical range of values observed in local, low-metallicity dwarf galaxies \citep[e.g.][]{perez-montero_impact_2009, berg_direct_2012, vale_asari_bond_2016, vincenzo_extragalactic_2018} as predicted by `primary' nitrogen enrichment, showcasing instead N/O consistent with the solar value.
A slightly higher N/O is inferred when including the tentative \NIVL[1483] detection in the G140M spectrum (hatched green symbol in the panel). 

A similar pattern characterised by high N/O abundance at low O/H and low C/O has been already observed in several high-redshift ($z>6$) galaxies observed with \emph{\jwst} \citep[e.g.,][]{cameron_gnz11_2023, isobe_CNO_2023, topping_z6_lens_2024, topping_deep_UV_2024, Napolitano_GHZ9_2024, castellano_ghz12_2024, Schaerer_nitrogen_z94_2024},
although we note that \targetshort does not reach the same level of N/O enhancement as the aforementioned systems, which are observed to be significantly super-solar in N/O.
Low C/N ratios are also often observed in the so-called `nitrogen-loud quasars' \citep{batra_baldwin_NL_AGN_2014}, as possibly produced by the enrichment from AGB stars or SNe occurring within the small, local volume of the narrow-line or even broad-line region of AGN \citep{maiolino_gnz11_2023}.
Such a scenario can produce a `chemical stratification' between central, denser regions and the rest of the galaxy, which is imprinted on the largely different N/O derived from UV lines (tracing dense, highly ionised regions) and optical lines (tracing the bulk of the galaxy) as observed e.g. in the $z\sim5.5$ active system GS-3073 \citep[][for which UV-based and optical-based measurements are marked by hatched and plain brown octagons in Figure~\ref{fig:CO_OH}]{ji_nitrogen_AGN_z5_2024}.
However, in the case of \targetshort the contribution from AGB stars is strongly disfavoured given the inferred age and SFH of the system, and moreover it would be difficult to reconcile with the simultaneous low O/H and C/O.

The N/O pattern seen in \targetshort also resembles that of 
nitrogen-enhanced metal-poor (NEMP) stars observed in the Galactic halo \citep{Belokurov_kravtsov_nitrogen_2023}, while even higher levels of N-enhancement (log(N/O)$\gtrsim0$) are seen in some dwarf stars within globular clusters \citep{Carretta_CNO_GC_2005, D_orazi_GC_dwarfs_2010} as well as in some old stellar populations in metal-poor ultra-faint dwarf galaxies (UFDs), systems formed at high redshift which are now completely gas depleted \citep{Alexander_UFDs_chem_2024}.
Indeed, it has been suggested that these objects share a similar chemical composition as some of the most extremely nitrogen-rich galaxies observed by \emph{\jwst} \citep[e.g. ][]{charbonnel_gnz11_2023, Senchyna_gnz11_2023,marques-chaves_Nitrogen_2024, topping_z6_lens_2024}, which in turn could represent their high-z progenitors.
%carrying signatures of nuclear burning of hydrogen in the CNO process.
 
Despite \targetshort showing a relatively lower N/O compared to such systems, its N-enrichment level compared to the N/O vs O/H plateau, coupled with the concentrated star-formation, could still be consistent with proto-globular cluster formation contributing, to some extent, to the observed integrated galaxy spectra.
Although the (mild) constraints on the density of the highly-ionised gas in \targetshort provided by UV-diagnostics (based on the low-significance detection of \NIVL[1483] in the absence of \NIVL[1486]) seems to exclude the presence of extremely-dense regions, the low S/N of the G140M spectrum, the \CIIIL doublet (an additional potential density diagnostics) unresolved in the PRISM spectrum, and the putative AGN contribution suggested by the presence of very high-ionization emission lines, prevents us from drawing more definitive conclusions in such sense. 

% Unfortunately, no direct constraints on the density of the highly ionised gas in \targetshort from UV-diagnostics (e.g. from \CIIIL or \NIIIL doublet ratios, either because unresolved in the PRISM spectrum, or undetected in R1000 spectra), as well as the putative AGN contribution suggested by the presence of very high-ionization emission lines, prevents us from drawing more definitive conclusions in such sense. 

\begin{figure}[!h]
    \centering
    \includegraphics[width=0.98\columnwidth]{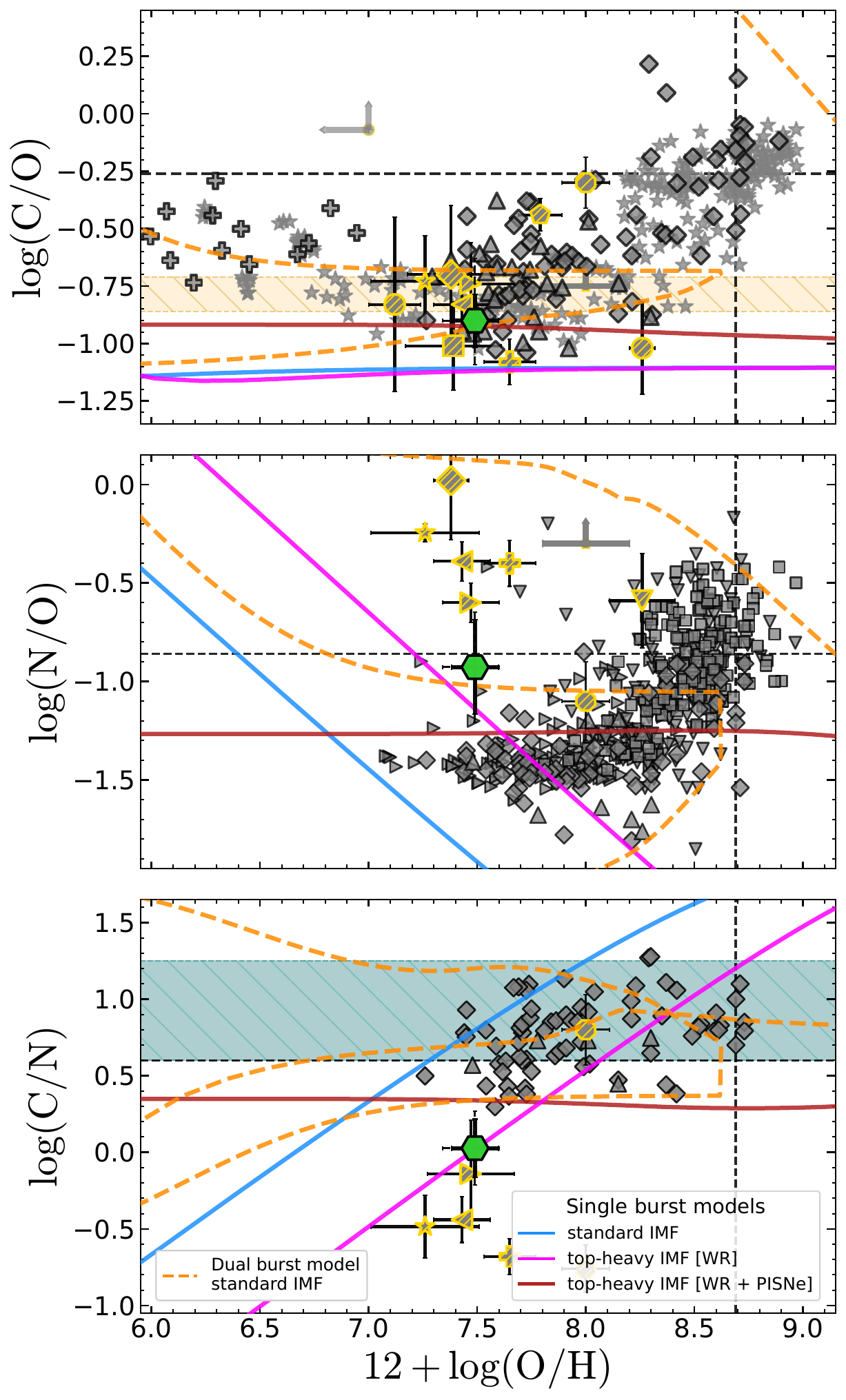}
    \caption{Evolution of CNO abundance ratios as a function of oxygen abundance for single starburst models with different IMFs, based on the framework of galactic chemical evolution models from \citealt{kobayashi_origin_2020}.
    Symbols are as in Figure~\ref{fig:CO_OH}, with \targetshort in green and other high-z objects observed by \emph{\jwst} from the literature highlighted in yellow.  
    Shaded regions in the C/O vs O/H and C/N vs O/H panels mark the range allowed by pure CCSNe based on the yields from \citealt{tominaga_2007} and \citealt{watanabe_2024}, respectively. 
    %from \citealt{kobayashi_ferrara_gnz11_2024} (solid lines). 
    The blue model assumes a `standard' IMF (with high-mass end slope $x= 1.3$), whereas the magenta and red curves assume a top-heavy IMF (with flat slope) for massive stars in range 30–120\MSun, and 100–280\MSun, respectively, with the latter including PISNe.
    % The red model includes WR stars with massive-end slope of \citealt{kroupa_distribution_1993} . The magenta curve assume a top-heavy IMF for Population III stars with a flat slope in the mass range 30–120\MSun, while the blue curve includes also stars in the range 100–280\MSun to add PISNe.
    For each model, the curve reaches the observed O/H in \targetshort after 4.05 Myr.
    The combined CNO pattern of \targetshort is better matched by models with underlying top-heavy IMF. 
    % For comparison, dashed lines show instead the fiducial `double burst model' tuned to reproduce highly nitrogen enhanced objects like GN-z11.    
    In order to reproduce the abundance patterns of \targetshort with a `standard' IMF, a more tailored `double burst model' (with two bursts separated by 100 Myr, dashed orange track) is needed, similar to what proposed by \citealt{kobayashi_ferrara_gnz11_2024} to explain the nitrogen enhancement in GN-z11.
    }
    \label{fig:CNO_models}
\end{figure}

Overall, in \targetshort we therefore possibly observe signatures of chemical enrichment mechanisms beyond that of pure CCSNe, as depicted by the relatively high N/O and low C/N.
Several scenarios have been explored to reproduce similar abundance patterns in high-z galaxies, including ejection of CNO-cycle processed material in stellar winds from Wolf-Rayet (WR), large mass-loss rates during the evolution of Very Massive Stars (VMS) or even Super Massive Stars (SMS), and contribution from fast-rotating Population III stars \citep[e.g.][]{Nagele_Umeda_N_gnz11_2023, watanabe_2024, vink_vms_2024, marques-chaves_Nitrogen_2024, Nandal_popIII_2024}.
Here, we compare the observed C/O, N/O, and C/N abundances in \targetshort with a set of GCE models from \cite{kobayashi_origin_2020, kobayashi_taylor_chemodyn_2023}, with updated carbon yields for WR stars.
In particular, we assume a single-burst model following similar prescriptions as in \cite{kobayashi_ferrara_gnz11_2024}, and explore the impact of different IMF shapes on the nucleosynthetic patterns of the most massive stars, and hence on the observed chemical abundances.
In Figure~\ref{fig:CNO_models}, we over-plot the tracks from
three different realisations of the model onto the C/O vs O/H, N/O vs O/H, and C/N vs O/H diagrams, respectively.
In particular, the blue track represents the predictions assuming a standard \cite{kroupa_IMF_2008} IMF (with high mass-end slope $x=1.3$), whereas the magenta and red tracks assume a top-heavy IMF ($x=0$) of Population III stars with upper mass limits of 120\MSun and 280\MSun, respectively. 
Each model track reaches the observed value of O/H for \targetshort after $4.05$~Myr.

Evolutionary tracks with underlying top-heavy IMF provide the best match to the position of \targetshort in all three diagrams simultaneously, with WR stars producing the required enrichment in N/O (magenta track). 
We note that the red track, which includes a contribution from pair-instability supernovae (PISNe) of massive PopIII progenitors ($>150$\MSun), also provides a good match to the observed abundances, maintaining the C/O to a level more consistent with that observed in \targetshort, while slightly underpredicting (overpredicting) N/O (C/N). 
Additional abundance measurements of elements like Sulphur, Argon, or Iron would be required to better constrain the possible signatures of PISNe and discriminate their relative contribution.

% \todo{TBC}
In addition, we also consider a `dual burst model' (dashed orange track in Figure~\ref{fig:CNO_models}) that reproduces the abundances in \targetshort without the need to invoke top-heavy IMFs, similarly to what has been developed by \cite{kobayashi_ferrara_gnz11_2024} to match the abundance patterns observed in GN-z11.
Contrary to `single burst models', this model assumes a more tailored star-formation history, which is halted for a period of $\sim 100$Myr between two strong bursts.
Such a model produces a quick enhancement of both C/O and N/O via WR stars just after the second burst, which occur within a pre-enriched ISM from the first star-formation episode. 
While N/O can be regulated (and lowered) by increasing the timescale of star-formation for the second burst (making it `less extreme'), this model slightly overpredicts the C/N level compared to what observed in \targetshort. 

Finally, we display the measured Ne/O versus the O/H abundances for \targetshort in the bottom right panel of Figure~\ref{fig:CO_OH}, where we compare our measurement at z$\sim9.4$ with the values inferred for the galaxy ERO-4590 at z$\sim8.5$ by \cite{arellano-corodva_2023}, as well as with measurements in blue compact dwarfs and extragalactic giant HII regions from \cite{vale_asari_bond_2016}, and in HII regions of local galaxies from the CHAOS sample as studied in \cite{berg_chaos_2020}. 
% All measurements in Figure~\ref{fig:CO_OH} are based on Te abundance determinations.
As neon is an $\alpha$-element, its production mechanisms and timescales are expected to closely follow that of oxygen, with no significant trend in Ne/O as a function of metallicity.
Indeed, the Ne/O abundance measured in \targetshort is fully in agreement with the low-metallicity tail in the diagram as probed by local blue compact dwarf galaxies. 
Although a possible evolution in the Ne/O abundance pattern at $z>3$ has been suggested on the basis of the observed scaling relation between stellar mass and the \NeIII/\OII line ratio \citep{shapley_mzr_2023}, and \cite{isobe_CNO_2023} find that significantly lower Ne/O ratios in some z$>6$ galaxies can be explained by models of CCSNe ejecta including massive (>30\MSun) progenitors,  
here we find no evidence for a deviation in the Ne/O abundance in \targetshort compared to the chemical abundance patterns of local, metal-poor compact galaxies.

%of a young galaxy at high-redshift.

% \begin{figure}
%     \centering
%     % \includegraphics[width=0.98\columnwidth]{figs/NO_plots_z94_te_opt.pdf}
%     \includegraphics[width=0.98\columnwidth]{figs/Ne_O_plots_z94_te_opt.pdf}
%     \caption{Ne/O abundance for \targetshort, compared to other \Te-based measurements in both local and high redshift galaxies.}
%     \label{fig:N_NE_O}
% \end{figure}

% \section{Modelling the Ly$\alpha$ break and implications for Reionization}
\section{Ly$\alpha$ emission  and implications for Reionization}
\label{sec:ly_break}

\begin{figure}
    \centering
    \includegraphics[width=0.95\columnwidth]{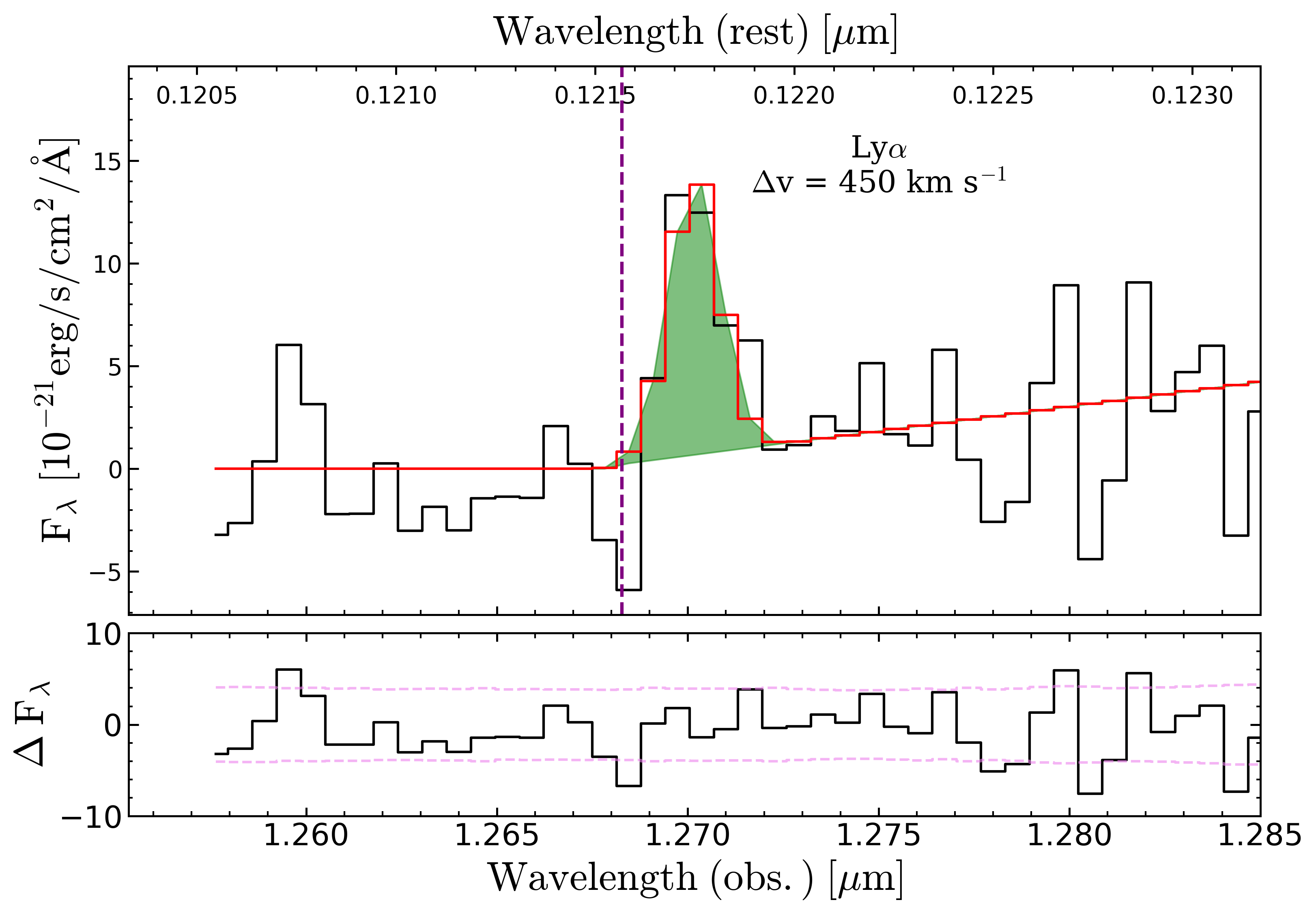}
    \caption{Zoom-in on the tentative ($\sim2.7\sigma$) detection of the \Lyalpha emission line in the G140M spectrum of \targetshort.
    The emission feature is observed shifted by $\sim450$~km s$^{-1}$ redward of the systemic \Lyalpha at redshift z$=9.4327$.}
    \label{fig:lya_g140m}
\end{figure}

% In the case of Gs-z9 however, we can try to retrieve additional information from the G140M spectrum, whose combined 1210+3215 data deliver enough signal-to-noise to attempt the fit of the \Lyalpha break at medium (R$\sim1000$) resolution.
Despite the MSA detector gap cutting the G140M spectrum just blueward of the systemic \Lyalpha wavelength, it still covers the onset of the \cite{Gunn_Peterson_IGM_1965} absorption trough.
% To mitigate the effect of the much noisier G140M spectrum compared to the prism spectrum (and the fact that it extends only to $\sim$1800$\AA$ rest-frame), we restrict the fit to the region around the break, i.e. up to $\sim$1500$\AA$.
% % we impose a stronger Gaussian prior on the UV $\beta$ slope (centred on the value inferred from the prism, with $\sigma$=0.2), while fixing \textbf{x$_{H_{I}}$}=1, R$_{\text{ion}}$=0, and letting N$_{HI}$ free to vary.
% The results of the G140M fit are shown in Figure XY.
% The best-fit solution again constrains the IGM to be fully neutral within the uncertainties, with negligible ionised bubble and no evidence for dense local absorbers. 
Notably, an emission feature stands above the continuum just redward of the systemic redshift \Lyalpha at $z=9.4327$.
If we interpret this feature as \Lyalpha emission, we can put more stringent constraints on the shape of the transmission curve, the escape fraction of \Lyalpha photons, and the possible existence of an ionised bubble around the system.
We fit this emission line with a single Gaussian component and model the underlying continuum with a broken linear component as shown in Fig.~\ref{fig:lya_g140m}, measuring a flux of 2.18 $\times$10$^{-19}$ erg~s$^{-1}$~cm$^{2}$, significant at $2.7\sigma$ ($2.3\sigma$ based on bootstrapping\footnote{the same feature is detected at $2.9\sigma$ in the 3-pixel boxcar extracted spectrum ($2.2\sigma$ with bootstrapping)}). The emission line centroid is redshifted by $450\pm145$ km/s with respect to the systemic redshift, the FWHM of the line is $380 \pm 130 $ km/s, and the estimated rest-frame EW of the line above the G140M continuum level is $31\pm16$~$\AA$.

Based on the theoretical ratio between (unattenuated) \Lyalpha and \Hbeta fluxes (\Lyalpha/\Hbeta$=25.1$ for case B recombination, T=20,000~K, n$_{e}$=650~cm$^{-3}$), and assuming the \Hbeta flux measured from the G395M spectrum, we derive an intrinsic \Lyalpha flux of 1.89 $\times$10$^{-17}$ erg~s$^{-1}$~cm$^{2}$, from which we infer that the \Lyalpha emission has been attenuated down to $\sim 1$ percent of its intrinsic flux.
Such a low \Lyalpha escape fraction, coupled with the presence of emission observed only significantly redward of the systemic wavelength, and the negligible dust attenuation, provides evidence against the presence of a large ionised region surrounding \targetshort. 
Such an ionised bubble in fact would likely allow a higher fraction of \Lyalpha photons to escape (given the intrinsic predicted flux) and, under the assumption of a broad, symmetric intrinsic \Lyalpha profile, would have also caused the line profile to spill across to the blue of the systemic redshift.

In addition to the detection of \Lyalpha emission, we can leverage the robust redshift determination enabled by the bright emission lines detected in \targetshort to remove the redshift-degeneracy in modelling the Ly$\alpha$ break as observed (at high signal-to-noise) in the PRISM spectrum, and get more insights into the properties of the surrounding medium, the possible extent of any ionised region, and the presence (or absence) of large column densities of neutral gas along the line of sight.
We model the attenuation due to neutral IGM along the line of sight following the prescriptions from \cite{inoue_igm_2014}, and include treatment of the the full optical depth (including damping-wing absorption) of Ly$\alpha$ photons following
\cite{Dijkstra_IGM_2014, Mason_Rion_2020} \citep[for further details, see][]{witstok_bubbles_2024}, under the assumptions of the gas in the IGM having mean cosmic density and T$=1$~K, while gas in the ionised bubble is fully ionised (residual neutral fraction x$_{HI}=10^{-8}$) with T$=10^4$~K. 
%the residual neutral fraction profile scales as $\propto r^2$, with normalisation x$_{HI}$(r$=0.1$~pMpc)$=10^{-8}$.
We apply such transmission curve to a power-law spectrum modelling the continuum blueward of 1500$\AA$ rest-frame\footnote{This introduces the slope and normalisation of the UV continuum as additional (nuisance) parameters.}, to avoid the region contaminated by emission lines, and we include the contribution from the \Lyalpha emission matching flux, velocity and width as observed in the G140M spectrum `a-posteriori' into the model (i.e., after attenuating the intrinsic spectrum), as it could possibly affect the observed profile.\footnote{We note that while such a weak \Lyalpha line is not expected to be seen directly at the PRISM resolution, where it becomes fully blended with the break, it does affect its spectral shape, e.g. \citet{jones_jades_lya_2024}.}
We then convolve the attenuated model by the PRISM LSF provided by \cite{de_graaff_jades_2024} and resample it to the PRISM wavelength grid before comparing to the data.
In order to properly account for the inter-correlation between adjacent wavelength bins in the fitting procedure, the $\chi^2$ goodness-of-fit statistics involves the covariance matrix empirically derived from 2000 bootstrapped realisations of the PRISM spectrum leveraging the 186 available individual exposures, as detailed in Jakobsen et al. (in prep.).
In particular, 
\begin{equation}
\chi^2 = {\vec R}^T  [{\bf Cov\kern1pt F_\lambda}]^{-1} {\vec R}
\label{eq:chi_squared}
\end{equation}
where $[{\bf Cov\kern1pt F_\lambda}]^{-1}$ is the inverted covariance matrix and ${\vec R}$={\bf F$_{\lambda}$} - {\bf F}$_{\text{model}}$ is the vector of residuals.

Here in particular we explore which constraints can be set on the size of the ionised bubble and the amount of neutral gas along the line of sight in a simple scenario where the intrinsic stellar spectrum (modelled by a `power-law') is attenuated by a fully neutral IGM.
In this model, the fraction of neutral hydrogen is set to unity (i.e. x$_{H_{I}}=1$), while the size of the surrounding ionised region R$_{ion}$ 
(expressed in physical Mpc) and the column density of neutral gas are free to vary.
We assume the neutral gas to pertain to the local galaxy ISM or the surrounding circum-galactic medium (CGM), similar to what is observed in DLA systems, hence we fix the redshift of such DLA component to the systemic redshift of \targetshort. 
The Ly$\alpha$ optical depth due to the DLA component is
modelled as a Voigt profile (following the approximation of \citealt{Tasitsiomi_lya_2006}) and is solely parametrised by the column density of neutral atomic hydrogen N$_{HI}$ (we assume an ISM temperature T=$10^4$~K, and no additional contribution due to a turbulent medium).

\begin{figure}
    \centering
    \includegraphics[width=0.95\columnwidth]{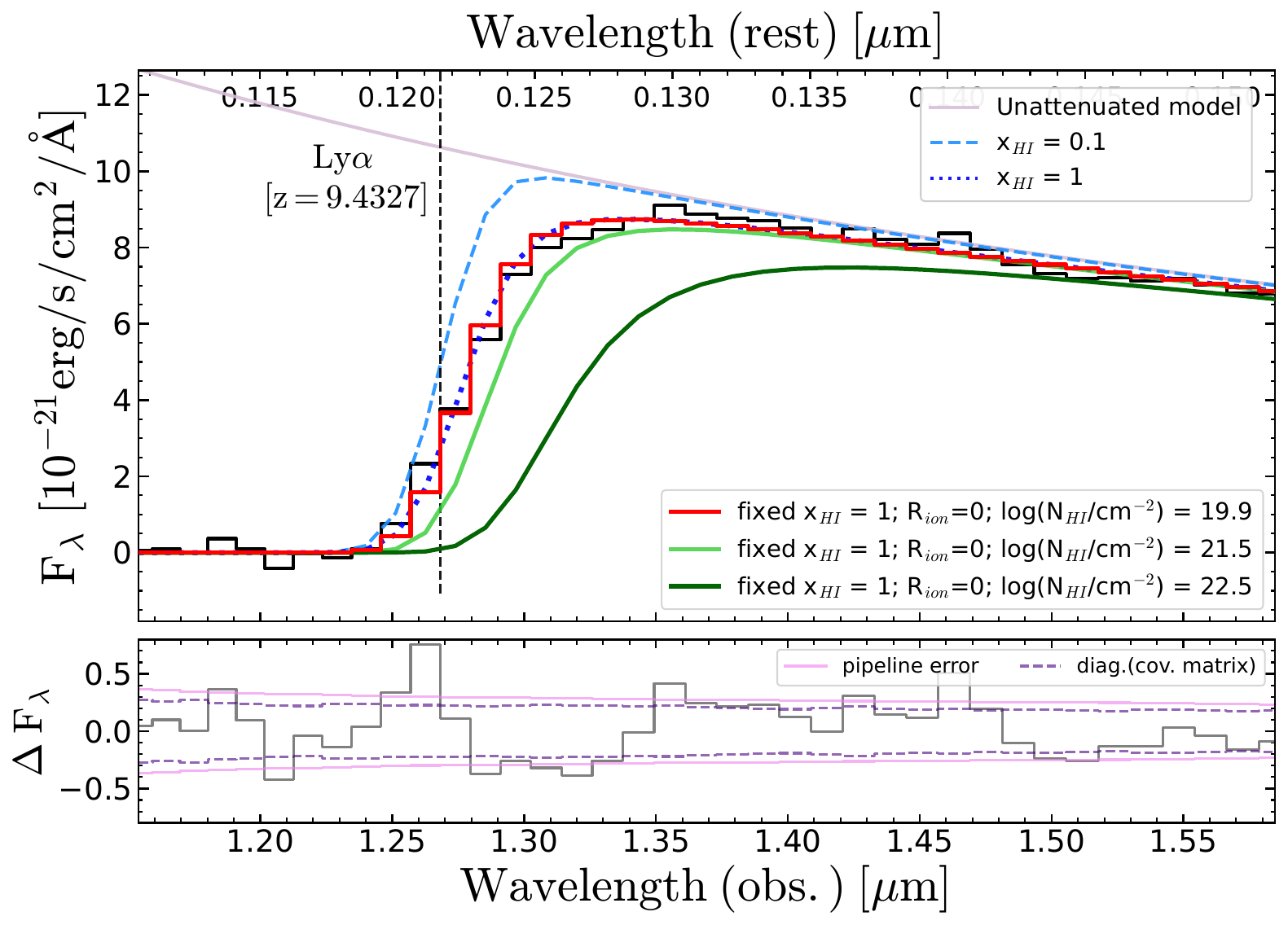}
    \includegraphics[width=0.48\columnwidth]{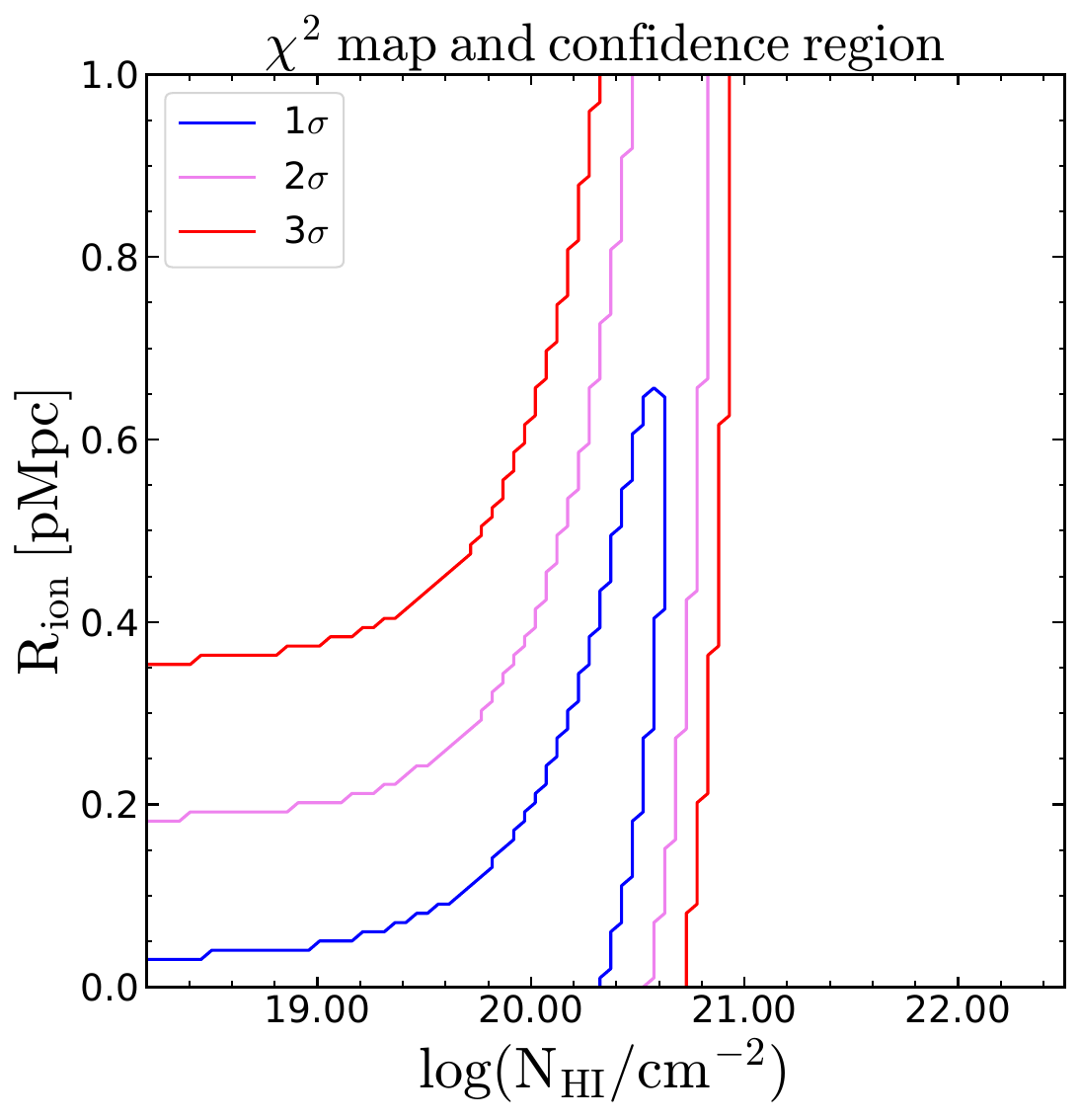}
    \includegraphics[width=0.48\columnwidth]{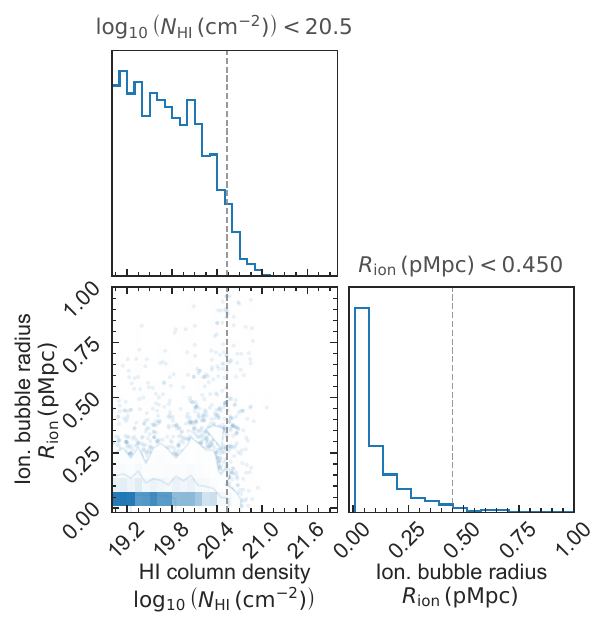}
    \caption{
    Model of the Ly$\alpha$ break in the PRISM spectrum.
    Upper panel: The red line represents the best-fit from a model in which we attenuate a power-law spectrum by applying the transmission curve from a fully neutral IGM (x$_{HI}$ fixed to 1) while fitting for the size of any ionised bubble (R$_{ion}$) surrounding \targetshort and for the column density of a local DLA system along the line of sight (N$_{HI}$). 
    The residuals from the fit are compared with both the pipeline error spectrum and the square root of the diagonal terms of the covariance matrix described in Section~\ref{sec:ly_break} 
    (Jakobsen et al., in prep.).
    The formal best-fit values provide R$_{ion}\sim0$ and log(N$_{HI}$/cm$^{-2}$)=19.9. We note that including attenuation from such a column density of neutral gas placed at the galaxy systemic redshift does not provide significant improvements to the model compared to only applying attenuation from purely neutral IGM (blue dotted line). For comparison, the light and dark green curves show the spectral shape obtained by including DLAs with log(N$_{\text{HI}}$/cm$^{-2}$)$=21.5$ and $22.5$, respectively.
    Bottom-left panel: Contour levels of the $\chi^2$-map on the log(N$_{\text{HI}}$) versus R$_{ion}$ parameter space, highlighting $1\sigma$, $2\sigma$, and $3\sigma$ confidence regions.
    The column density of any DLA at the systemic redshift is constrained to log(N$_{\text{HI}}$/cm$^{-2}$)$<20.75$ at $99.7\%$ confidence. 
    Bottom-right panel: Joint contours and marginalised posterior probability density function for the same parameters. 
    The quoted upper limits report the $95$th percentiles of the posterior distributions.
    }
    \label{fig:LYA_break}
\end{figure}

The best-fit model (red curve in Figure~\ref{fig:LYA_break}, $\chi^2 = 54.57$ for $34$ degrees of freedom) predicts log(N$_{HI}$/cm$^{-2}$)$=19.92$ and R$_{ion}=0$ as formal best-fit values, delivering a good match to the observed spectral profile.
We explore the confidence region for the two main parameters by mapping the $\chi^2$ over the R$_{ion}=0$ versus N$_{HI}$ parameter space, taking the $\Delta \chi^2 = (\chi^2-\chi^2_{\text{best}}) = 1,4,9$ contour levels as the $1\sigma$, $2\sigma$ and $3\sigma$ confidence intervals, respectively. 
As shown in the bottom left panel of Figure~\ref{fig:LYA_break}, any local (i.e. at systemic redshift) column density of neutral gas is strongly constrained to be lower than log(N$_{HI}$/cm$^{-2}$)$\lesssim20.75$ at $99.7\%$ confidence.
Very similar constraints are obtained by employing the \textsc{multinest} algorithm \citep{feroz_multinest_2009}
to sample the posterior probability density function of each parameter\footnote{Our likelihood model includes the covariance matrix as in Equation~\ref{eq:chi_squared}, and we assume flat linear prior on log(N$_{HI}$/cm$^{-2}$), while log-uniform prior on R$_{ion}$ between linear values of $10^{-2}$ and $10^{0.5}$}. The bottom-right panel of Figure~\ref{fig:LYA_break} shows the corner plot for R$_{ion}$ and log(N$_{HI}$/cm$^{-2}$)), from which we quote an upper limit on log(N$_{HI}$/cm$^{-2}$)$<20.5$ as estimated from the $95$th percentile of the marginalised posterior distribution.

Interestingly, our inferred upper limits are much lower than what other recent observations of $z\gtrsim10$ galaxies have suggested, reporting detection of local column densities of neutral gas as high as log(N$_{HI}$/cm$^{-2}$)=22.5 which have been interpreted as evidence for the presence of large reservoirs of cold gas available to sustain early star-formation and galaxy growth \citep[e.g.][]{d_eugenio_gsz12_2023, heintz_DLA_2023, Heintz_primal_2024, hainline_zgtr10_jades_2024, carniani_z14_2024}.
Such large column densities in fact would push the onset of the \Lyalpha absorption trough redward than the systemic redshift and significantly soften the observed spectral break: this is illustrated for instance by the light and dark green curves in Figure~\ref{fig:LYA_break}, which show how the transmitted spectrum would look like in the presence of local DLAs of log(N$_{HI}$/cm$^{-2}$)=21.5 and 22.5, respectively, resulting in very different spectral shapes than observed in the PRISM spectrum of \targetshort.

To test the amount of degeneracy between the column density of neutral gas in the CGM/ISM and the hydrogen neutral fraction in the IGM and its impact on the results of our fit, we have performed a run by leaving also x$_{H_{I}}$ free to vary between 0 and 1, finding that the fit still favours a solution with high x$_{H_{I}}$ ($0.97\substack{+0.02 \\ -0.05}$), little R$_{ion}$ ($0.16 \substack{+0.34 \\ -0.11}$~pMpc), and low N$_{HI}$ (log(N$_{HI}$/cm$^{-2}$)=$19.37 \substack{+0.95 \\ -1.61}$), i.e. the transmission curve from the IGM absorption can not be exactly mimicked by local ISM/CGM absorption in the form of a DLA component at any column density.
We note that in such a scenario the inferred value for x$_{H_{I}}$ is slightly higher than (though consistent with) that measured by \cite{Umeda_damping_wing_2024} on an average, stacked JWST/PRISM spectrum of z~$>9$ galaxies. 
Furthermore, we have tested that only a fine-tuned model with the DLA component placed at a redshift z[DLA]~$\sim9.3$ and large column density (log(N$_{HI}$/cm$^{-2}$)$\gtrsim 22.3$) can mimic the transmission curve of a fully neutral IGM (delivering a comparable goodness of fit), but this would likely imply the presence of a foreground system which is not seen in NIRCAM imaging (see also Jakobsen et al., in prep.). 

For what concerns the constraints inferred from our simple model on the size of a possible ionised region around \targetshort, the bottom left panel of Figure~\ref{fig:LYA_break} shows that for any value of log(N$_{HI}$/cm$^{-2}$)$\lesssim20$ R$_{ion}$ is constrained to be $\lesssim 0.1$pMpc at $1\sigma$ confidence ($\lesssim 0.4$pMpc at $3\sigma$ confidence), whereas only within a rather fine-tuned interval of log(N$_{HI}$/cm$^{-2}$) values ($20.3 \lesssim$log(N$_{HI}$/cm$^{-2}$)$\lesssim 20.8$) the $1\sigma$ confidence region extends up to $0.6$~pMpc (
with $2\sigma$ and $3\sigma$ intervals extending beyond $1$pMpc).
Again, similar constraints are obtained from the marginalised posterior PDF for R$_{ion}$, from which we quote R$_{ion}<0.447$pMpc as the $95th$ percentile of the distribution.

We can compare the constraints on R$_{ion}$ derived from our modelling of the \Lyalpha spectral break with what we could expect for a galaxy like \targetshort on the basis of simple arguments involving the age of the system and its ionizing efficiency.
First, we estimate the ionizing photons production efficiency of \targetshort leveraging the measured flux of \Hbeta. %and under the assumption of case B recombination.
We follow e.g. \cite{saxena_lya_2023,simmonds_xi_ion_2023, witstok_lyA_z8_2024} to infer
\begin{equation}
    \xi_{\text{ion}} = 3 \times 10^{12}\ \text{erg}^{-1}\ L_{\Hbeta}/L_{\nu,1500}\ ,
\end{equation}
measuring the UV luminosity density at $1500\AA$ rest-frame and under the assumption of case B recombination and zero escape fraction of Lyman continuum photons ($f_{\text{esc,LyC}}$=0, hence yielding a lower limit), finding log($\xi_{\text{ion}}$/erg$^{-1}$Hz)=25.64, in agreement with typical values for bright (M$_\text{UV}$<-20) galaxies in the EoR \citep[e.g.][]{endsley_sfh_2023}. 
Following e.g., \cite{Mason_Rion_2020, witstok_bubbles_2024}, we can then estimate the size of the surrounding ionised region that would be produced by a galaxy like \targetshort from
\begin{equation}
\label{eq:rion_growth}
\text{R}_{ion}(t) \approx \biggl(\frac{3 f_{\text{esc,LyC}}\dot{N}_{\text{ion}}t}{4\pi \bar{n}_{H}(z)} \biggr) \,
\end{equation}
where $\bar{n}_{H}$(z) is the mean hydrogen number density at redshift z and $\dot{N}_{\text{ion}}$ the production rate of ionising photons.
Assuming that the ionising photon production is powered by a very recent burst of star-formation occurred over the past $5$~Myr, as suggested by both the steeply rising SFH derived in Section~\ref{sec:SED_fitting} and the chemical abundance patterns discussed in Section~\ref{sec:discussion}, and a $f_{\text{esc,LyC}}=5\%$ \citep{Finkelstein_LyC_escape_2019} would produce an ionised bubble with size R$_{ion}\approx0.1$pMpc ($0.05$-$0.12$pMpc if assuming $f_{\text{esc,LyC}}$ ranging between $1\%$ and $10\%$), in good agreement with the $1\sigma$ confidence limits inferred from our modelling of the \Lyalpha break for the majority of allowed N$_{HI}$ values.
These values are also in line with what predicted for \Lyalpha emitters (LAEs) with similar physical properties to \targetshort discovered at slightly lower redshift \citep[see e.g. $\sim7-8$ objects in][]{witstok_bubbles_2024, witstok_lyA_z8_2024}, whose (sometimes significantly) larger inferred sizes of ionised bubbles (constrained by large observed \Lyalpha/\Hbeta ratios and small velocity offsets) requires to include the contribution from either faint neighbour companions or older galaxies, to assume bursty SFHs, or to invoke particularly favourable geometrical configurations. 
Summarising, we have found indications for \targetshort to be surrounded by a small ionised bubble, as suggested by the low escape fraction of \Lyalpha photons ($\sim 1\%$), the relatively large velocity offset ($\sim 450$ km/s), and the constraints set by modelling the damping wing of the \Lyalpha break.
Despite the high production efficiency of ionising photons powered by its recent SFH, this galaxy is therefore unlikely to have contributed significantly to Reionization at the epoch of observations.
Nonetheless, it is possible that the presence of outflowing gas (as also suggested by the velocity shift observed in the \CIV emission) is helping to create channels through which \Lyalpha photons can escape from a fully neutral IGM; the presence of such outflows has been indeed suggested as almost ubiquitous in early galaxies with high sSFR \citep{ferrara_BM_2024a, ferrara_BM_2024b}. 

Finally, we note that, despite providing a reasonably good match to the data, our simplified baseline model assuming x$_{H_{I}}$=1 is not capable of perfectly reproducing the observed spectral shape of \targetshort in the PRISM spectrum, as clear residuals are observed for instance between $\sim1.3$--$1.35\mu$m, where the profile softens before turning-over (see Figure~\ref{fig:LYA_break}). 
Indeed, we stress that in order to place more stringent constraints on the neutral fraction of the IGM it is critical to have a better understanding of the systematics involved in modelling the intrinsic UV continuum and the exact shape of the \Lyalpha emission that emerges from the galaxy, as well as the impact of possible fluctuations in the hydrogen density associated to large scale structures at these redshifts \citep[e.g.][]{smith_thesan_lya_LSS_2022}.
Moreover, a proper modelling of the contribution to attenuation due to local absorbers would require the precise knowledge of the geometry and density profiles of the intervening gas clouds \citep{gronke_lya_break_2017, hutter_astraeus_reionisation_2023}.
A more detailed discussion on the damping wing of the \Lyalpha absorption profile in \targetshort, the inferred constraints on x$_{H_{I}}$, and the implications on the history of cosmic Reionization, is presented in Jakobsen et al., (in prep.).

\section{Summary}
\label{sec:conclusions}
We have presented ultra-deep \jwst/NIRSpec observations of \targetshort, a luminous (M$_{\text{UV}}$=-20.43) galaxy at redshift $z\sim9.4327$. Combining datasets from JADES programmes PID 1210 and PID 3215 provides a total of 72 hours of observations in PRISM/CLEAR mode, 44 hours in G395M/F290LP mode, and 16 hours in G140M/F070LP mode.
Leveraging the high signal-to-noise of the spectrum (Figure~\ref{fig:plot_spectra}) and the detection of multiple emission lines in both rest-frame optical and UV regimes (Figure~\ref{fig:prism_fit} and \ref{fig:G395M_fit}), we characterise the ISM properties and the star-formation history of this source, observed only $\sim500$ Myr after the Big Bang.

The main results discussed in the paper are summarised as follows : 

\begin{itemize}
    \item The system is young (light-weighted age $\sim3$~Myr, mass-weighted age of $\sim30$ Myr), compact (R$_{e} \sim 110$~pc), and exhibit a steeply rising star-formation history over the past $3$--$10$~Myr with high star-formation rate surface density ($\Sigma_{\text{SFR}}\sim 72 $~\MSun~yr$^{-1}$~kpc$^{-2}$), suggesting that the majority of mass assembly has occurred during a recent and concentrated burst of star-formation (Figure~\ref{fig:bagpipes}).
    Negligible dust attenuation is inferred from both SED fitting and Balmer lines (Figure~\ref{fig:BD}).
    \item Emission line ratios from the rest-frame UV are indicative of a hard radiation field powered by young, massive, and metal-poor stars (Figure~\ref{fig:diagnostic_diagrams}), in particular when the abundance patterns of the model grids (e.g. in C/O) are matched to those inferred independently from the emission lines. 
    However, the tentative detection of the extremely high-ionization \NeVL line (Figure~\ref{fig:prism_fit}) suggests that contribution to ionization from the NLR of an AGN, or from shocks driven by SNe explosions, can not be excluded (Figure~\ref{fig:diagnostics_N3}).
    \item The measured C/O abundance (log(C/O)$=-0.90$, [C/O]$=-0.64$) lies at the lowest envelope of the distribution of local and high-redshift galaxies at similar metallicity (12+log(O/H)$\sim7.5$), in agreement with a rapid, recent history of chemical enrichment dominated by yields of core-collapse Supernovae from massive stars progenitors (Figure~\ref{fig:CO_OH}).
    \item However, the detection of \NIIIL emission in the PRISM spectrum suggests low C/N and hence over-abundant N/O (log(N/O)$=-0.93$, [N/O]$=-0.07$) compared to the average plateau occupied by local, low-metallicity galaxies, similar to (though not as extreme as) nitrogen-enriched galaxies at $z>6$ recently observed by \emph{\jwst} (Figure~\ref{fig:CO_OH}). This is corroborated also by the marginal detection of \NIVL[1483] in the 3-pixel extracted G140M spectrum. 
    \item The observed CNO abundance patterns are well reproduced by `single-burst' chemical evolution models including contribution from very massive stars (M$_{\star}\gtrsim 100$M$_{\odot}$), possibly unveiling the short timescale signatures of enrichment from Wolf-Rayet stars or pair-instability Supernovae progenitors, in line with a top-heavy IMF scenario regulating the process of star-formation in the earliest galaxies (Figure~\ref{fig:CNO_models}).
    More `fine-tuned' SFHs (e.g. characterised by two bursts separated by a $\sim100$~Myr period of quiescence) are required to match the CNO abundance patterns assuming a standard IMF.
    \item By modelling the \Lyalpha spectral break under the assumption of a fully neutral IGM, we rule out the presence of very-high column density local absorbers (log(N$_{HI}$/cm$^{-2}$)$\lesssim20.75$ at $99.7\%$ confidence, Figure~\ref{fig:LYA_break}). 
    We also report tentative detection of \Lyalpha emission in the G140M spectrum (Figure~\ref{fig:lya_g140m}), redshifted by $\sim450$~km/s from systemic, from which we infer an escape fraction of \Lyalpha photons from IGM attenuation of $\sim1$ percent.
    Together with the constraints from the \Lyalpha break, and in line with the young age of the system, this suggests the size of any ionised region surrounding \targetshort to be $\lesssim0.1$--$0.2$~pMpc and that, despite its high ionising photons production efficiency, this system have not yet significantly contributed to cosmic Reionization at the time of observations.
\end{itemize}

Sources like \targetshort currently represent the best opportunities to directly investigate the physics underlying the early phases of galaxy assembly, their evolution, and the environment they live in.
The richness of spectral features and the peculiar chemical abundances observed in this and other high-redshift systems challenge our understanding of the nature of massive stars powering the extreme emission lines seen in the spectra and possibly hint at the role of early accretion onto (super-)massive black holes.
In this sense, the present analysis also highlights the wealth of information delivered by deep observations that simultaneously probe rest-frame UV and optical spectra of $z\sim6$--$10$ galaxies. 
However, despite the enormous advancements already brought by \jwst, the intrinsic faintness of most rest-frame UV stellar and nebular features currently hinders their detailed characterisation within the context of the more general high-redshift galaxy population.
Alongside the identification and follow-up of new, bright high-redshift candidates, deeper spectroscopy at medium-high resolution probing the UV continuum and emission line features over large samples will therefore be key to gradually move from the study of the most luminous systems towards a broader understanding of the physical properties of galaxies in the early Universe.

\begin{acknowledgements}
We thank the anonymous referee for the insightful comments that contributed to strengthen the analysis and improved the clarity of the paper.
We are grateful to Danielle Berg, Matilde Mingozzi, Anna Feltre, Bethan James, John Chisholm, Evan Skillman, Alessandro Marconi, Charles Steidel, Andrea Ferrara, and Jorick Vink for enlightening conversations.

MC acknowledges support by the European Southern Observatory (ESO) Fellowship Programme.
JW, FDE, RM, JS, \&  WB  acknowledge support by the Science and Technology Facilities Council (STFC), ERC Advanced Grant 695671 ``QUENCH", and by the UKRI Frontier Research grant RISEandFALL. RM also acknowledges funding from a research professorship from the Royal Society.
The Cosmic Dawn Center (DAWN) is funded by the Danish National Research Foundation under grant DNRF140.
ECL acknowledges support of an STFC Webb Fellowship (ST/W001438/1).
JC, AJB \& AC acknowledge funding from the ``FirstGalaxies" Advanced Grant from the European Research Council (ERC) under the European Union’s Horizon 2020 research and innovation programme (Grant agreement No. 789056).
SC \& GV acknowledge support by European Union’s HE ERC Starting Grant No. 101040227 - WINGS.
SA acknowledges grant PID2021-127718NB-I00 funded by the Spanish Ministry of Science and Innovation/State Agency of Research (MICIN/AEI/ 10.13039/501100011033).
This research is supported in part by the Australian Research Council Centre of Excellence for All Sky Astrophysics in 3 Dimensions (ASTRO 3D), through project number CE170100013.
DJE, BDJ, BR, and CNAW are supported by JWST/NIRCam contract to the University of Arizona NAS5-02015.
DJE is also supported as a Simons Investigator.
BER acknowledges support from the NIRCam Science Team contract to the University of Arizona, NAS5-02015, and JWST Program 3215.
ST acknowledges support by the Royal Society Research Grant G125142.
H{\"U} gratefully acknowledges support by the Isaac Newton Trust and by the Kavli Foundation through a Newton-Kavli Junior Fellowship.
The research of CCW is supported by NOIRLab, which is managed by the Association of Universities for Research in Astronomy (AURA) under a cooperative agreement with the National Science Foundation.
\end{acknowledgements}

%-------------------------------------------------------------------

% \bibliography{hzm_bib}{}
\bibliography{aa}{}
\bibliographystyle{aa}

\end{document}